\documentclass[onecolumn]{aastex631}

\submitjournal{ApJ}

\shorttitle{SSC CRs}
\shortauthors{Fraija, Dainotti}

\usepackage{epstopdf}
\usepackage{ulem}
\usepackage{amssymb}
\usepackage{times}
\usepackage[usenames,dvipsnames]{pstricks}
\usepackage{psfrag}
\usepackage{lipsum}
\usepackage{natbib}
\usepackage{textcase}

\usepackage{booktabs}
\usepackage{comment}
\usepackage{graphicx,subfigure}
\usepackage{textgreek}
\usepackage{hyperref}

\def\apjl{ApJ}
\def\apjs{ApJS}
\def\ao{Appl.~Opt.}
\def\aap{A\&A}

\newcommand{\be}{\begin{equation}}
\newcommand{\ee}{\end{equation}}
\newcommand{\bary}{\begin{eqnarray}}
\newcommand{\eary}{\end{eqnarray}}

\graphicspath{{images/}}

\begin{document}

\title{Off-axis Afterglow Closure Relations and Fermi-LAT Detected Gamma-Ray Bursts}

\author[0000-0002-0173-6453]{N. Fraija}
\affiliation{Instituto de Astronom\' ia, Universidad Nacional Aut\'onoma de M\'exico,\\ Circuito Exterior, C.U., A. Postal 70-264, 04510 M\'exico City, M\'exico}

\author[0000-0003-4442-8546]{M. G. Dainotti}
\affiliation{National Astronomical Observatory of Japan, 2-21-1 Osawa, Mitaka, Tokyo 181-8588, Japan}
\affiliation{Space Science Institute, Boulder, CO, USA}
\affiliation{The Graduate University for Advanced Studies, SOKENDAI, Shonankokusaimura, Hayama, Miura District, Kanagawa 240-0193, Japan}

\author{D. Levine}
\affiliation{Department of Astronomy, University of Maryland, College Park, MD 20742, USA}

\author[0000-0002-2516-5739]{B. Betancourt Kamenetskaia}
\affiliation{Technical University of Munich, TUM School of Natural Sciences, Physics Department, James-Franck-Str 1, 85748 Garching, Germany}
\affiliation{Max-Planck-Institut f\"ur Physik (Werner-Heisenberg-Institut), F\"ohringer Ring 6, 80805 Munich, Germany}

\author{A. Galvan-Gamez}
\affiliation{Instituto de Astronom\' ia, Universidad Nacional Aut\'onoma de M\'exico,\\ Circuito Exterior, C.U., A. Postal 70-264, 04510 M\'exico City, M\'exico}

\begin{abstract}

Gamma-ray bursts (GRBs) are one of the most promising transient events for studying multi-wavelength observations in extreme conditions. Observation of GeV photons from bursts would provide crucial information on GRB physics, including the off-axis emission. The Second Gamma-ray Burst Catalog (2FLGC) was announced by the Fermi Large Area Telescope (Fermi-LAT) Collaboration. This catalog includes 29 bursts with photon energy higher than 10 GeV.  While the synchrotron forward-shock model has well explained the afterglow data of GRBs, photon energies greater than 10 GeV are very difficult to interpret within this framework. To study the spectral and temporal indices of those bursts described in 2FLGC,  \cite{2022ApJ...934..188F} proposed the closure relations (CRs) of the synchrotron self-Compton (SSC)  emitted from an on-axis jet which decelerates in stellar-wind and the constant-density medium. In this paper,   we extend the CRs of the SSC afterglow from an on-axis scenario to an off-axis, including the synchrotron afterglow radiation that seems off-axis. In order to investigate the spectral and temporal index evolution of those bursts reported in 2FLGC, we consider the hydrodynamical evolution with energy injection in the adiabatic and radiative regime for an electron distribution with a spectral index of $1<p<2$ and $2 < p$. The results show that the most likely scenario for synchrotron emission corresponds to the stellar wind whether or not there is energy injection and that the most likely scenario for SSC emission corresponds to the constant density when there is no energy injection and to the stellar wind when there is energy injection.

\end{abstract}


\section{Introduction}
One of the most energetic transients in the universe are gamma-ray bursts (GRBs). These transients, which can range in duration from a few milliseconds to hours, are categorized according to how long the prompt episode is \citep{1993ApJ...413L.101K, 1999PhR...314..575P, 2015PhR...561....1K}. The prompt emission modelled by the empirical ``Band" function \citep{1993ApJ...413..281B} is usually detected in the keV-MeV energy range. When a relativistic outflow produced by the core engine as predicted in the fireball scenario interacts with the external environment and transmits a significant portion of its energy to this external medium, it is known as ``afterglow" and is detectable from radio to GeV bands after the prompt gamma-ray emission \citep{1997ApJ...476..232M, 1999PhR...314..575P}.   The synchrotron radiation generating photons ranging in energy from radio to gamma rays \citep{1998ApJ...497L..17S,2009MNRAS.400L..75K, 2010MNRAS.409..226K,2013ApJ...763...71A, 2016ApJ...818..190F} and the synchrotron self-Compton (SSC) process, which produces photons at very high energies \citep[VHE $\geq$ 10 GeV;][]{2019ApJ...885...29F, 2019ApJ...883..162F,2021ApJ...918...12F, 2017ApJ...848...94F,2019arXiv191109862Z}, are the fundamental cooling mechanisms used to model the evolution of the temporal and spectral indexes of the multi-wavelength observations detected during the afterglow phase. The evolution of these indexes should follow the closure relations (CRs) of these fundamental cooling mechanisms.  These simple power-law (SPL) equations, denoted by the notation $F_\nu \propto t^{-\alpha}\nu^{-\beta}$, describe the relationship between the temporal index, $\alpha$, and the spectral index, $\beta$.  Each one of these equations, defined in each cooling regime ``fast" or ``slow" is derived in the radiative and adiabatic regimes, with and without energy injection for the on-axis emission, using assumptions of possible astrophysical environments, including the constant-density interstellar medium (ISM) or stellar-wind environment. \\  

Covering the first 10 years of operations (from 2008 to 2018 August 4),  \cite{Ajello_2019} presented the Second Gamma-ray Burst Catalog (2FLGC) with high-energy emission higher than $\ge100$~ MeV.   A sample of 169 GRBs with temporarily-extended emission and photon energies greater than a few GeV are included in the database.  In 2FLGC, the temporally extended emission was modeled by either a broken power-law (BPL) function with a temporal break of several hundred seconds or a SPL. Several authors have used the CRs expected in  different mechanisms and scenarios to explain the evolution of the LAT light curves. For instance, \cite{2020ApJ...905..112F} modelled the LAT light curve of GRB 160509A in the external shock scenario.  The evolution of temporal and spectral indexes that exhibited that temporally-extended LAT emission  was interpreted in the synchrotron forward-shock model and temporal break as the passage of the synchrotron cooling break through the Fermi-LAT band ($h\nu_{\rm LAT}=100\,{\rm MeV}$).  \cite{2019ApJ...883..134T} performed a comprehensive examination of the CRs in a sample of 59 chosen LAT-detected bursts while taking into account temporal and spectral indexes. The authors found that the classical synchrotron forward-shock emission often explains the spectral and temporal indices. However, this model still does not explain a large number of bursts.   Moreover, \cite{2019ApJ...883..134T} found that a large portion of bursts satisfies the CRs when the synchrotron scenario evolves in the constant-density medium, lies in the slow-cooling regime ($\nu^{\rm sync}_{\rm m}<\nu_{\rm LAT}<\nu^{\rm sync}_{\rm c}$) and requires a modest unusual value for the magnetic microphysical parameter of $\epsilon_B<10^{-7}$. The frequencies $\nu^{\rm sync}_{\rm c}$ and $\nu^{\rm sync}_{\rm m}$ corresponds to the cooling and characteristic spectral breaks, respectively.  \cite{2010MNRAS.403..926G}, on the other hand, analyzed the high-energy emission of eleven Fermi LAT-detected bursts until October of 2009. They came to the conclusion that the LAT light curves evolved as $\sim t^{-\frac{10}{7}}$, and therefore concluded that afterglows lied in the radiative regime for a hard value of spectral index $p\sim 2$.  \cite{2021ApJS..255...13D} investigated the presence of plateau phase in 2FLGC, which is naturally accounted for by a continuous injection of energy into the blastwave from the central engine \citep{2006Sci...311.1127D, 2006ApJ...636L..29P, 2006MNRAS.370L..61P, 2006ApJ...642..354Z, 2005Sci...309.1833B}.  The most favourable scenario to explain this phase was a slow-cooling regime in the ISM. Finally, \cite{2022ApJ...934..188F} presented the CRs of SSC forward-shock scenario in the adiabatic and radiative regime with and without energy injection into the blastwave for the stellar-wind and ISM. They found that the CRs of the SSC model could explain a significant fraction of bursts that can hardly be described with the synchrotron model scenario.   The analysis showed that ISM is preferred for the scenario without energy injection and the stellar-wind medium for an energy injection scenario. \\

In order to test the evolution of LAT light curves,  CRs have been extensively explored in the cases of radiative and adiabatic regime, with and without energy injection, and in ISM and stellar-wind environments only in the case of the on-axis emission, as mentioned above. However, now several authors seem to be interested in finding out the off-axis emission  given the multi-wavelength detection and theoretical modeling of GRB 170817A \citep{troja2017a, 2017Sci...358.1559K, 2017MNRAS.472.4953L, 2018MNRAS.478..733L, 2018ApJ...867...57R, 2017ApJ...848L..20M, 2017ApJ...848L...6L, 2019ApJ...884...71F, 2018ApJ...867...95H, 2019ApJ...871..200F}, short GRBs detected by Gamma-ray Burst Monitor \citep[GBM;][]{2019ApJ...876...89V} and Burst Alert Telescope \citep[BAT;][]{2020MNRAS.492.5011D} on board of the Fermi ad Swift satellite, respectively, with characteristics similar to GRB 170817A 
and a group of bursts exhibiting comparable off-axis emission features such as GRB 080503 \citep{2009ApJ...696.1871P,2015ApJ...807..163G}, GRB 140903A \citep{2016ApJ...827..102T, 2017ApJ...835...73Z}, GRB 150101B \citep{2018NatCo...9.4089T, 2018ApJ...863L..34B}, GRB 160821B \citep{2019MNRAS.489.2104T}, GRB 170817A \citep{2017Sci...358.1559K, 2017MNRAS.472.4953L, mooley, 2018ApJ...867...95H, 2019ApJ...884...71F} and SN2020bvc \citep{2020A&A...639L..11I}.\\ 

We extend the CRs of SSC afterglow model in the adiabatic and radiative scenario with and without energy injection emitted from an on-axis \citep{2022ApJ...934..188F} to off-axis jet. We analyze the SSC off-axis model in stellar-wind and constant-density medium, and the CRs as functions of the radiative parameter, energy injection index, and electron spectral index for $1<p<2$ and $ 2 < p$. The entire bursts modeled with a SPL and BPL function in the 2FLGC are selected. Keeping this in mind, the following is the structure of the paper: In Section \S\ref{sec2} we derive the CRs of synchrotron and SSC off-axis models evolving in the ISM and stellar-wind medium. In Section \S\ref{sec3}, we introduce Fermi-LAT data and the methodical approach. Section \S\ref{sec4} presents the discussion, and finally, in Section \S\ref{sec5}, we provide a summary of our work and some concluding remarks. From now on, we will call them ``unprimed" and ``prime" in the observer and in the comoving frame. We will adopt the convention $Q_{\rm x}=\frac{Q}{10^{\rm x}}$ in cgs units and assume for the cosmological constants a spatially flat universe $\Lambda$CDM model with  $H_0=69.6\,{\rm km\,s^{-1}\,Mpc^{-1}}$, $\Omega_{\rm M}=0.286$ and  $\Omega_\Lambda=0.714$ \citep{2016A&A...594A..13P}. 


\vspace{1cm}

\section{Closure relations of off-axis forward-shock emission}
\label{sec2}
When a relativistic GRB jet decelerates in the circumstellar medium, the afterglow emission is produced. We assume that the circumstellar medium has a density profile defined by $n(r) \propto r^{\rm -k}$, where the density-profile indexes ${\rm k=0}$ 
and ${\rm k=2}$ correspond to ISM \cite[i.e, see][]{1998ApJ...497L..17S} and the stellar-wind environment \cite[i.e, see][]{2000ApJ...536..195C}, respectively. During the deceleration phase, the relativistic jet transmits a large fraction of its energy to the surrounding external medium.  This energy density is given to accelerate electrons during the forward shock ($\epsilon_e$) and amplify a magnetic field ($\epsilon_B$). Accelerated electrons are distributed in accordance with their Lorentz factors ($\gamma_e$) and are described by the electron power index $p$ as $N(\gamma_e)\,d\gamma_e \propto \gamma_e^{-p}\,d\gamma_e$ for $\gamma_m\leq \gamma_{\rm e}$, where $\gamma_m=\frac{m_{\rm p}}{m_{\rm e}}g(p)\varepsilon_{\rm e}(\Gamma-1)$ is the electron Lorentz factor of the lowest energy with $\Gamma$ the bulk Lorentz factor, $g(p)=\frac{p-2}{p-1}$, and $m_{\rm p}$ and $m_{\rm e}$ the proton and electron mass, respectively.   The cooling electron Lorentz factor is $\gamma_{\rm c}=\frac{6\pi m_e c}{\sigma_T}(1+Y)^{-1}\Gamma^{-1}B'^{-2}t^{-1}$, where $\sigma_T$ is the cross-section in the Thomson regime, $t$ is the observer time and $Y$ is the Compton parameter \citep{2001ApJ...548..787S, 2010ApJ...712.1232W}. The comoving magnetic field in the blastwave $\frac{B'^2}{8\pi}=\varepsilon_B\,U_B$  is derived from the energy density $U_B=\frac{\hat\gamma\Gamma +1}{\hat\gamma - 1}(\Gamma -1)n(r) m_pc^2$ with $\hat\gamma$ the adiabatic index \citep{1999MNRAS.309..513H}. The synchrotron spectral breaks and the synchrotron radiation power per electron in the comoving frame are given by $\nu'^{\rm sync}_{\rm i}=\frac{q_e}{2\pi m_ec}\gamma^{2}_{\rm i}B'$ and $P'_{\nu'_m}\simeq \frac{\sqrt{3}q_e^3}{m_ec^2}B'$, respectively,  with hereafter the subindex ${\rm i=m}$ and ${\rm c}$ for the characteristic and cooling break, and the constants $q_e$ and $c$ the elementary charge and the speed of light, respectively \citep[e.g., see][]{1998ApJ...497L..17S, 2015ApJ...804..105F}.   The total number of emitting electrons is $N_e=\frac{\Omega}{4\pi}\, n(r) \frac{4\pi}{3-k} r^3$ with $r=\frac{\delta_D}{1+z} \Gamma\beta c t$ the shock radius.   The transformation laws between both frames are $\nu_{\rm i}=\frac{\delta_D}{1+z}\nu'_{\rm i}$ for the spectral breaks, $\Omega= \frac{\Omega'}{\delta^2_D}$ for  the solid angle and $P_{\nu_m}=\frac{\delta_D}{1+z} P'_{\nu'_m}$ for the radiation power with $\delta_D=\frac{1}{\Gamma(1-\mu\beta)}$ the Doppler factor. The SSC process takes place when the same electron population that radiates synchrotron photons up-scatters them up to higher energies as $h\nu^{\rm ssc}_{\rm i}\simeq \gamma^2_{\rm i} h\nu^{\rm syn}_{\rm i}$, with ${\rm i=m, c}$ \citep[e.g., see][]{2001ApJ...548..787S}.  The terms $\mu=\cos \Delta \theta$, $\beta=v/c$ with $v$ the velocity of the material, and $\Delta \theta=\theta_{\rm obs} - \theta_{\rm j}$ is given by the viewing angle ($\theta_{\rm obs}$) and the half-opening angle of the jet ($\theta_{\rm j}$) with  $\theta_{\rm obs} > \theta_{\rm j}$.\\

Considering the point source limit ($\theta_{\rm obs} \gtrsim 2\theta_{\rm j}$) for which the model gives reasonable results \citep[e.g., see][]{2002ApJ...570L..61G}, the maximum synchrotron flux for a point source that has a difference between the observation angle and the opening angle $\Delta \theta$ becomes $F^{\rm  syn}_{\rm max}=\frac{(1+z)^2\delta^3_D}{4\pi d_z^2}N_eP'_{\nu'_m}$. It is worth noting that the integration over the solid angle in this approximation can be ignored \citep{2017ApJ...850L..24G, 2004ApJ...614L..13E}.  The maximum flux of the SSC process is estimated as $F^{\rm ssc}_{\rm max}\sim \sigma_T n(r)\,r\, F^{\rm syn}_{\rm max}$.\\

\subsection{Radiative/adiabatic scenario}

\begin{figure*}
\centering
\includegraphics[width=0.48\textwidth]{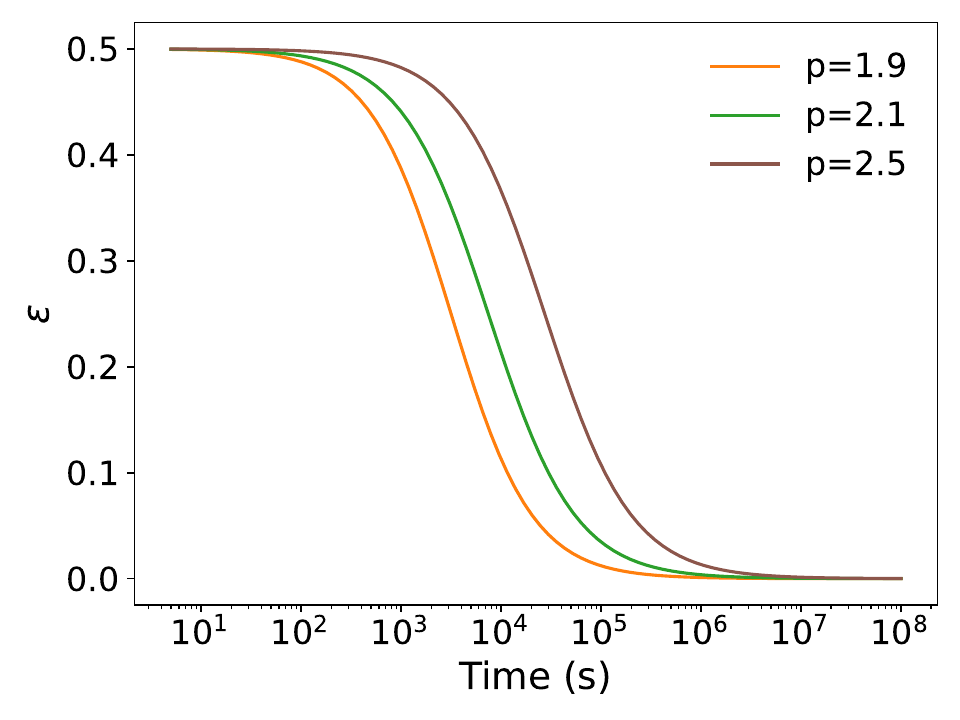}
\includegraphics[width=0.48\textwidth]{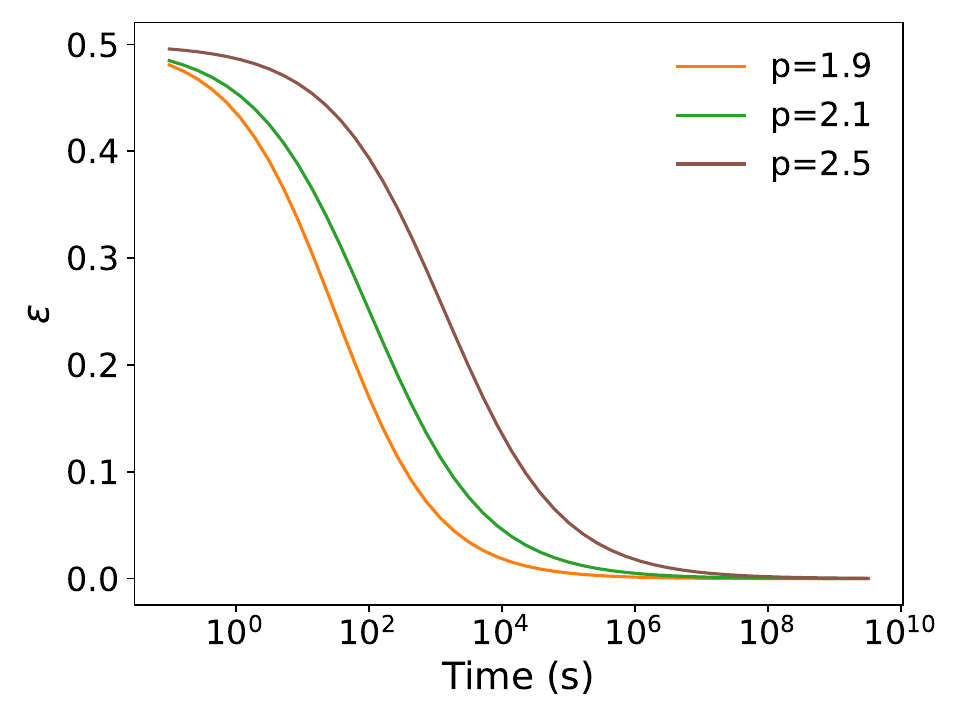}\\
\includegraphics[width=0.48\textwidth]{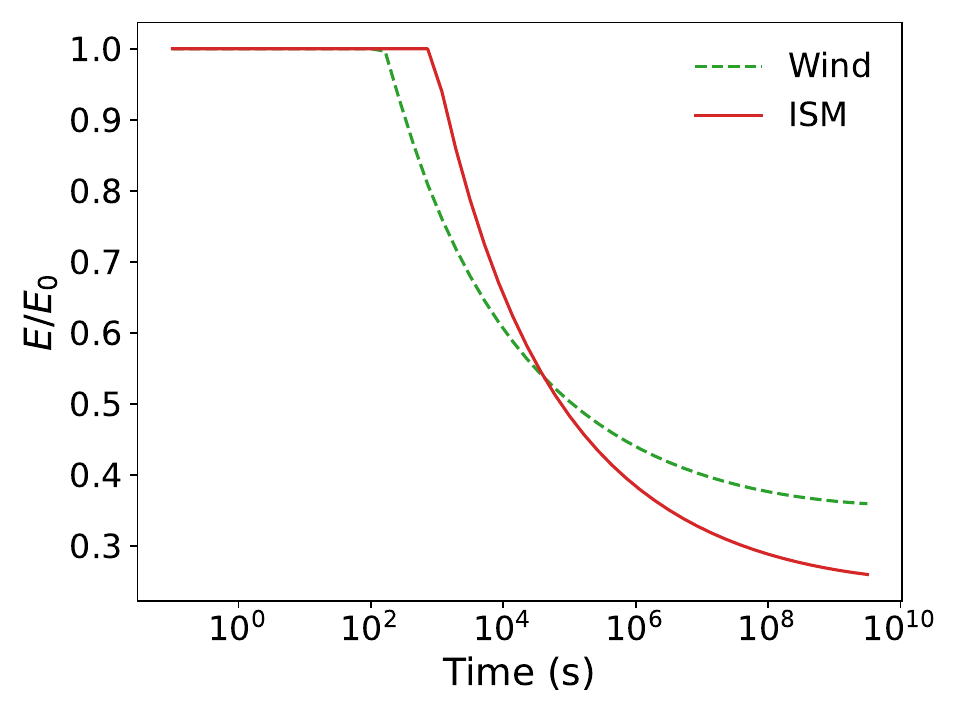}
\includegraphics[width=0.48\textwidth]{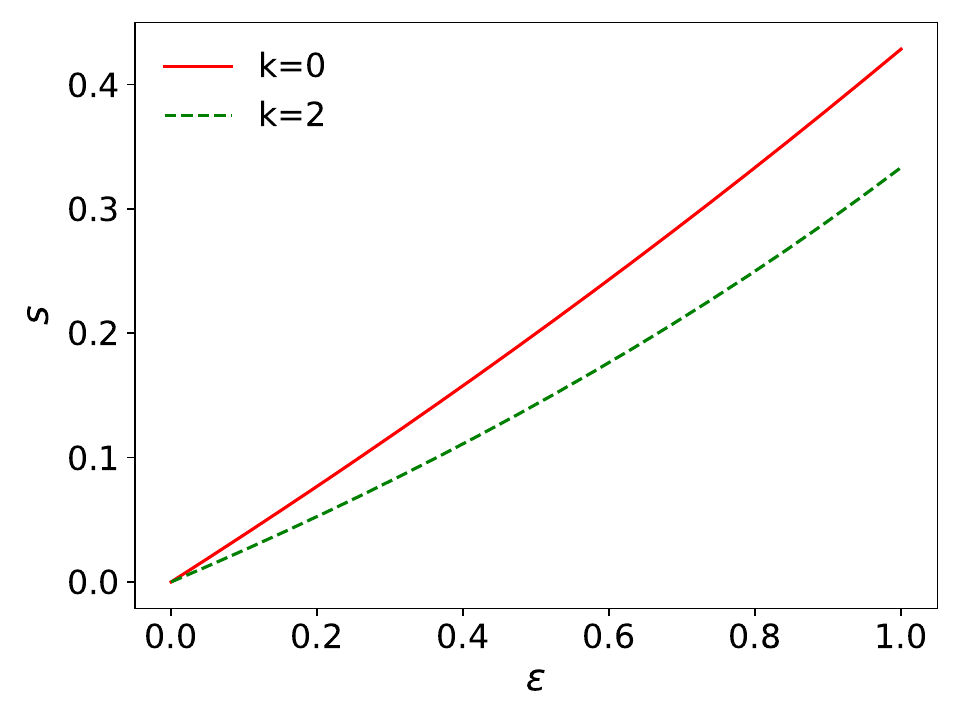}\\
\caption{Upper panels: Evolution of the radiative parameter in the stellar-wind (left) and uniform-density (right) medium considering the synchrotron and expansion timescales. Lower panels show the evolution of energy radiated away (left) and relation between the radiative parameters $\epsilon$ and $s$ (right). The uniform-density and stellar-wind environments are shown in red solid and green dashed lines, respectively. The term $E_0$ corresponds to the initial equivalent kinetic energy.  We use the values of $\Gamma_0=60$, $E_0=10^{52}\,{\rm erg}$, $\varepsilon_e=0.5$, $\varepsilon_B=5\times10^{-2}$, $\chi_{\rm e}=1$, $A_{W}=0.1$, $n=1\,{\rm cm^{-3}}$.}
\label{k_eps}
\end{figure*}

The external medium decelerates the GRB fireball more when it is radiative than when it is adiabatic. In the fireball scenario, it is believed that the forward shock created during the deceleration phase is totally adiabatic; however, it might be partly or fully radiative \citep[e.g., see][]{1998MNRAS.298...87D, 1998ApJ...497L..17S, 2000ApJ...532..281B, 2010MNRAS.403..926G}. Consequently, the synchrotron forward-shock light curves and the energetics of the shock are affected when they evolve in the radiative regime \citep{2000ApJ...532..281B, 2005ApJ...619..968W, 2000ApJ...529..151M}.  In the radiative regime, the isotropic-equivalent kinetic energy evolves as $E= E_0 \left( \frac{\Gamma}{\Gamma_0}\right)^\epsilon$ \citep{2000ApJ...532..281B, 2005ApJ...619..968W}, where $E_0$ is the initial isotropic-equivalent kinetic energy,  $\Gamma_0$ is the initial Lorentz factor and $\epsilon$ is the parameter that provides the details of the hydrodynamic evolution in the entirely radiative ($\epsilon=1$), entirely adiabatic ($\epsilon=0$),  or intermediate ($0 <\epsilon < 1$) regimes.  In most cases, the radiative parameter of the blast wave is determined by relating the synchrotron (${t'}_{\rm syn}$) and the expansion (${t'}_{\rm ex}$) timescale as $\epsilon = \epsilon_{e}\frac{{t'}^{-1}_{\rm syn}}{{t'}^{-1}_{\rm syn}+{t'}^{-1}_{\rm ex}}$  \citep[e.g., see][]{2000ApJ...543...90H}.  Considering the limiting cases between ${t'}_{\rm syn}$ and ${t'}_{\rm exp}$, we can estimate the $\epsilon$-parameter. For $t'_{\rm syn}\ll t'_{\rm exp}$, this parameter $\epsilon = \varepsilon_{e}(1+t'_{\rm syn}/t'_{\rm exp})^{-1}\approx \varepsilon_{e}$ lies in the fully radiative regime if $\varepsilon_{e}\approx 1$ and for $t'_{\rm exp}\ll t'_{\rm syn}$, the radiative parameter $\epsilon = \varepsilon_{e}\, t'_{\rm exp}/t'_{\rm syn} (1+t'_{\rm exp}/t'_{\rm syn})^{-1}\approx 0$ lies in the fully adiabatic regime. The radiative parameter in a realistic environment is partly radiative/adiabatic ($0< \epsilon <1$).  The upper panels in Figure \ref{k_eps} show the evolution of the radiative parameter in the stellar-wind (left) and constant-density (right) medium considering the synchrotron and expansion timescales. These upper panels display that irrespective of the circumburst medium, when $p$ increases for $2<p$, both scenarios become similar. The curves for $p=1.9$ are only exhibited in the expansion timescale scenario given that they diverge in the other scenario. One can notice that for $2<p$, the parameter $\epsilon$ decreases slowly as the value of $p$ does not deviate from $2$.  As expected, for $p\to 2$, the radiative parameter goes to the same value $\epsilon\simeq \varepsilon_e$.   The lower left-hand panel in Figure \ref{k_eps} shows the evolution of the energy radiated away in the stellar-wind (dashed green line) and constant-density (solid red line) medium for $p=2.1$. This panel shows that the equivalent kinetic energy is dissipated during the fast-cooling regime, opposite to the slow-cooling regime.\\

Some authors have introduced the radiative parameter $s$ instead of $\epsilon$ via the explicit temporal variation of the equivalent kinetic energy as \citep{2000ApJ...532..281B, 2005ApJ...619..968W,2010MNRAS.403..926G} 
\begin{eqnarray}\label{Ef}
E\left(\frac{t}{t_{\rm dec}}\right)^{-s}\propto \cases{
(2-\epsilon)\, n\, \Gamma^2\,r^3\,, \hspace{1cm}{\rm for\,\,\,\, k=0} \cr
(2-\epsilon)\, A_{\rm w}\, \Gamma^2\,r\,,\hspace{0.9cm} {\rm for\,\,\,\, k=2}.\cr
}
\end{eqnarray}
%
In this scenario, the radiative parameters ($\epsilon, s$) are related as

\begin{eqnarray}
(3+s)(8-\epsilon)-24=0,\,\,\,\,\,\,\,\,&&{\rm for\,\,\,\, k=0}\cr
(1+s)(4-\epsilon)-4=0,\,\,\,\,\,\,\,\,\,\,\,&&{\rm for\,\,\,\, k=2}\,,
\end{eqnarray}

where the parameter s lies in the range of $0\leq s \leq 3/7$ and $0\leq s \leq 1/3$ for constant-density and stellar-wind medium, respectively.  The lower right-hand panel in Figure \ref{k_eps} shows  the relation between the radiative parameters $\epsilon$ and $s$ for constant-density and stellar-wind environments. This panel displays that the radiative parameters are directly proportional to each other, and that the constant-density scenario is more susceptible to variations in the parameters.

\subsubsection{Constant-density environment (${\rm k=0}$)}\label{sec_211}
Considering the isotropic-equivalent kinetic energy $E= E_0 \left( \frac{\Gamma}{\Gamma_0}\right)^\epsilon$ and the evolution of the forward shock with the Blandford-McKee solution $E=\frac{2\pi}{3}(2-\epsilon) m_pc^2 n r^{3} \Gamma^2$ \citep{1976PhFl...19.1130B}, the temporal evolution of the bulk Lorentz factor in the radiative regime becomes {\small $\Gamma\simeq 1.0\times 10^{-3}\left(\frac{1+z}{0.022}\right)^{\frac{3}{2-\epsilon}}\,\Delta\theta_{2}^{\frac{6}{2-\epsilon}}\,\theta_{j,1}^{-\frac{2}{2-\epsilon}}\, n^{-\frac{1}{2-\epsilon}} E^\frac{1}{2-\epsilon}_{0_j,50}\Gamma^{-\frac{\epsilon}{2-\epsilon}}_{0,2.78} t_{3}^{-\frac{3}{2-\epsilon}}$} with $\theta_{j,1}=\frac{\theta_{j}}{1^\circ}$ and $\Delta\theta_{2}=\frac{\Delta\theta}{2^\circ}$. We have taken the approximation  $\delta_D\simeq\frac{2\Gamma}{1+\left(\Gamma\Delta\theta\right)^2} \simeq\frac{2}{\Gamma\Delta\theta^2}$ for $\Gamma\Delta\theta\gg 1$ \citep{2002ApJ...570L..61G, 2018PTEP.2018d3E02I} and $E_0=\frac{E_{0_j}}{1-\cos\theta_{j}}\approx \frac{2E_{0_j}}{\theta^2_{j}}$ for the jet dynamics \citep{2018MNRAS.481.1597G}.  It is worth noting that for $\epsilon=0$, the evolution of the bulk Lorentz factor $\Gamma\propto t^{-\frac32}$ is obtained \cite[e.g., see][]{2018MNRAS.478..407N, 2022ApJ...940..189F}.   The Lorentz factors of the electrons with the lowest energy and above which they cool effectively are

{\small
\begin{eqnarray}\nonumber
\label{ele_Lorent_ism}
\gamma_{\rm m}= \cases{
3.7\times10^4\left(\frac{2-p}{p-1}\right)^{\frac{1}{p-1}}\left(\frac{1+z}{1.022}\right)^{\frac{3(4-p)}{2(p-1)(2-\epsilon)}}\, \varepsilon^{\frac{1}{p-1}}_{e,-1}  \varepsilon^{\frac{-(p-2)}{4(p-1)}}_{B,-2}  n^{-\frac{4 + \epsilon(2-p)}{4(p-1)(2-\epsilon)}}\, \theta_{j,1}^{-\frac{4-p}{(p-1)(2-\epsilon)}}\Delta\theta_{2}^{\frac{3(4-p)}{(p-1)(2-\epsilon)}}\, E_{0_j,50}^{\frac{4-p}{2(p-1)(2-\epsilon)}}\,\cr  \hspace{7cm}\Gamma^{-\frac{\epsilon(4-p)}{2(p-1)(2-\epsilon)}}_{0,2.78}   t_{3}^{-\frac{3(4-p)}{2(p-1)(2-\epsilon)}}  \hspace{2.5cm} {\rm for} \hspace{0.2cm} { 1<p<2 }\cr
9.3\times10^4\left(\frac{p-2}{p-1}\right)\left(\frac{1+z}{1.022}\right)^{\frac{3}{2-\epsilon}} n^{-\frac{1}{2-\epsilon}} \varepsilon_{e,-1}\theta_{j,1}^{-\frac{2}{2-\epsilon}}\Delta\theta_{2}^{\frac{6}{2-\epsilon}}E_{0_j,50}^{\frac{1}{2-\epsilon}}\Gamma^{-\frac{\epsilon}{2-\epsilon}}_{0,2.78} t_{3}^{-\frac{3}{2-\epsilon}} \hspace{3.25cm} {\rm for} \hspace{0.2cm} {2 < p }\cr
}
\end{eqnarray}
}
{\small
\bary\label{nus_ism}
\gamma_{\rm c}&=&5.2\times10\left(\frac{1+z}{1.022}\right)^{-\frac{1+\epsilon}{2-\epsilon}} n^{-\frac{1-\epsilon}{2-\epsilon}} (1+Y)^{-1}\varepsilon_{B,-2}^{-1}  \theta_{j,1}^{\frac{2}{2-\epsilon}}  \Delta\theta_{2}^{-\frac{2(1+\epsilon)}{2-\epsilon}}E_{0_j, 50}^{-\frac{1}{2-\epsilon}} \Gamma^{\frac{\epsilon}{2-\epsilon}}_{0,2.78}t_{3}^{\frac{1+\epsilon}{2-\epsilon}},\hspace{4.5cm}
\eary
}

respectively. Hereafter, we assume the values of spectral indexes of the electron distribution $p=1.95$ and $2.15$ for $1<p<2$ and $2 < p$, respectively.   The characteristic (for $1<p<2$ and $2 < p$) and cooling  spectral breaks, and the maximum flux become 

{\small
\begin{eqnarray}\nonumber
\label{ele_Lorent_ism}
\nu^{\rm sync}_{\rm m}= \cases{
7.6\times 10\,{\rm eV}\,\left(\frac{2-p}{p-1}\right)^{\frac{2}{p-1}}\left(\frac{1+z}{1.022}\right)^{\frac{14 - p(5-\epsilon) - \epsilon}{(p-1)(2-\epsilon)}}\,  \varepsilon^{\frac{2}{p-1}}_{e,-1}  \varepsilon^{\frac{1}{2(p-1)}}_{B,-2}  n^{-\frac{2(3-p) + \epsilon}{2(p-1)(2-\epsilon)}}\, \theta_{j,1}^{-\frac{2(4-p)}{(p-1)(2-\epsilon)}}\Delta\theta_{2}^{\frac{2[14 - p(5-\epsilon) - \epsilon]}{(p-1)(2-\epsilon)}}    \, E_{0_j,50}^{\frac{4-p}{2(p-1)(2-\epsilon)}}\,\cr  \hspace{7cm}\Gamma^{-\frac{\epsilon(4-p)}{(p-1)(2-\epsilon)}}_{0,2.78}   t_{3}^{-\frac{3(4-p)}{(p-1)(2-\epsilon)}}  \hspace{2.5cm} {\rm for} \hspace{0.2cm} { 1<p<2 }\cr
4.9\times 10^2\,{\rm eV}\,\left(\frac{p-2}{p-1}\right)^{2}\left(\frac{1+z}{1.022}\right)^{\frac{4+\epsilon}{2-\epsilon}} n^{-\frac{2+\epsilon}{2(2-\epsilon)}} \varepsilon_{e,-1}^2 \varepsilon_{B,-2}^{\frac{1}{2}}\theta_{j,1}^{-\frac{4}{2-\epsilon}}\Delta\theta_{2}^{\frac{2(4+\epsilon)}{2-\epsilon}}E_{0_j,50}^{\frac{2}{2-\epsilon}} \Gamma^{-\frac{2\epsilon}{2-\epsilon}}_{0,2.78} t_{3}^{-\frac{6}{2-\epsilon}} \hspace{1.70cm} {\rm for} \hspace{0.2cm} {2 < p }\cr
}
\end{eqnarray}
}
{\small
\bary\label{nus_ism}
\nu^{\rm sync}_{\rm c}&=& 1.6\times10^{-4}\,{\rm eV}\,\left(\frac{1+z}{1.022}\right)^{-\frac{\epsilon+4}{2-\epsilon}} n^{\frac{3\epsilon-2}{2(2-\epsilon)}} (1+Y)^{-2}\varepsilon_{B,-2}^{-\frac32}\theta_{j,1}^{\frac{4}{2-\epsilon}}\Delta\theta_{2}^{-\frac{2(\epsilon+4)}{2-\epsilon}}E_{0_j,50}^{-\frac{2}{2-\epsilon}} \Gamma^{\frac{2\epsilon}{2-\epsilon}}_{0,2.78} t_{3}^{\frac{2(\epsilon+1)}{2-\epsilon}}\,\cr
F^{\rm sync}_{\rm max}&=& 1.3\times10^{5}\,{\rm mJy}\,\left(\frac{1+z}{1.022}\right)^{\frac{\epsilon-8}{2-\epsilon}} n^{\frac{10-3\epsilon}{2(2-\epsilon)}} \varepsilon_{B,-2}^{\frac{1}{2}}d_{z,27.8}^{-2}\theta_{j,1}^{\frac{4}{2-\epsilon}}\Delta\theta_{2}^{-\frac{12(3-\epsilon)}{2-\epsilon}}E_{0_j,50}^{-\frac{2}{2-\epsilon}} \Gamma^{\frac{2\epsilon}{2-\epsilon}}_{0,2.78} t_{3}^{\frac{3(4-\epsilon)}{2-\epsilon}}\,,\hspace{4.cm}
\eary
}
where {\small $d_{\rm z}=(1+z)\frac{c}{H_0}\int^z_0\,\frac{d\tilde{z}}{\sqrt{\Omega_{\rm M}(1+\tilde{z})^3+\Omega_\Lambda}}$}  \citep{1972gcpa.book.....W}  is the luminosiy distance. Based on the fact that spectral breaks and maximum flux of synchrotron off-axis model in ISM evolve throughout time as Eq. \ref{nus_ism}, we derive the CRs shown in Table \ref{table1}.\\

Given the electron Lorentz factors (Eq. \ref{ele_Lorent_ism}), the synchrotron spectral breaks and the maximum synchrotron flux,  the characteristic (for $1<p<2$ and $2 < p $) and cooling  spectral breaks, and the maximum flux in the synchrotron scenario become 

{\small
\begin{eqnarray}\nonumber
\label{break_ssc_hom}
h\nu^{\rm ssc}_{\rm m} = \cases{ 
2.5\times 10^{2}\,{\rm GeV}\,\left(\frac{2-p}{p-1}\right)^{\frac{4}{p-1}}\left(\frac{1+z}{1.022}\right)^{\frac{26 - p(8-\epsilon) - \epsilon}{(p-1)(2-\epsilon)}}\,  \varepsilon^{\frac{4}{p-1}}_{e,-1}  \varepsilon^{\frac{3-p}{2(p-1)}}_{B,-2}  n^{-\frac{10 + 3\epsilon -p(2+ \epsilon)}{2(p-1)(2-\epsilon)}}\, \theta_{j,1}^{-\frac{4(4-p)}{(p-1)(2-\epsilon)}}\Delta\theta_{2}^{\frac{2[26 - p(8-\epsilon) - \epsilon]}{(p-1)(2-\epsilon)}}    \, E_{0_j,50}^{\frac{2(4-p)}{(p-1)(2-\epsilon)}}\,\cr  \hspace{7cm}\Gamma^{-\frac{2\epsilon(4-p)}{(p-1)(2-\epsilon)}}_{0,2.78}   t_{3}^{-\frac{6(4-p)}{(p-1)(2-\epsilon)}}  \hspace{3.5cm} {\rm for} \hspace{0.2cm} { 1<p<2 }\cr
8.7\times 10^{3}\,{\rm GeV}\,\left(\frac{p-2}{p-1}\right)^4 \left(\frac{1+z}{1.022}\right)^{\frac{10+\epsilon}{2-\epsilon}}\, \varepsilon^4_{e,-1} \varepsilon^{\frac12}_{B,-2} n^{-\frac{6+\epsilon}{2(2-\epsilon)}}\, \theta_{j,1}^{-\frac{8}{2-\epsilon}}\Delta\theta_{2}^{\frac{2(10+\epsilon)}{2-\epsilon}}     E_{0_j,50}^{\frac{4}{2-\epsilon}}\, \Gamma^{-\frac{4\epsilon}{2-\epsilon}}_{0,2.78}   t_{3}^{-\frac{12}{2-\epsilon}} \hspace{1.2cm} {\rm for} \hspace{0.2cm} {2 < p }\cr
}
\end{eqnarray}
}

{\small
\bary \label{ssc_br_hom_v1}
h\nu^{\rm ssc}_{\rm c}&=& 5.8\times 10^{-5}\,{\rm GeV} \left(\frac{1+z}{1.022}\right)^{-\frac{3(2+\epsilon)}{2-\epsilon}}\, \varepsilon^{-\frac72}_{B,-2}(1+Y)^{-4} n^{\frac{7\epsilon-6}{2(2-\epsilon)}}\,  \theta_{j,1}^{\frac{8}{2-\epsilon}}\Delta\theta_{2}^{-\frac{6(2+\epsilon)}{2-\epsilon}}            E_{0_j,50}^{-\frac{4}{2-\epsilon}}\, \Gamma^{\frac{4\epsilon}{2-\epsilon}}_{0,2.78} \,t_{3}^{\frac{4(\epsilon+1)}{2-\epsilon}}\cr
F^{\rm ssc}_{\rm max}&=& 2.1\times 10^{-3}\,{\rm mJy}\left(\frac{p-2}{p-1}\right)^{-1} \left(\frac{1+z}{1.022}\right)^{\frac{2(\epsilon-5)}{2-\epsilon}}\, \varepsilon^\frac12_{B,-2} d_{\rm z,26.5}^{-2}\,n^{\frac{14-5\epsilon}{2(2-\epsilon)}}\,  \theta_{j,1}^{\frac{4}{2-\epsilon}}\Delta\theta_{2}^{\frac{2(7\epsilon-20)}{2-\epsilon}}\,             E_{0_j,50}^{-\frac{2}{2-\epsilon}}\, \Gamma^{\frac{2\epsilon}{2-\epsilon}}_{0,2.78}   t_{3}^{\frac{2(7-2\epsilon)}{2-\epsilon}}\,,
\eary
}

respectively.  Given the temporal evolution of the spectral breaks and the maximum flux of SSC off-axis model in ISM (Eq. \ref{ssc_br_hom_v1}), we estimate the CRs as listed in Table \ref{table2}.

\subsubsection{Stellar-wind environment (${\rm k=2}$)}
 Relativistic electrons are accelerated in the forward shock and cooled efficiently by synchrotron and SSC processes. In general,  at a considerable distance from the progenitor, the evolution in the adiabatic and radiative regime of the bulk Lorentz factor is $\Gamma=2.2\times 10^2 \left(\frac{1+z}{1.022}\right)^{\frac{1}{2-\epsilon}}\, A_{\rm w,-1}^{-\frac{1}{2-\epsilon}}\theta^{-\frac{2}{2-\epsilon}}_{\rm j,5}\Delta\theta^{\frac{2}{2-\epsilon}}_{1}\,E_{0_j,50}^{\frac{1}{2-\epsilon}}\Gamma_{0,2.78}^{-\frac{\epsilon}{2-\epsilon}}t_{3}^{-\frac{1}{2-\epsilon}}$, where the term $A_{\rm w}$ corresponds to the density parameter, which is associated with the stellar wind density $n(r) = \frac{\dot{M}_{\rm W}}{4\pi m_p\,v_{\rm W}} A_{\rm W} r^{-2} =3.0\times 10^{35}\,{\rm cm^{-1}}\,A_{\rm W}r^{-2}$ \citep{2000ApJ...536..195C}, with $m_p$ the proton mass.\footnote{The values of the wind velocity ($v_{\rm W}$) and the mass loss rate ($\dot{M}_{\rm W}$) used are  given for a Wolf-Rayet \citep[see][]{2000ApJ...536..195C}.} The Lorentz factors of the electrons with the lowest energy, and above which they cool effectively become

{\small
\begin{eqnarray}\nonumber
\label{ele_win}
\gamma_{\rm m}= \cases{ 
6.7\times 10^3\,\left(\frac{2-p}{p-1}\right)^{\frac{1}{p-1}}\left(\frac{1+z}{1.022}\right)^{\frac{8 - 3p - \epsilon(2-p)}{2(p-1)(2-\epsilon)}}\,  \varepsilon^{\frac{1}{p-1}}_{e,-1}  \varepsilon^{\frac{-(p-2)}{4(p-1)}}_{B,-2}  A_{\rm w,-1}^{-\frac{4 + \epsilon(2-p)}{4(p-1)(2-\epsilon)}}\, \theta_{j,1}^{-\frac{4-p}{(p-1)(2-\epsilon)}}\Delta\theta_{2}^{\frac{8 - 3p - \epsilon(2-p)}{(p-1)(2-\epsilon)}}\, E_{0_j,50}^{\frac{4-p}{2(p-1)(2-\epsilon)}}\,\cr  \hspace{7cm} \Gamma^{-\frac{\epsilon(4-p)}{2(p-1)(2-\epsilon)}}_{0,2.78}   t_{3}^{-\frac{8 - 3p - \epsilon(2-p)}{2(p-1)(2-\epsilon)}}  \hspace{2.5cm} {\rm for} \hspace{0.2cm} { 1<p<2 }\cr
2.1\times 10^4\,\left(\frac{p-2}{p-1}\right)\left(\frac{1+z}{1.022}\right)^{\frac{1}{2-\epsilon}} A_{\rm w,-1}^{-\frac{1}{2-\epsilon}} \varepsilon_{e,-1}\theta_{j,1}^{-\frac{2}{2-\epsilon}}\Delta\theta_{2}^{\frac{2}{2-\epsilon}} E_{0_j,50}^{\frac{1}{2-\epsilon}} \Gamma^{-\frac{\epsilon}{2-\epsilon}}_{0,2.78}t_{3}^{-\frac{1}{2-\epsilon}} \hspace{3.25cm} {\rm for} \hspace{0.2cm} {2 < p }
}
\end{eqnarray}
}
{\small
\bary\label{nus_wind}
\gamma_{\rm c}&=& 1.8\times 10\left(\frac{1+z}{1.022}\right)^{\frac{\epsilon-3}{2-\epsilon}} A_{\rm w,-1}^{-\frac{1-\epsilon}{2-\epsilon}} (1+Y)^{-1}\varepsilon_{B,-2}^{-1}  \theta_{j,1}^{\frac{2}{2-\epsilon}} \Delta\theta_{2}^{-\frac{2(3-\epsilon)}{2-\epsilon}}E_{0_j,50}^{-\frac{1}{2-\epsilon}} \Gamma^{\frac{\epsilon}{2-\epsilon}}_{0,2.78}t_{3}^{\frac{3-\epsilon}{2-\epsilon}},\hspace{4.7cm}
\eary
}

respectively, with $t$ the observer time.  The characteristic (for $1<p<2$ and $2 < p$) and cooling  spectral breaks, and the maximum flux become

{\small
\begin{eqnarray}\nonumber
\label{ele_win}
\nu^{\rm sync}_{\rm m}= \cases{ 
9.4\,{\rm eV}\,\left(\frac{2-p}{p-1}\right)^{\frac{2}{p-1}}\left(\frac{1+z}{1.022}\right)^{\frac{8 - 3p  + \epsilon(p-2)}{(p-1)(2-\epsilon)}}\,  \varepsilon^{\frac{2}{p-1}}_{e,-1}  \varepsilon^{\frac{1}{2(p-1)}}_{B,-2}  A_{\rm w,-1}^{-\frac{2(3-p) + \epsilon}{2(p-1)(2-\epsilon)}}\, \theta_{j,1}^{-\frac{2(4-p)}{(p-1)(2-\epsilon)}}\Delta\theta_{2}^{\frac{16 - 6p + 2\epsilon(p-2)}{(p-1)(2-\epsilon)}}    \, E_{0_j,50}^{\frac{4-p}{(p-1)(2-\epsilon)}}\,\cr  \hspace{7cm}\Gamma^{-\frac{\epsilon(4-p)}{(p-1)(2-\epsilon)}}_{0,2.78}   t_{3}^{-\frac{6 - p - \epsilon}{(p-1)(2-\epsilon)}}  \hspace{2.5cm} {\rm for} \hspace{0.2cm} { 1<p<2 }\cr
8.8\times 10^1\,{\rm eV}\,\left(\frac{p-2}{p-1}\right)^{2}\left(\frac{1+z}{1.022}\right)^{\frac{2}{2-\epsilon}} A_{\rm w,-1}^{-\frac{2+\epsilon}{2(2-\epsilon)}} \varepsilon_{e,-1}^2 \varepsilon_{B,-2}^{\frac{1}{2}}\theta_{j,1}^{-\frac{4}{2-\epsilon}}\Delta\theta_{2}^{\frac{4}{2-\epsilon}}E_{0_j,50}^{\frac{2}{2-\epsilon}}\Gamma^{-\frac{2\epsilon}{2-\epsilon}}_{0,2.78} t_{3}^{-\frac{4-\epsilon}{2-\epsilon}} \hspace{2.10cm} {\rm for} \hspace{0.2cm} {2 < p }
}
\end{eqnarray}
}
{\small
\bary\label{nus_wind}
\nu^{\rm sync}_{\rm c}&=& 6.9\times 10^{-5}\,{\rm eV}\, \left(\frac{1+z}{1.022}\right)^{-\frac{2(3-\epsilon)}{2-\epsilon}} A_{\rm w,-1}^{-\frac{2-3\epsilon}{2(2-\epsilon)}} (1+Y)^{-2}\varepsilon_{B,-2}^{-\frac{3}{2}}\theta_{j,1}^{\frac{4}{2-\epsilon}}\Delta\theta_{2}^{-\frac{4(3-\epsilon)}{2-\epsilon}}E_{0_j,50}^{-\frac{2}{2-\epsilon}}\Gamma^{\frac{2\epsilon}{2-\epsilon}}_{0,2.78} t_{3}^{\frac{4-\epsilon}{2-\epsilon}},\cr
F^{\rm syn}_{\rm max}&=& 1.2\times10^{8}\,{\rm mJy}\,\left(\frac{1+z}{1.022}\right)^{\frac{2(1-\epsilon)}{2-\epsilon}} A_{\rm w,-1}^{\frac{10-3\epsilon}{2(2-\epsilon)}} \varepsilon_{B,-2}^{\frac{1}{2}}d_{z,26.5}^{-2}\theta_{j,1}^{\frac{4}{2-\epsilon}}\Delta\theta_{2}^{-\frac{2(8-3\epsilon)}{2-\epsilon}}E_{0_j,50}^{-\frac{2}{2-\epsilon}} \Gamma^{\frac{2\epsilon}{2-\epsilon}}_{0,2.78}t^{\frac{2}{2-\epsilon}}_{3}.\hspace{3cm}
\eary
}

Based on the fact that spectral breaks and maximum flux of synchrotron off-axis model in stellar wind evolve throughout time as Eq. \ref{nus_wind}, we derive the CRs shown in Table \ref{table1}.

Given the electron Lorentz factors (Eq. \ref{nus_wind}), the synchrotron spectral breaks and the maximum synchrotron flux,  the characteristic (for $1<p<2$ and $2 < p$) and cooling  spectral breaks, and the maximum flux in the SSC scenario become 

{\small
\begin{eqnarray}\nonumber
\label{break_ssc_win}
h\nu^{\rm ssc}_{\rm m}= \cases{ 
8.5\times 10^{-1}\,{\rm GeV}\,\left(\frac{2-p}{p-1}\right)^{\frac{4}{p-1}}\left(\frac{1+z}{1.022}\right)^{\frac{2(8-3p) + 2\epsilon (p-2)}{(p-1)(2-\epsilon)}}\, \varepsilon^{\frac{4}{p-1}}_{e,-1}  \varepsilon^{\frac{3-p}{2(p-1)}}_{B,-2}  A_{\rm w,-1}^{-\frac{10 + 3\epsilon-p(2+ \epsilon)}{2(p-1)(2-\epsilon)}}\, \theta_{j,1}^{-\frac{4(4-p)}{(p-1)(2-\epsilon)}}\Delta\theta_{2}^{\frac{4(8-3p) + 4\epsilon (p-2)}{(p-1)(2-\epsilon)}}    \, E_{0_j,50}^{\frac{2(4-p)}{(p-1)(2-\epsilon)}}\,\cr  \hspace{7cm}\Gamma^{-\frac{2\epsilon(4-p)}{(p-1)(2-\epsilon)}}_{0,2.78}   t_{3}^{-\frac{2(7-2p) + \epsilon(p-3) }{(p-1)(2-\epsilon)}}  \hspace{2.5cm} {\rm for} \hspace{0.2cm} { 1<p<2 }\cr
7.4\times 10\,{\rm GeV}\,\left(\frac{p-2}{p-1}\right)^{4}\left(\frac{1+z}{1.022}\right)^{\frac{4}{2-\epsilon}}\, \varepsilon^{\frac12}_{B,-2} \varepsilon^{4}_{e,-1}\,  A_{\rm w,-1}^{-\frac{6+\epsilon}{2(2-\epsilon)}}\,\theta_{j,1}^{-\frac{8}{2-\epsilon}}\Delta\theta_{2}^{\frac{8}{2-\epsilon}}    \, E_{0_j,50}^{\frac{4}{2-\epsilon}}\, \Gamma^{-\frac{4\epsilon}{2-\epsilon}}_{0,2.78} t_{3}^{\frac{\epsilon-6}{2-\epsilon}} \hspace{1.30cm} {\rm for} \hspace{0.2cm} {2 < p }\cr
}
\end{eqnarray}
}
{\small
\bary \label{ssc_br_win_eps}
h\nu^{\rm ssc}_{\rm c}&=& 4.7\times 10^{-11}\,{\rm GeV} (1+Y)^{-4} \left(\frac{1+z}{1.022}\right)^{\frac{4(\epsilon-3)}{2-\epsilon}}\, \varepsilon^{-\frac72}_{B,-2}   A_{\rm w,-1}^{-\frac{6-7\epsilon}{2(2-\epsilon)}}\,\theta^{\frac{8}{2-\epsilon}}_{j,1} \Delta\theta^{-\frac{8(3-\epsilon)}{2-\epsilon}}_2 E_{0_j,50}^{-\frac{4}{2-\epsilon}}\, \Gamma^{\frac{4\epsilon}{2-\epsilon}}_{0,2.78}   t_{3}^{\frac{10-3\epsilon}{2-\epsilon}}\cr
F^{\rm ssc}_{\rm max}&=& 2.6\times 10\,{\rm mJy}\left(\frac{p-2}{p-1}\right)^{-1} \left(\frac{1+z}{1.022}\right)^{\frac{4-3\epsilon}{2-\epsilon}}\,\, \varepsilon^{\frac12}_{B,-2}\, d_{\rm z,26.5}^{-2} A_{\rm w,-1}^{\frac{14-5\epsilon}{2(2-\epsilon)}}\,\theta^{\frac{4}{2-\epsilon}}_{j,1} \Delta\theta^{-\frac{4(3-\epsilon)}{2-\epsilon}}_2 E_{0_j,50}^{-\frac{2}{2-\epsilon}}\,\Gamma^{\frac{2\epsilon}{2-\epsilon}}_{0,2.78}   t_{3}^{\frac{\epsilon}{2-\epsilon}}\,.\hspace{2cm}
\eary
}

Given the evolution of the spectral breaks and the maximum flux (Eqs. \ref{ssc_br_win_eps}), we estimate the CRs of SSC model in stellar-wind environment as listed in Table \ref{table2}. 


\subsection{Injection of energy into the blastwave}\label{sec_22}

Continuous energy injection by the central engine on the GRB afterglow can produce refreshed shocks. A continuous luminosity of the central engine can be described by  $L_{\rm inj}(t)\propto L_0\, t^{\rm -q}$  where $q$ is the energy injection index and $L_{\rm inj}$ is the luminosity injected into the blastwave \citep[e.g.][]{2006ApJ...642..354Z}. It is worth noting that for this section we consider $\epsilon=0$ (adiabatic case), and then $E=E_0$. The equivalent kinetic energy can be estimated as $E =\int L_{\rm inj}\, dt\propto L_0 t^{-q+1}$.

\subsubsection{Constant-density medium (${\rm k=0}$)}

The evolution of the bulk Lorentz factor in ISM becomes  {\small $\Gamma=1.5\times 10^3\left(\frac{1+z}{1.022}\right)^{\frac{3}{2}}\,  n^{-\frac{1}{2}}\theta^{-5}_{j,1}\,\Delta\theta_{2}\,E_{0_j,50}^{\frac{1}{2}} t_{3}^{-\frac{q+2}{2}}$}. The Lorentz factors of the electrons with the lowest energy, above which they cool effectively are

{\small
	\begin{eqnarray}\nonumber
		\gamma_{\rm m}= \cases{
			1.7\times10^{4}\,\left(\frac{2-p}{p-1}\right)^{\frac{1}{p-1}}\left(\frac{1+z}{1.022}\right)^{\frac{3(4-p)}{4(p-1)}} n^{-\frac{1}{2(p-1)}} \varepsilon_{e,-1}^{\frac{1}{p-1}}\varepsilon_{B,-2}^{\frac{2-p}{4(p-1)}}\theta_{j,1}^{-\frac{4-p}{2(p-1)}}\Delta\theta_{2}^{\frac{3(4-p)}{2(p-1)}}E_{0_j,50}^{\frac{4-p}{4(p-1)}} t_{3}^{\frac{(p-4)(2+q)}{4(p-1)}}  \hspace{0.2cm} {\rm for} \hspace{0.2cm} { 1<p<2 }\,\,\,\,\,\,\,\cr
			4.7\times10^4\,\left(\frac{p-2}{p-1}\right)\left(\frac{1+z}{1.022}\right)^{\frac{3}{2}} n^{-\frac{1}{2}} \varepsilon_{e,-1}\theta_{j,1}^{-1}\Delta\theta_{2}^3E_{0_j,50}^{\frac{1}{2}} t_{3}^{-\frac{q+2}{2}} \hspace{5.6cm} {\rm for} \hspace{0.2cm} {2 < p }\cr
		}
	\end{eqnarray}
}

{\small
\begin{eqnarray}\label{gamma_q}
\gamma_{\rm c}&=& 5.6\times 10\,\left(\frac{1+z}{1.022}\right)^{-\frac{1}{2}} n^{-\frac{1}{2}} (1+Y)^{-1}\varepsilon_{B,-2}^{-1}  \theta_{j,1}\Delta\theta_{2}^{-1}E_{0_j,50}^{-\frac{1}{2}} t_{6.7}^{\frac{q}{2}},\hspace{7.5cm}
\end{eqnarray}
}
respectively. The characteristic (for $1<p<2$ and $2 < p $) and cooling  spectral breaks, and the maximum flux become

{\small
	\begin{eqnarray}\nonumber  
		\nu^{\rm sync}_{\rm m}= \cases{
  1.7\times 10\,{\rm eV} \left(\frac{2-p}{p-1}\right)^{\frac{2}{p-1}}
			\left(\frac{1+z}{1.022}\right)^{\frac{14-5p}{2(p-1)}} n^{-\frac{3-p}{2(p-1)}} \varepsilon_{e,-1}^{\frac{2}{p-1}}\varepsilon_{B,-2}^{\frac{1}{2(p-1)}}\theta_{j,1}^{-\frac{4-p}{p-1}}\Delta\theta_{2}^{\frac{14-5p}{p-1}}E_{0_j,50}^{\frac{4-p}{2(p-1)}} t_{3}^{\frac{(p-4)(2+q)}{2(p-1)}}  \hspace{0.2cm} {\rm for} \hspace{0.2cm} { 1<p<2 }\,\,\,\,\,\,\,\cr
			1.3\times 10^2\,{\rm eV}\left(\frac{p-2}{p-1}\right)^2\left(\frac{1+z}{1.022}\right)^2 n^{-\frac{1}{2}} \varepsilon_{e,-1}^2 \varepsilon_{B,-2}^{\frac{1}{2}}\theta_{j,1}^{-2}\Delta\theta_{2}^4E_{0_j,50} t_{3}^{-(2+q)} \hspace{3.50cm} {\rm for} \hspace{0.2cm} {2 < p }\cr
		}
	\end{eqnarray}
}

{\small
\begin{eqnarray}\label{nu_c_q}
\nu^{\rm sync}_{\rm c}&=& 1.8\times 10^{-4}\,{\rm eV} \left(\frac{1+z}{1.022}\right)^{-2} n^{-\frac{1}{2}} (1+Y)^{-2}\varepsilon_{B,-2}^{-\frac{3}{2}}\theta_{j,1}^{2}\Delta\theta_{2}^{-4}E_{0_j,50}^{-1} t_{3}^{q}\cr
F^{\rm syn}_{\rm max}&=& 5.7\times 10^{4}\,{\rm mJy}\left(\frac{1+z}{1.022}\right)^{-4} n^{\frac{5}{2}} \varepsilon_{B,-2}^{\frac{1}{2}}d_{z,26.5}^{-2}\theta_{j,1}^{2}\Delta\theta_{2}^{-18}E_{0_j,50}^{-1} t_{3}^{q+5}\,.\hspace{7cm}
\end{eqnarray}
}

Based on the fact that spectral breaks and maximum flux of synchrotron off-axis model in stellar wind evolve throughout time as Eq. \ref{nus_wind}, we derive the CRs shown in Table \ref{table1}.    We derive the CRs reported in Table \ref{table3} from the time-dependent variation in the spectral breaks and maximum flux of the synchrotron off-axis model in ISM, as given in Eqs. \ref{nu_c_q}.

Given the electron Lorentz factors (Eq. \ref{gamma_q}), the synchrotron spectral breaks and the maximum synchrotron flux,  the characteristic (for $1<p<2$ and $2 < p $) and cooling  spectral breaks, and the maximum flux in the SSC scenario become 

{\small
	\begin{eqnarray}\nonumber
		\label{syn_esp_win}
		h\nu^{\rm ssc}_{\rm m} = \cases{ 
			1.1\,\times 10\,{\rm GeV}\,\left(\frac{2-p}{p-1}\right)^{\frac{4}{p-1}}\left(\frac{1+z}{1.022}\right)^{\frac{13-4p}{(p-1)}} n^{-\frac{5-p}{2(p-1)}} \varepsilon_{e,-1}^{\frac{4}{p-1}}\varepsilon_{B,-2}^{\frac{3-p}{2(p-1)}}\theta_{j,1}^{-\frac{2(4-p)}{p-1}}\Delta\theta_{2}^{\frac{2(13-4p)}{p-1}}E_{0_j,50}^{\frac{4-p}{p-1}} t_{3}^{\frac{(p-4)(2+q)}{(p-1)}}  \hspace{0.7cm} {\rm for} \hspace{0.2cm} { 1<p<2 }\cr
			5.5\times 10^2\,{\rm GeV} \left(\frac{p-2}{p-1}\right)^{4}\left(\frac{1+z}{1.022}\right)^{5} n^{-\frac{3}{2}} \varepsilon_{e,-1}^4 \varepsilon_{B,-2}^{\frac{1}{2}}\theta_{j,1}^{-4}\Delta\theta_{2}^{10}E_{0_j,50}^2 t_{3}^{-2(2+q)} \hspace{5.8cm} {\rm for} \hspace{0.2cm} {2 < p  }\cr
		}
	\end{eqnarray}
}

{\small
\bary \label{ssc_q_0}
h\nu^{\rm ssc}_{\rm c}&=& 1.2\times 10^{-9}\,{\rm GeV} \left(\frac{1+z}{1.022}\right)^{-3} n^{-\frac{3}{2}} (1+Y)^{-4}\varepsilon_{B,-2}^{-\frac{7}{2}}\theta_{j,1}^{4}\Delta\theta_{2}^{-6}E_{0_j,50}^{-2} t_{3}^{2q}\cr
F^{\rm ssc}_{\rm max}&=& 1.8\times 10^{-3}\,{\rm mJy}     \left(\frac{p-2}{p-1}\right)^{-1}\left(\frac{1+z}{1.022}\right)^{-5} n^{\frac{7}{2}} \varepsilon_{B,-2}^{\frac{1}{2}}d_{z,26.5}^{-2}\theta_{j,1}^{2}\Delta\theta_{2}^{-20}E_{0_j,50}^{-1} t_{3}^{6+q}\,,
\eary
}

respectively.  Based on the temporal evolution of the spectral breaks and the maximum flux (Eq. \ref{ssc_q_0}), we estimate the CRs of the SSC of off-axis model in ISM as listed in Table \ref{table4}.

\subsubsection{Stellar-wind environment (${\rm k=2}$)}

The evolution of the bulk Lorentz factor during the deceleration phase in stellar-wind medium is {\small $\Gamma= 4.3\times 10^2\left(\frac{1+z}{1.022}\right)^{\frac{1}{2}}\,   A_{\rm w,-1}^{-\frac{1}{2}}\theta^{-1}_{j,1}\Delta\theta_{2} E_{0_j,50}^{\frac{1}{2}}t_{3}^{-\frac{q}{2}}$}.   The Lorentz factors of the electrons with the lowest energy, above which they cool effectively become

{\small
	\begin{eqnarray}\nonumber 
		\gamma_{\rm m}= \cases{
			4.0\times 10^3\,\left(\frac{2-p}{p-1}\right)^{\frac{1}{p-1}}\left(\frac{1+z}{1.022}\right)^{\frac{8-3p}{4(p-1)}} A_{\rm w,-1}^{-\frac{1}{2(p-1)}} \varepsilon_{e,-1}^{\frac{1}{p-1}}\varepsilon_{B,-2}^{\frac{2-p}{4(p-1)}}\theta_{j,1}^{-\frac{4-p}{2(p-1)}}\Delta\theta_{2}^{\frac{8-3p}{2(p-1)}}E_{0_j,50}^{\frac{4-p}{4(p-1)}} t_{3}^{\frac{4-(4-p)(2+q)}{4(p-1)}}  \hspace{0.2cm} {\rm for} \hspace{0.2cm} { 1<p<2 }\,\,\,\,\,\,\,\cr
			1.3\times 10^4\,\left(\frac{p-2}{p-1}\right)\left(\frac{1+z}{1.022}\right)^{\frac{1}{2}} A_{\rm w,-1}^{-\frac{1}{2}} \varepsilon_{e,-1}\theta_{j,1}^{-1}\Delta\theta_{2}E_{0_j,50}^{\frac{1}{2}} t_{3}^{-\frac{q}{2}} \hspace{6.0cm} {\rm for} \hspace{0.2cm} {2 < p }\cr
		}
	\end{eqnarray}
}

{\small
\begin{eqnarray}\label{nu_c_i}
\gamma_{\rm c}&=& 6.3\times 10 \,\left(\frac{1+z}{1.022}\right)^{-\frac{3}{2}} A_{\rm w,-1}^{-\frac{1}{2}} (1+Y)^{-1}\varepsilon_{B,-2}^{-1}  \theta_{j,1}\Delta\theta_{2}^{-3}E_{0_j,50}^{-\frac{1}{2}} t_{3}^{\frac{2+q}{2}},\hspace{7cm}
\end{eqnarray}
}
respectively. The characteristic (for $1<p<2$ and $2 < p $) and cooling  spectral breaks, and the maximum flux become

{\small
	\begin{eqnarray}\label{gamma_q_wind}  
		\nu^{\rm sync}_{\rm m}= \cases{
			1.7\,{\rm eV}\left(\frac{2-p}{p-1}\right)^{\frac{2}{p-1}}\left(\frac{1+z}{1.022}\right)^{\frac{8-3p}{2(p-1)}} A_{\rm w,-1}^{-\frac{3-p}{2(p-1)}} \varepsilon_{e,-1}^{\frac{2}{p-1}}\varepsilon_{B,-2}^{\frac{1}{2(p-1)}}\theta_{j,1}^{-\frac{4-p}{p-1}}\Delta\theta_{2}^{\frac{8-3p}{p-1}}E_{0_j,50}^{\frac{4-p}{2(p-1)}} t_{3}^{-\frac{2 + q(4-p)}{2(p-1)}}  \hspace{0.2cm} {\rm for} \hspace{0.2cm} { 1<p<2 }\,\,\,\,\,\,\,\cr
			1.8\times 10\,{\rm eV}\left(\frac{p-2}{p-1}\right)^{2}\left(\frac{1+z}{1.022}\right) A_{\rm w,-1}^{-\frac{1}{2}} \varepsilon_{e,-1}^2 \varepsilon_{B,-2}^{\frac{1}{2}}\theta_{j,1}^{-2}\Delta\theta_{2}^{2}E_{0_j,50} t_{3}^{-(q+1)} \hspace{3.35cm} {\rm for} \hspace{0.2cm} {2 < p }\cr
		}
	\end{eqnarray}
}

{\small
\begin{eqnarray}\label{nu_c_i}
\nu^{\rm sync}_{\rm c}&=& 4.1\times 10^{-4} \,{\rm eV}\,\left(\frac{1+z}{1.022}\right)^{-3} A_{\rm w,-1}^{-\frac{1}{2}} (1+Y)^{-2}\varepsilon_{B,-2}^{-\frac{3}{2}}\theta_{j,1}^{2}\Delta\theta_{2}^{-6}E_{0_j,50}^{-1} t_{3}^{q+1}\cr
F^{\rm syn}_{\rm max}&=&  4.3\times 10^6\,{\rm mJy}                  \left(\frac{1+z}{1.022}\right) A_{\rm w,-1}^{\frac{5}{2}} \varepsilon_{B,-2}^{\frac{1}{2}}d_{z,26.5}^{-2}\theta_{j,1}^{2}\Delta\theta_{2}^{-8}E_{0_j,50}^{-1} t_{3}^{q}\,.\hspace{7cm}
\end{eqnarray}
}

Our derivation of the CRs of synchrotron off-axis model in stellar wind is reported in Table \ref{table3}. It was derived based on the spectral breaks and the maximum flux (Eq. \ref{nu_c_i}).\\

Given the electron Lorentz factors (Eq. \ref{gamma_q_wind}), the synchrotron spectral breaks and the maximum synchrotron flux,  the characteristic (for $1<p<2$ and $2 < p$) and cooling  spectral breaks, and the maximum flux in the SSC scenario become

{\small
	\begin{eqnarray}\nonumber
		\label{syn_esp_win}
		h\nu^{\rm ssc}_{\rm m} = \cases{ 
			5.5\times 10^{-2}\,{\rm GeV}\,\left(\frac{2-p}{p-1}\right)^{\frac{4}{p-1}}\left(\frac{1+z}{1.022}\right)^{\frac{8-3p}{p-1}} A_{\rm w,-1}^{-\frac{5-p}{2(p-1)}} \varepsilon_{e,-1}^{\frac{4}{p-1}}\varepsilon_{B,-2}^{\frac{3-p}{2(p-1)}}\theta_{j,1}^{-\frac{2(4-p)}{p-1}}\Delta\theta_{2}^{\frac{2(8-3p)}{p-1}}E_{0_j,50}^{\frac{4-p}{p-1}} t_{3}^{-\frac{(3-p) + q(4-p)}{p-1}}  \hspace{0.7cm} {\rm for} \hspace{0.2cm} { 1<p<2 }\cr
			5.9\,{\rm GeV}\,\left(\frac{p-2}{p-1}\right)^{4}\left(\frac{1+z}{1.022}\right)^2 A_{\rm w,-1}^{-\frac{3}{2}} \varepsilon_{e,-1}^4 \varepsilon_{B,-2}^{\frac{1}{2}}\theta_{j,1}^{-4}\Delta\theta_{2}^{4}E_{0_j,50}^2 t_{3}^{-(2q+1)} \hspace{6cm} {\rm for} \hspace{0.2cm} {2 < p }\cr
		}
	\end{eqnarray}
}

{\small
\bary \label{ssc_q_2}
h\nu^{\rm ssc}_{\rm c}&=& 3.2\times 10^{-9} \,{\rm GeV}\,\left(\frac{1+z}{1.022}\right)^{-6} A_{\rm w,-1}^{-\frac{3}{2}} (1+Y)^{-4}\varepsilon_{B,-2}^{-\frac{7}{2}}\theta_{j,1}^{4}\Delta\theta_{2}^{-12}E_{0_j,50}^{-2} t_{3}^{3+2q}\cr
F^{\rm ssc}_{\rm max}&=& 4.4\times 10^{-1}\,{\rm mJy}  \left(\frac{p-2}{p-1}\right)^{-1}\left(\frac{1+z}{1.022}\right)^{2} A_{\rm w,-1}^{\frac{7}{2}} \varepsilon_{B,-2}^{\frac{1}{2}}d_{z,26.5}^{-2}\theta_{j,1}^{2}\Delta\theta_{2}^{-6}E_{0_j,50}^{-1} t_{3}^{q-1}\,.
\eary
}

We estimate the CRs of the SSC off-axis model in stellar-wind medium using the temporal evolution of the spectral breaks and the maximum flux (Eq. \ref{ssc_q_2}), as shown in Table \ref{table4}.

\section{The methodical approach}
\label{sec3}
We select GRBs from the 2FLGC \citep{Ajello_2019}: 86 with temporally extended emission with durations lasting from 31 s to 34366 s \citep{2009MNRAS.400L..75K, 2010MNRAS.409..226K, 2016ApJ...818..190F}. In 2FLGC, the temporal emission is fitted with a SPL function, and GRBs with four or more flux measurements (that are not simply upper limits) are additionally fitted with a BPL.

For 21 bursts for which it was possible to fit the PL and the BPL, we use the value of $\alpha_2$ from the BPL fit, as the BPL fit allows us to investigate the CR later.   For the analysis of the CRs, we follow our previous strategy presented in \cite{2021PASJ...73..970D, 2021ApJS..255...13D, 2020ApJ...903...18S}.  The errorbars quoted in the plots are in the 1$\sigma$ range, and we consider the errorbars among the $\alpha$ and $\beta$ values to be dependent; thus, Figures \ref{fig1}, \ref{fig2}, \ref{fig3} and \ref{fig4} in the result sections are shown with ellipses. 

 When analyzing the CRs, we use the temporal and spectral indices according to the PL function $F_{\rm \nu} \propto t^{-\alpha} \nu^{-\beta}$. We use the temporal index $\alpha$ from the 2FLGC, and the spectral index $\beta$ is taken as the photon index from the catalog minus one.   The distributions of $\alpha$ and $\beta$ are roughly Gaussian, with most values of $\alpha$ ranging between $0.25$ and $2.75$, and most values of $\beta$ ranging between $0.6$ and $1.8$.

We have derived and listed in Tables \ref{table1}, \ref{table2}, \ref{table3} and \ref{table4} the CRs of synchrotron and SSC afterglow model seen off-axis in the adiabatic and radiative scenario, and with and without energy injection.  We consider the SSC afterglow model evolving in ISM and stellar-wind medium, and the CRs as function of the radiative parameter $\epsilon$, the energy injection index $q$, and the electron spectral index for $1<p<2$ and $ 2 < p$. 

\section{Results and Discussion}
\label{sec4}

Based on the CRs of synchrotron and SSC off-axis afterglow models in the adiabatic and radiative scenario with and without energy injection and  for ISM and stellar-wind, we here discuss the results in terms of the bursts reported in 2FLGC.

\subsection{The radiative/adiabatic and energy injection scenarios}

\begin{figure}
{\centering
\resizebox*{\textwidth}{0.4\textheight}
{\includegraphics{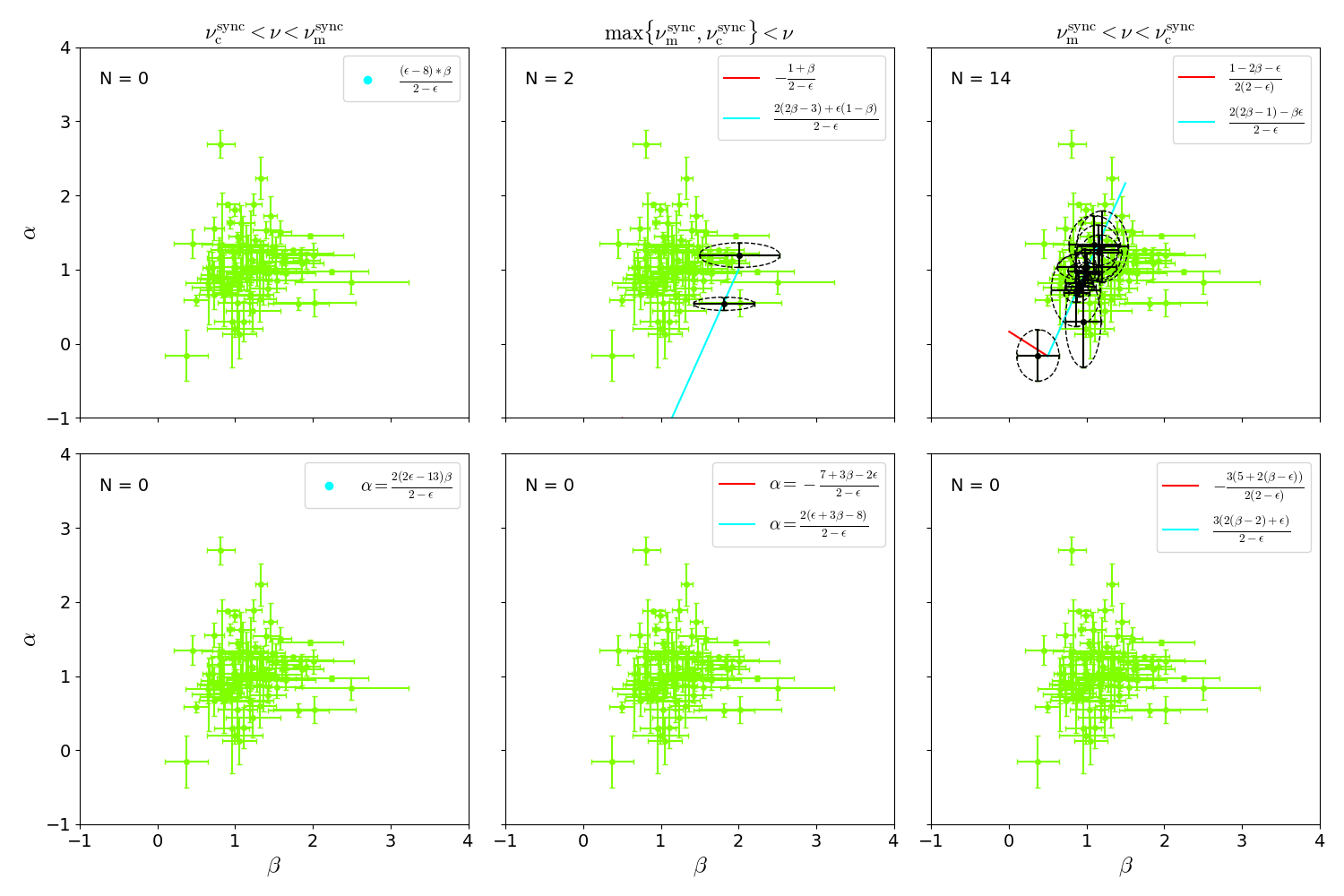}}
} 
\caption{The panels show the CR of synchrotron off-axis scenario in all regimes for constant-density (above) and stellar-wind (below) environment for $\epsilon=0.5$ without energy injection with the spectral and temporal indexes from 2FLGC.   The purple ellipses are displayed when the CR are satisfied within 1 sigma error bars and lines in  red and cyan colors correspond to the expected CRs.}
\label{fig1}
\end{figure}

\begin{figure}
{\centering
\resizebox*{\textwidth}{0.4\textheight}
{\includegraphics{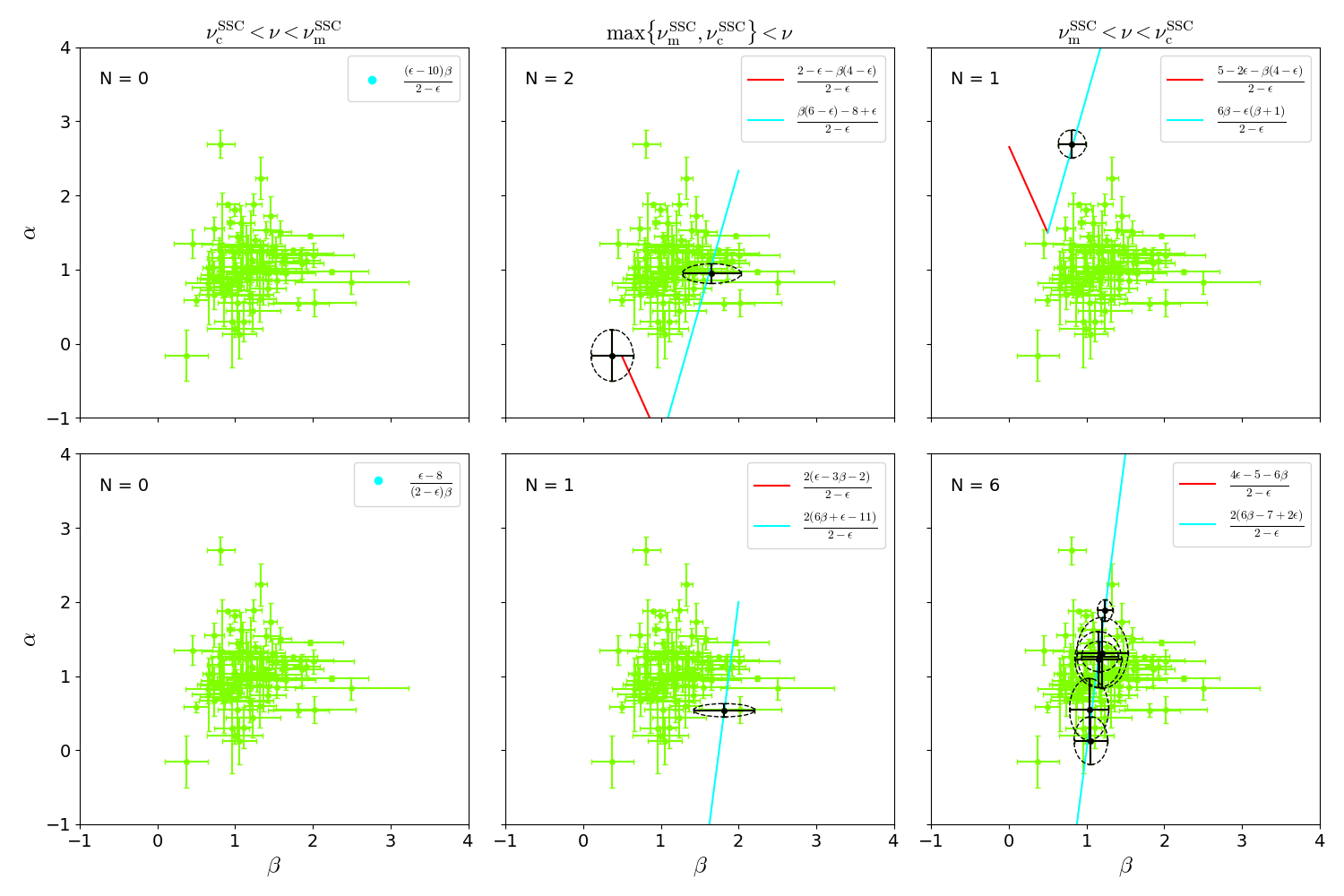}}
} 
\caption{The same as Fig. \ref{fig1}, but for SSC off-axis scenario.}
\label{fig2}
\end{figure}

The radiative scenario with no-energy injection is presented in Figures \ref{fig1} and \ref{fig2} and Tables \ref{table5} and \ref{table6}.   Figure \ref{fig1} shows CRs of synchrotron off-axis model in each cooling condition for slow and fast cooling regime in ISM (panels above) and stellar wind (panels below) for $\epsilon=0.5$ and with spectral index of $1<p<2$ and $2 < p $. Table \ref{table5} shows the number and proportion of GRBs satisfying each cooling condition for the synchrotron off-axis model. The most preferred scenario corresponds to stellar wind for slow cooling regime (${\rm \nu_m^{sync} < \nu_{\rm LAT} < \nu_c^{sync} }$; 16.09\%) followed by fast/slow cooling regime (${\rm max\{\nu_m^{sync},\nu_c^{sync} \} < \nu_{\rm LAT}}$; 2.29\%). The occurrence of 16.09\% and 2.29\% corresponds to 14 and 2 GRBs, respectively.  It is worth noting that no case was found for ISM scenario neither fast not slow cooling regime.  Figure \ref{fig2} shows CRs of SSC off-axis model in each cooling condition for slow and fast cooling regime in ISM (panels above) and stellar wind (panels below) for $\epsilon=0.5$ and with spectral index of $1<p<2$ and $2 < p $. Table \ref{table6} shows the number and proportion of GRBs satisfying each cooling condition for the SSC off-axis model.  The most preferred scenario corresponds to ISM for slow cooling regime  (${\rm \nu_m^{ssc} < \nu_{\rm LAT} < \nu_c^{ssc} }$; 6.89\%) followed by stellar wind for fast/slow cooling regime (${\rm max\{\nu_m^{ssc},\nu_c^{ssc} \} < \nu_{\rm LAT}}$; 2.29\%). The occurrence of 6.89\% and 2.29\% corresponds to 6 and 2 GRBs, respectively. \\
 
\begin{figure}
{\centering
\resizebox*{\textwidth}{0.4\textheight}
{\includegraphics{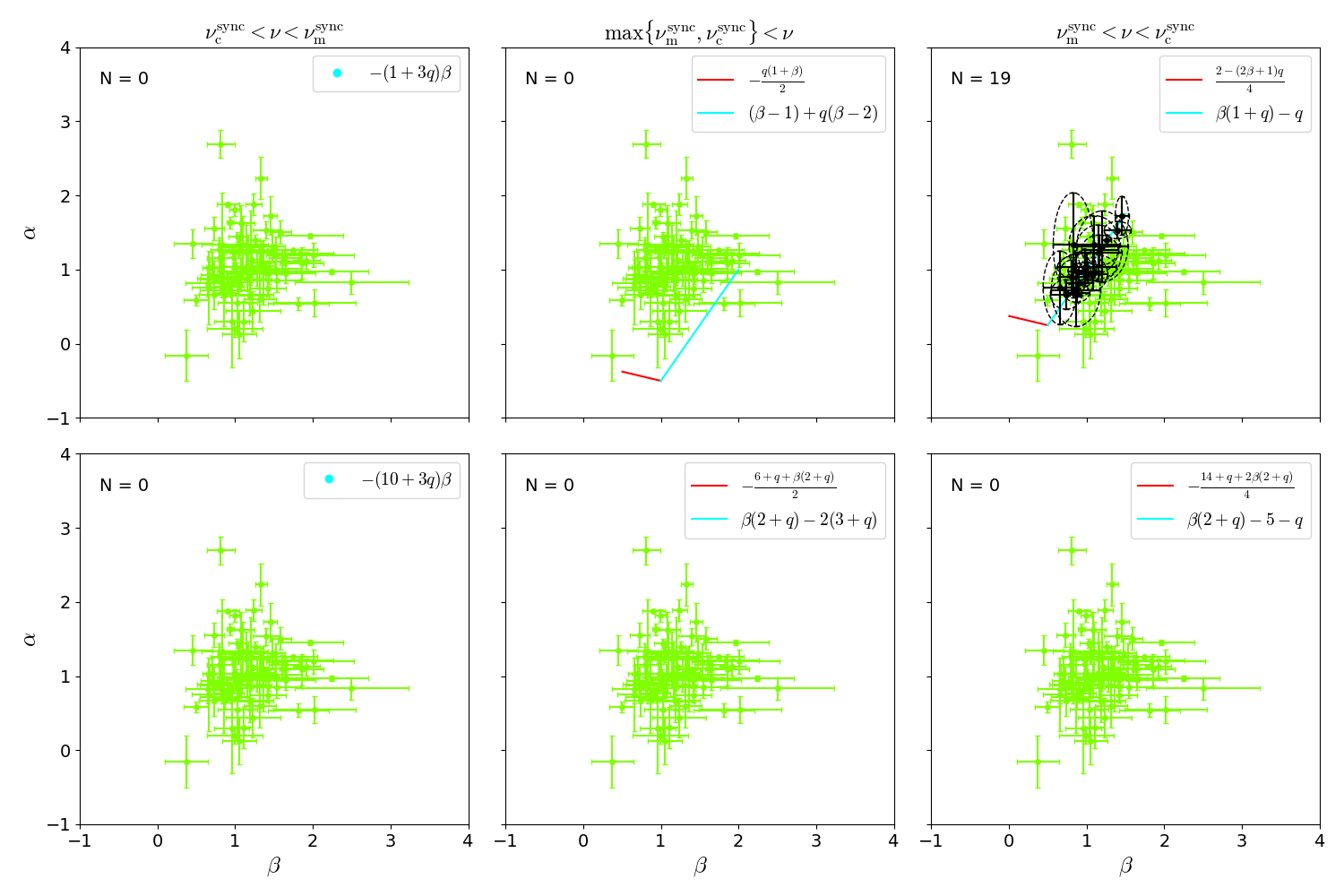}}
} 
\caption{The panels show the closure relations of synchrotron off-axis model in all regimes for ISM (above) and wind (below) environment in the adiabatic regime  with energy injection ($q=0.5$) with the spectral and temporal indexes from 2FLGC.   The purple ellipses are displayed when the CR are satisfied within 1 sigma error bars and lines in  
red and cyan colors correspond to the expected CRs.}
\label{fig3}
\end{figure}

\begin{figure}
{\centering
\resizebox*{\textwidth}{0.4\textheight}
{\includegraphics{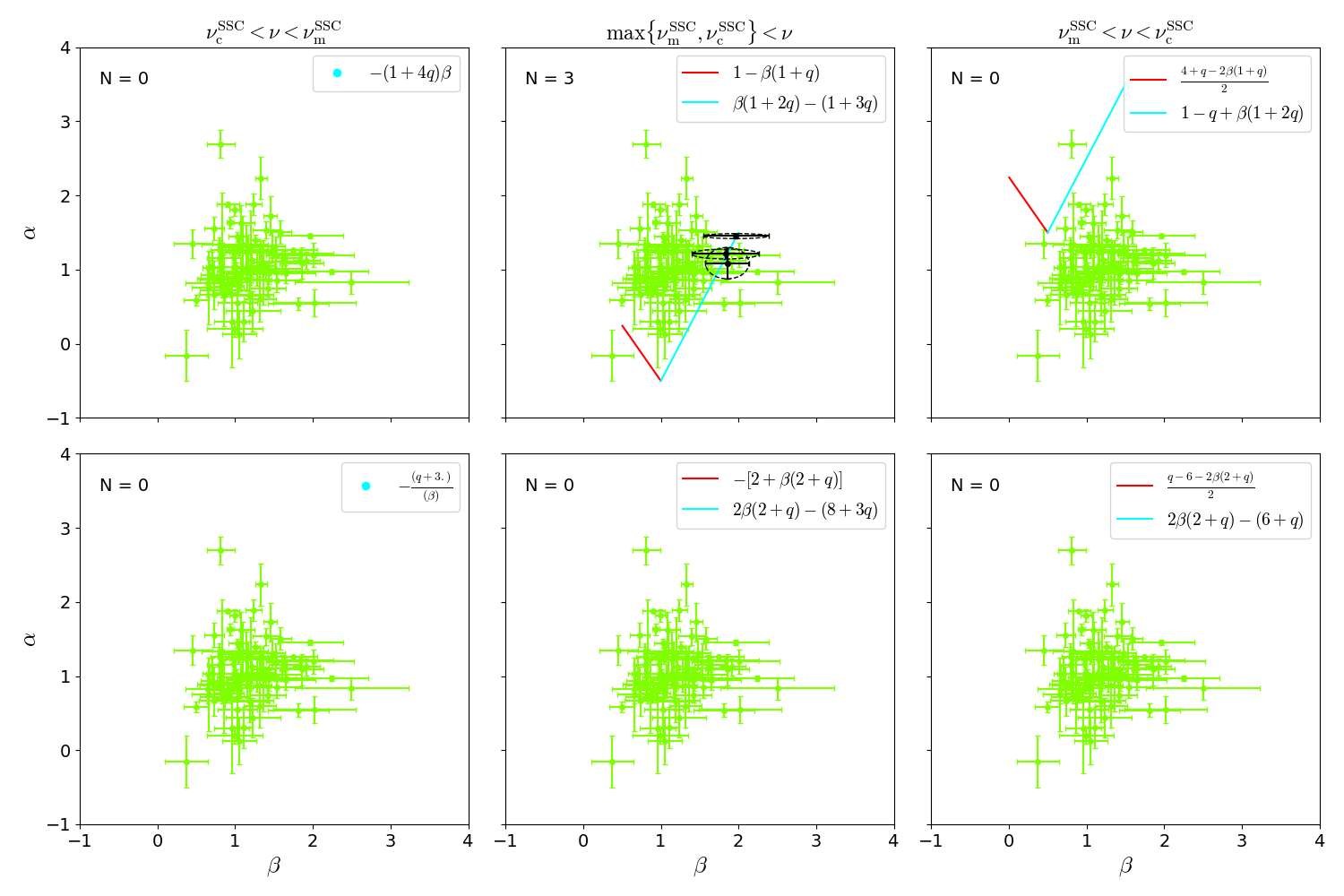}}
} 
\caption{The same as Fig \ref{fig3}, but for SSC off-axis scenario.}
\label{fig4}
\end{figure}

The adiabatic scenario with energy injection is presented in Figures \ref{fig3} and \ref{fig4} and Tables \ref{table7} and \ref{table8}.  Figure \ref{fig3} displays CRs of synchrotron off-axis model in each cooling condition for slow and fast cooling regime in ISM (panels above) and stellar-wind (panels below) medium for $q=0.5$ and with spectral index of $1<p<2$ and $2 < p $. Table \ref{table7} shows the number and proportion of GRBs satisfying each cooling condition for the synchrotron off-axis model. The most preferred scenario corresponds to wind for slow cooling regime (${\rm \nu_m^{sync} < \nu_{\rm LAT} < \nu_c^{sync} }$; 21.83\%), which corresponds to 19 bursts. It is worth noting that none case was found for ISM scenario neither fast not slow cooling regime. Figure \ref{fig4} shows CRs of SSC off-axis model in each cooling condition for slow and fast cooling regime in ISM (panels above) and stellar wind (panels below) for $q=0.5$ and with spectral index of $1<p<2$ and $2 < p $. Table \ref{table8} shows the number and proportion of GRBs satisfying each cooling condition for the SSC off-axis model.  The most preferred scenario corresponds to wind for fast/slow cooling regime (${\rm max\{\nu_m^{sync},\nu_c^{sync} \} < \nu_{\rm LAT}}$; 3.34\%), which corresponds to 3 bursts. It is worth noting that none case was found for ISM scenario neither fast not slow cooling regime.

Regarding the closure relations, which provide the function $\alpha(\beta(p);\epsilon;q)$, we note that uncertainty in the particle energy power-law index ($p$), the radiative parameter ($\epsilon$) and the presence or lack of the energy injection ($q$) as well as whether the emission is on- or off-axis will propagate into an uncertainty in the value of $\alpha$. This uncertainty is included in the large error bars that the observational parameters of $\alpha$ and $\beta$ carry. We also notice that these uncertainties lead to degeneracies in $\alpha$, namely that different combinations of $\beta,\epsilon,q$ will lead to allowed values of $\alpha$. However, the impact of these degeneracies on our analysis is negligible given our choices of $\epsilon=0.5$ and $q=0.5$, which correspond to the average value of $\epsilon$ and $q$.

\subsection{Stellar-wind/density-constant medium}

For both short and long GRB progenitors, the external surroundings' density profiles are used. Short GRBs originate from the merging of binary compact objects, such as a black hole (BH) and a neutron star (NS) or NS-NS \citep{1992ApJ...392L...9D, 1992Natur.357..472U, 1994MNRAS.270..480T, 2011MNRAS.413.2031M}, whereas long GRBs originate from the death of massive stars with differing mass-loss development and the stellar-wind medium \citep{1993ApJ...405..273W,1998ApJ...494L..45P}. A low density constant medium is expected for short GRBs, and the stellar-wind environment is expected for long GRBs associated to massive stars that reach the end of their existence in a variety of ways \citep{2005ApJ...631..435R,  2006A&A...460..105V}.   Note that a wind-to-ISM afterglow transition may be predicted in the scenario of core-collapse of large stars when the wind expelled by its progenitor does not extend beyond the deceleration ratio \citep[e.g., see][]{2017ApJ...848...15F,2019ApJ...879L..26F}.  For synchrotron scenario, we found that 19 and 18 GRBs within stellar-wind medium satisfy at least one CR in the scenario with and without injection, respectively. We note that for this model none GRB within ISM satisfies at least one CR with and without energy injection.  For the SSC model without energy injection, 8 GRBs within ISM and 5 within the stellar-wind environment fulfill at least  one CR. With energy injection, 6 bursts within stellar-wind medium satisfy at least one CR and non with ISM.

Irrespective of the circumburst medium the synchrotron light curve as a function of the density evolves as $F_\nu \propto n^{\frac{9}{4}}$ for ${\rm \nu_c^{sync} < \nu_{\rm LAT} < \nu_m^{sync} }$, $\propto n^{\frac{7+p}{4}}$ for  ${\rm \nu_m^{sync} < \nu_{\rm LAT} < \nu_c^{sync} }$ and $\propto n^{\frac{6+p}{4}}$ for ${\rm max\{\nu_m^{sync},\nu_c^{sync} \} < \nu_{\rm LAT}}$ for $1<p<2$, and  $F_\nu \propto n^{\frac{9}{4}}$ for ${\rm \nu_c^{sync} < \nu_{\rm LAT} < \nu_m^{sync} }$, $\propto n^{\frac{11-p}{4}}$ for  ${\rm \nu_m^{sync} < \nu_{\rm LAT} < \nu_c^{sync} }$ and $\propto n^{\frac{10-p}{4}}$ for ${\rm max\{\nu_m^{sync},\nu_c^{sync} \} < \nu_{\rm LAT}}$ for $2 < p $ with and without energy injection. It indicates that a variation of the flux during the wind-to-ISM afterglow transition is not expected in any scenario, and the flux is more sensible to the density for $2 < p $.

\subsection{Consequences of the viewing angle}

\subsubsection{The maximum energy of synchrotron model}

It is possible to determine the maximal energy emitted by the synchrotron process by equalling the acceleration and synchrotron timescales.    The maximum Lorentz factor of the electron distribution is $\gamma_{\rm max}=\left( \frac{3q_e}{\xi\sigma_T B'_{\rm }}\right)^{\frac12}$ with $\xi$ the Bohm parameter\footnote{In the Bohm limit, this parameter becomes $\xi\sim 1$.} and $q_e$ the electron charge. The maximum energy radiated by the synchrotron model is achieved when the bulk Lorentz factor lies in the line-of-sight ($\delta_{\rm D,max}\approx 2\Gamma$) \citep[for discussion, see][]{2010ApJ...718L..63P}.  In constant-density and stellar-wind medium, the maximal photon energy emitted by the synchrotron process, $h\nu^{\rm syn}_{\rm max}= \frac{\Gamma} {(1+z)} \frac{3q^2_e}{2\pi \sigma_T m_e c}$, can be written as

\begin{eqnarray}
\label{ene_max_eps}
h\nu^{\rm syn}_{\rm max} \approx \cases{ 
46.4\, {\rm MeV}\left(\frac{1+z}{1.022}\right)^{-\frac{5-\epsilon}{8-\epsilon}}\,  n^{-\frac{1}{8-\epsilon}} E_{0,52}^{\frac{1}{8-\epsilon}}\Gamma_{0,2.78}^{-\frac{\epsilon}{8-\epsilon}}t_{3}^{-\frac{3}{8-\epsilon}} \hspace{1.3cm} {\rm for} \hspace{0.2cm} {\rm k=0 }\cr
35.7\, {\rm MeV}\left(\frac{1+z}{1.022}\right)^{-\frac{3-\epsilon}{4-\epsilon}}\,    A_{\rm w,-1}^{-\frac{1}{4-\epsilon}} E_{0,52}^{\frac{1}{4-\epsilon}}\Gamma_{0,2.78}^{-\frac{\epsilon}{4-\epsilon}}t_{3}^{-\frac{1}{4-\epsilon}} \hspace{1.4cm} {\rm for} \hspace{0.2cm} {\rm k=2}\,.
}
\end{eqnarray}

for the radiative scenario, and as

\begin{eqnarray}
\label{ene_max_q}
h\nu^{\rm syn}_{\rm max} \approx \cases{ 
87.2\, {\rm MeV}\left(\frac{1+z}{1.022}\right)^{-\frac{5}{8}}\,    n^{-\frac{1}{8}}E_{0,52}^{\frac{1}{8}}\,t_{3}^{-\frac{2+q}{8}} \hspace{1.5cm} {\rm for} \hspace{0.2cm} {\rm k=0 }\cr
101.3\, {\rm MeV}\left(\frac{1+z}{1.022}\right)^{-\frac{3}{4}}\,    A_{\rm w,-1}^{-\frac{1}{4}}E_{0,52}^{\frac{1}{4}}\,t_{3}^{-\frac{q}{4}} \hspace{1.5cm} {\rm for} \hspace{0.2cm} {\rm k=2}\,,
}
\end{eqnarray}

for the energy injection scenario.   Even while the temporal and spectral indices of GRBs recorded by Fermi-LAT may be understood in the synchrotron scenario, it cannot explain photons with energy exceeding the synchrotron limit. In light of this, a different mechanism, such as SSC, is most suggested rather than the synchrotron scenario.

While the photo-hadronic interactions may produce photons with energies beyond the synchrotron limit, the non-coincidences of neutrinos with GRBs reported by the IceCube team place severe constraints on the number of hadrons engaged in these processes \citep{2012Natur.484..351A, 2016ApJ...824..115A, 2015ApJ...805L...5A}. Thus, synchrotron scenario explanations may work for LAT photons below the synchrotron limit, whereas SSC explanations would apply beyond this limit \citep[e.g., see][]{1997PhRvL..78.2292W}.

Table \ref{table7} of 2FLGC shows 29 Fermi-LAT bursts with photon energy exceeding $> 10\,{\rm GeV}$, all of which must satisfy the CRs of SSC afterglow scenarios, at least in a relatively short amount of time. The highest energy produced by the SSC process is much higher than that of the LAT ($h\nu^{\rm ssc}_{\rm max}=\gamma^2_{\rm max} h\nu^{\rm syn}_{\rm max}\gg 300\,{\rm GeV}$).

Our results are consistent with those bursts detected with evidence of energy injection and off-axis emission at X-ray, optical and radio bands but not at the LAT energy range.  For instance, GRB 160821B and GRB 170817A were observed at X-ray, optical and radio bands and upper limits were derived by Fermi-LAT instrument \citep{2041-8205-848-2-L12, 2017ifs..confE..84P}. In addition, both bursts provide evidence of energy injection. Over 3.3 years after the merger, \cite{2021arXiv210402070H} analyzed the latest X-ray data of GRB 170817A taken with the Chandra X-ray Observatory. These observations were consistent with X-ray radiation from accretion processes on a compact-object remnant. On the other hand,  after describing GRB 160821B with internal energy dissipation, \cite{2017ApJ...835..181L} proposed that the progenitor was consistent with a new born supramassive magnetar.  Both GRBs were modelled with synchrotron off-axis emission by deceleration of a relativistic jet  with large viewing angles, $\theta_{\rm obs}=10.299^{+1.125}_{-1.135}$ degrees \citep{2022ApJ...940..189F} and  ($5 - 10$) degrees \citep{troja2017a, 2017Sci...358.1559K, 2017MNRAS.472.4953L, 2018MNRAS.478..733L, 2018ApJ...867...57R, 2017ApJ...848L..20M}, for GRB 160821B and GRB 170817A, respectively. Therefore, with the best-fit values the maximum synchrotron energies are below the LAT energy range. In general, we want to highlight that high-energy emission at the LAT energy range is not expected from large viewing angles which might occurred at timescales of days.\\

Detail analysis of the prompt episode revealed a delayed onset of the LAT temporally-extended component. In the 2FLGC were reported 38 (7) bursts with the component beginning at time longer than $10^2$ ($10^3$) s. These bursts, in principle, agree with the synchrotron off-axis emission emitted from a second relativistic jet with viewing angles much less than 1 degree.

\subsubsection{The microphysical parameter $\epsilon_B$}

\cite{2019ApJ...883..134T} systematically analyzed the closure relations in a sample of 59 chosen LAT-detected bursts, taking into account temporal and spectral indices. 
They found that while the traditional synchrotron emission explains the spectral and temporal indices in the vast majority of instances, there is still a significant percentage of bursts that cannot be explained with this model. Furthermore, numerous GRBs were identified to fulfill the closure relations of the slow-cooling regime, although with a very low value of the magnetic microphysical parameter ($\epsilon_B<10^{-7}$).  An important point to emphasize is that afterglow modelling studies have yielded values between $3.5\times10^{-5}\leq\epsilon_B\leq0.33$ \citep[e.g., see][]{2014ApJ...785...29S}, and not quite as small as \cite{2019ApJ...883..134T} reported.\\

Here, we constrain the value of $\epsilon_B$ that allows the Fermi-LAT observations to lie in the cooling condition $\nu^{\rm j}_{\rm m}<\nu_{\rm LAT}<\nu^{\rm j}_{\rm c}$ or $\nu^{\rm i}_{\rm m}<\nu^{\rm j}_{\rm c}<\nu_{\rm LAT}$ with ${\rm j=ssc}$ or ${\rm sync}$ for typical values of GRB afterglows.  For the case of synchrotron radiation for ISM and stellar-wind medium with and without energy injection, we have 

{\small
\begin{eqnarray}
\label{eps_B2_sync}
\varepsilon_B \lesssim  \cases{ 
2.2\times10^{-4}\left(\frac{1+z}{1.022}\right)^{-\frac{4}{3}}\, (1+Y(\gamma_c))^{-\frac43}  n^{-\frac{1}{3}}\theta_{j,1}^{\frac43}\Delta\theta^{-\frac{8}{3}}_1E_{0_j,50}^{-\frac{2}{3}}t_{3}^{\frac{2q}{3}} \left(\frac{h\nu^{\rm sync}_c}{100\,{\rm MeV}}\right)^{-\frac23} \hspace{1.6cm} {\rm for} \hspace{0.2cm} {\rm k=0 }\cr
6.3\times10^{-5}\left(\frac{1+z}{1.022}\right)^{-2}\, (1+Y(\gamma_c))^{-\frac43}  A_{\rm w,-1}^{-\frac{1}{3}}\theta_{j,1}^{\frac43}\Delta\theta^{-4}_1E_{0_j,50}^{-\frac{2}{3}}t_{3}^{\frac{2(1+q)}{3}} \left(\frac{h\nu^{\rm sync}_c}{100\,{\rm MeV}}\right)^{-\frac23} \hspace{1.0cm} {\rm for} \hspace{0.2cm} {\rm k=2 }\,,
}
\end{eqnarray}
}

and

{\small
\begin{eqnarray}
\label{eps_B1_sync}
\varepsilon_B \lesssim \cases{ 
2.5\times10^{-4}\left(\frac{1+z}{1.022}\right)^{-\frac{2(4+\epsilon)}{3(2-\epsilon)}}\, (1+Y(\gamma_c))^{-\frac43}  n^{\frac{3\epsilon-2}{3(2-\epsilon)}}\theta_{j,1}^{\frac{8}{3(2-\epsilon)}}\Delta\theta^{-\frac{4(4+\epsilon)}{3(2-\epsilon)}}_1E_{0_j,50}^{-\frac{4}{3(2-\epsilon)}}\Gamma_{0,2.78}^{\frac{4\epsilon}{3(2-\epsilon)}}t_{3}^{\frac{4(\epsilon+1)}{3(2-\epsilon)}} \left(\frac{h\nu^{\rm sync}_c}{100\,{\rm MeV}}\right)^{-\frac23} \hspace{1.3cm} {\rm for} \hspace{0.2cm} {\rm k=0 }\cr
1.7\times10^{-5}\left(\frac{1+z}{1.022}\right)^{-\frac{4(3-\epsilon)}{3(2-\epsilon)}}\, (1+Y(\gamma_c))^{-\frac43}  A_{\rm  w,-1}^{-\frac{2-3\epsilon}{3(2-\epsilon)}}\theta_{j,1}^{\frac{8}{3(2-\epsilon)}}\Delta\theta^{-\frac{8(3-\epsilon)}{3(2-\epsilon)}}_1E_{0_j,50}^{-\frac{4}{3(2-\epsilon)}}\Gamma_{0,2.78}^{\frac{4\epsilon}{3(2-\epsilon)}}t_{3}^{\frac{2(4-\epsilon)}{3(2-\epsilon)}} \left(\frac{h\nu^{\rm sync}_c}{100\,{\rm MeV}}\right)^{-\frac23} \hspace{1.05cm} {\rm for} \hspace{0.2cm} {\rm k=2 },
}
\end{eqnarray}
}

respectively.   For the case of SSC radiation for ISM and stellar-wind medium with and without energy injection, we have 

{\small
\begin{eqnarray}
\label{eps_B2}
\varepsilon_B \lesssim  \cases{ 
2.2\times10^{-4}\left(\frac{1+z}{1.022}\right)^{-\frac{6}{7}}\, (1+Y(\gamma_c))^{-\frac87}  n^{-\frac{3}{7}}\theta_{j,1}^{\frac87}\Delta\theta^{-\frac{12}{7}}_1E_{0_j,50}^{-\frac{4}{7}}t_{3}^{\frac{4q}{7}} \left(\frac{h\nu^{\rm ssc}_c}{100\,{\rm MeV}}\right)^{-\frac27} \hspace{2.0cm} {\rm for} \hspace{0.2cm} {\rm k=0 }\cr
6.3\times10^{-5}\left(\frac{1+z}{1.022}\right)^{-\frac{12}{7}}\, (1+Y(\gamma_c))^{-\frac87}  A_{\rm w,-1}^{-\frac{3}{7}}\theta_{j,1}^{\frac87}\Delta\theta^{-\frac{24}{7}}_1 E_{0_j,50}^{-\frac{4}{7}}t_{3}^{\frac{2(3+2q)}{7}} \left(\frac{h\nu^{\rm ssc}_c}{100\,{\rm MeV}}\right)^{-\frac27} \hspace{1.0cm} {\rm for} \hspace{0.2cm} {\rm k=2 }\,,
}
\end{eqnarray}
}

and

{\small
\begin{eqnarray}
\label{eps_B1}
\varepsilon_B \lesssim \cases{ 
2.5\times10^{-4}\left(\frac{1+z}{1.022}\right)^{-\frac{6(2+\epsilon)}{7(2-\epsilon)}}\, (1+Y(\gamma_c))^{-\frac87}  n^{\frac{7\epsilon-6}{7(2-\epsilon)}}\theta_{j,1}^{\frac{16}{7(2-\epsilon)}}\Delta\theta^{-\frac{12(2+\epsilon)}{7(2-\epsilon)}}_1E_{0_j,50}^{-\frac{8}{7(2-\epsilon)}}\Gamma_{0,2.78}^{\frac{8\epsilon}{7(2-\epsilon)}}t_{3}^{\frac{8(\epsilon+1)}{7(2-\epsilon)}} \left(\frac{h\nu^{\rm ssc}_c}{100\,{\rm MeV}}\right)^{-\frac27} \hspace{1.3cm} {\rm for} \hspace{0.2cm} {\rm k=0 }\cr
1.7\times10^{-5}\left(\frac{1+z}{1.022}\right)^{\frac{8(\epsilon-3)}{7(2-\epsilon)}}\, (1+Y(\gamma_c))^{-\frac87}  A_{\rm w,-1}^{-\frac{6-7\epsilon}{7(2-\epsilon)}}\theta_{j,1}^{\frac{16}{7(2-\epsilon)}}\Delta\theta^{-\frac{16(3-\epsilon)}{7(2-\epsilon)}}_1E_{0_j,50}^{-\frac{8}{7(2-\epsilon)}}\Gamma_{0,2.78}^{\frac{8\epsilon}{7(2-\epsilon)}}t_{3}^{\frac{2(10-3\epsilon)}{7(2-\epsilon)}} \left(\frac{h\nu^{\rm ssc}_c}{100\,{\rm MeV}}\right)^{-\frac27} \hspace{1.0cm} {\rm for} \hspace{0.2cm} {\rm k=2 },
}
\end{eqnarray}
}
respectively.

The Compton parameter is non-constant when the Klein-Nishina (KN) effects are significant. In this case, this parameter can be estimated with 

{\small
\begin{eqnarray}
\label{Yth_kn}
Y(\gamma_c)[Y(\gamma_c)+1] = \frac{\varepsilon_{e}}{\varepsilon_{B}}  \left(\frac{\gamma_{\rm m}}{\gamma_{\rm c}}\right)^{p-2}\,\cases{ 
1\, \hspace{4.40cm} {\rm for} \hspace{0.2cm} {\nu^{\rm syn}_{\rm c} < \nu^{\rm sync}_{\rm KN}(\gamma_{\rm c})}\cr
\left(\frac{\nu^{\rm sync}_{\rm KN}(\gamma_{\rm c})}{\nu^{\rm syn}_c}\right)^{-\frac{p-3}{2}} \, \hspace{2.3cm} {\rm for} \hspace{0.2cm} { \nu^{\rm sync}_{\rm m} < \nu^{\rm syn}_{\rm KN}(\gamma_{\rm c}) < \nu^{\rm sync}_{\rm c}}\cr
\left( \frac{\nu^{\rm sync}_{\rm m}}{\nu^{\rm sync}_{\rm c}} \right)^{-\frac{p-3}{2}}\,\left( \frac{\nu^{\rm sync}_{\rm KN}(\gamma_{\rm c})}{\nu^{\rm sync}_{\rm m}}\right)^\frac43 \, \hspace{1cm} {\rm for} \hspace{0.2cm} {\nu^{\rm sync}_{\rm KN}(\gamma_{\rm c}) <\nu^{\rm sync}_{\rm m} }\,,
}
\end{eqnarray}
}

where $h\nu^{\rm sync}_{\rm KN} (\gamma_c) \simeq\frac{\delta_D}{(1+z)}\,\frac{m_e c^2}{\gamma_c}$ for $\nu^{\rm sync}_{\rm c} > \nu^{\rm sync}_{\rm m}$ \citep[see][]{2009ApJ...703..675N, 2010ApJ...712.1232W}.

Determining the Lorentz factor of electrons that may release high-energy photons through the synchrotron process, including a new spectrum break, and recalculating the Compton value are all necessary to characterize the LAT data above 100 MeV. The new spectral break is given by $h\nu^{\rm sync}_{\rm KN}(\gamma_*)$.
The new Compton parameter ($Y(\gamma_{*})$), for instance, is defined as {\small $Y(\gamma_*)=Y(\gamma_c) \left(\frac{\nu_{*}}{\nu_c}\right)^{\frac{p-3}{4}}   \left(\frac{\nu^{\rm sync}_{\rm KN}(\gamma_{\rm c})}{\nu^{\rm sync}_c}\right)^{-\frac{p-3}{2}}$} for $ \nu^{\rm sync}_{\rm m} <  h\nu^{\rm sync}_{\rm KN}(\gamma_*)=100\,{\rm MeV} < \nu^{\rm sync}_{\rm c} < \nu^{\rm sync}_{\rm KN}(\gamma_{\rm c})$  \citep[for details, see][]{2010ApJ...712.1232W}.\\

Note that for $Y(\gamma_c)\ll 1$, the original Eqs. \ref{eps_B1} and \ref{eps_B2} remain intact, and  for $Y(\gamma_c)\gg 1$, $\epsilon_B$ increases. For $Y(\gamma_c)\gg 1$ and with the spectral breaks in the order $ \nu^{\rm sync}_{\rm m} <  h\nu^{\rm sync}_{\rm KN}(\gamma_*) < \nu^{\rm sync}_{\rm c} < \nu^{\rm sync}_{\rm KN}(\gamma_{\rm c})$, the new Compton parameter is $Y(\gamma_*)\simeq 1$ in terms of usual parameters of GRB afterglow \citep[see ][]{2010ApJ...712.1232W}.

\begin{figure}
{\centering
\resizebox*{\textwidth}{0.6\textheight}
{\includegraphics{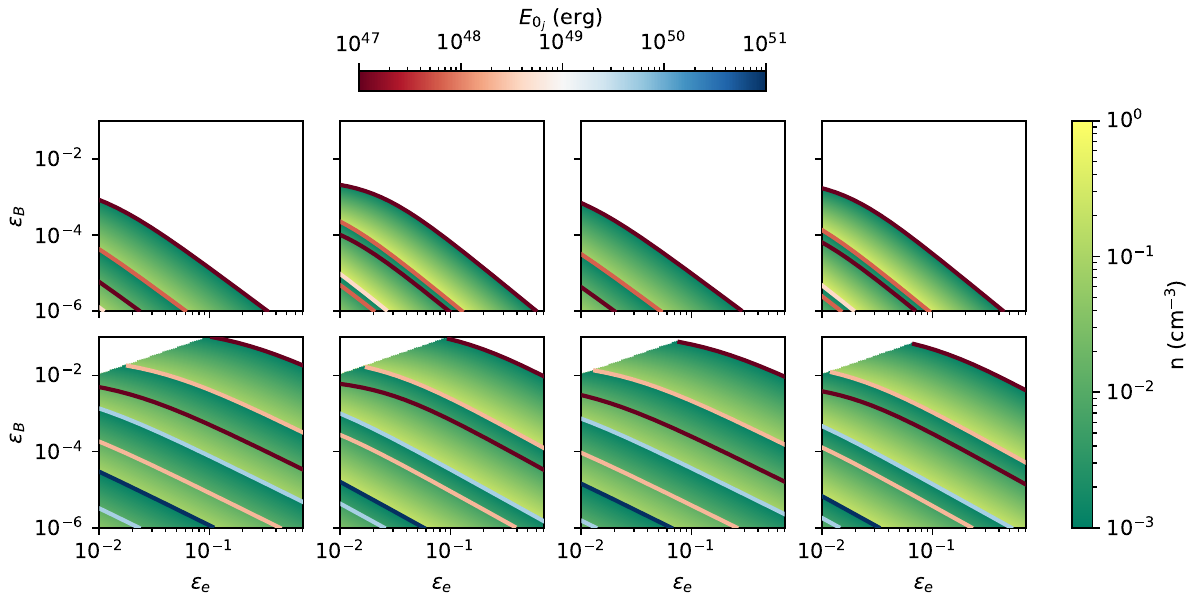}}
} 
\caption{The 4-D parameter space (for the forward shock model with $\rm k=0$) of the expected values of $\varepsilon_{\rm B}$ as a function of typical values of $\varepsilon_{\rm e}$, $n$ and $E_{0,j}$ with two different values of $p$, for the cooling break of synchrotron (row 1) and SSC (row 2).  We consider typical values in the ranges $10^{-2} \leq \varepsilon_{\rm e}\leq 1$, $10^{-3}\leq n \leq 1\,{\rm cm^{-3}}$ and $10^{47}\leq E_{\rm 0,j} \leq 10^{51}\,{\rm erg}$ for $p=2.2$ (columns one and three) and $2.6$ (columns two and four).  Columns one and two corresponds to $q=0.7$ and columns three and four to $q=0.3$.}
\label{fig9}
\end{figure}

\begin{figure}
{\centering
\resizebox*{\textwidth}{0.6\textheight}
{\includegraphics{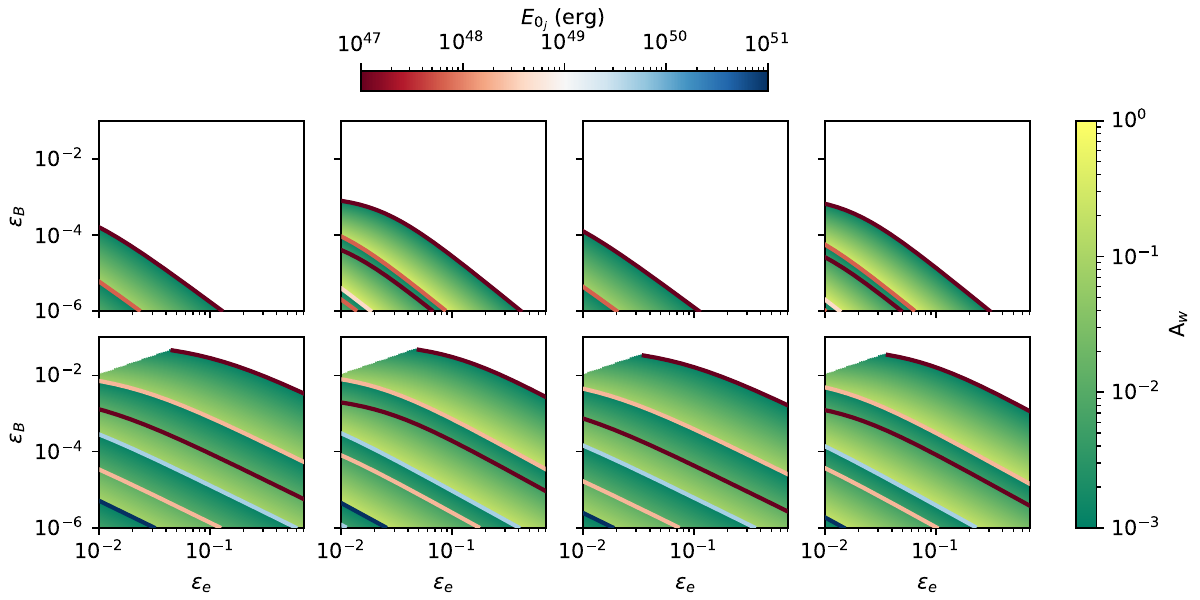}}
} 
\caption{The same as Fig. \ref{fig9}, but for wind-like scenario ($\rm k=2$) with $10^{-3}\leq A_{\rm W}\leq 1$.}
\label{fig10}
\end{figure}

In Figures \ref{fig9} and \ref{fig10}, we display the constraints on the value of $\varepsilon_{\rm B}$ in the case of adiabatic evolution (Eqs. \ref{eps_B2_sync} and \ref{eps_B2}) for ${\rm k=0}$ and ${\rm k=2}$, respectively. We plot the constraints as a function of $\epsilon_{\rm e}$, $n$ ($A_{\rm w}$) and $E_{0,j}$ over a large range of values for $t=5\times 10^3\,{\rm s}$. The first row of both figures uses synchrotron constraints (eq. \ref{eps_B2_sync}) while the second row uses SSC (eq. \ref{eps_B2}). We explore two different values for the parameters $p$ and $q$, namely in the two leftmost columns we choose $q=0.7$ and in the two rightmost columns we pick $q=0.3$, while the first and third columns correspond to $p=2.2$ and the second and fourth columns adopt the value $p=2.6$.  In general, by comparison between both Figures we notice that the case ${\rm k=2}$ leads to better constraints in $\varepsilon_{\rm B}$, that is, the value of $\varepsilon_{\rm B}$ is smaller as ${\rm k}$ is increased. We also notice that for both values of ${\rm k}$, the synchrotron scenario leads to smaller values of $\varepsilon_{\rm B}$ in contrast to the SSC case.
Regarding the value of the density parameter, we note that, in all panels, an increase in this parameter leads to a decrease in $\varepsilon_{\rm B}$. This can be readily observed from Eqs. \ref{eps_B2_sync} and \ref{eps_B2}, as they all have negative powers in the density. By the same argument, we make the same conclusions for $E_{0,j}$.   The spectral index $p$ does not appear explicitly in Eqs. \ref{eps_B2_sync} and \ref{eps_B2}. It is implicit in the Compton parameter given by eq. \ref{Yth_kn}. In the synchrotron case, we note that for a larger value of $p$ we obtain a larger value of $\varepsilon_{\rm B}$. For the SSC scenario, the relation between both parameters is not the same and we observe that it depends non-trivially on the value of $E_{0,j}$. In a similar fashion, the dependence of $\varepsilon_{\rm B}$ on $\varepsilon_{\rm e}$ is also implicit in the Compton parameter, but in this case the figures show in general that as $\varepsilon_{\rm e}$ increases, $\varepsilon_{\rm B}$ decreases. Finally, in the case of the parameter $q$, variation does not lead to a large difference. In general, for smaller $q$, we obtain a smaller $\varepsilon_{\rm B}$ but the difference is difficult to perceive in the both figures.  From Figures \ref{fig9} and \ref{fig10} we display that when the synchrotron and SSC off-axis afterglow models are taken into account, $\varepsilon_B$ always fall within the range of values given by \cite{2014ApJ...785...29S}.

\section{Modelling,  Timescale analysis and  Dynamics of on-axis emission}

For the on-axis emission, the Doppler factor can be approximated as $\delta_D\simeq\frac{2\Gamma}{1+\left(\Gamma\Delta\theta\right)^2}\simeq 2\Gamma$ for $\Gamma\Delta\theta\ll1$ \citep{2002ApJ...570L..61G, 2018PTEP.2018d3E02I}. We consider the radiative/adiabatic and the energy injection scenarios.

\subsection{Radiative/adiabatic scenario}

\subsubsection{Constant-density medium (${\rm k=0}$)}
Given the evolution of the bulk Lorentz factor before afterglow emission enters the observer's field of view $\Gamma\propto t^{-\frac{3}{8- \epsilon}}$ \citep{1998MNRAS.298...87D, 1998ApJ...497L..17S, 2000ApJ...532..281B}, the Lorentz factor of the electrons with the lowest energy and the Lorentz factor of the electrons with energy above which they cool effectively evolve as $\gamma_m\propto t^{-\frac{6(4-p)}{4(p-1)[8-\epsilon]}}$ ($t^{-\frac{3}{8 - \epsilon}}$) for $1<p<2$ ($2<p$)  and $\gamma_c\propto t^{\frac{1+\epsilon }{8-\epsilon}}$, respectively. The synchrotron spectral breaks evolve as $\nu^{\rm sync}_{\rm m}\propto t^{-\frac{6(p+2)}{2(p-1)[8-\epsilon]}}$ ($t^{-\frac{24}{2[8-\epsilon]}}$) for $1<p<2$ ($2<p$) and $\nu^{\rm sync}_{\rm c}\propto t^{\frac{4(\epsilon-2)}{2[8-\epsilon]}}$, and the maximum synchrotron flux as $F^{\rm sync}_{\rm max}\propto t^{-\frac{3\epsilon}{8-\epsilon}}$. The SSC spectral breaks and maximum flux for SSC emission evolve as  $\nu^{\rm ssc}_{\rm m}\propto t^{-\frac{36}{2(p-1)[8-\epsilon]}}$ ($t^{-\frac{18}{8-\epsilon}}$) for $1<p<2$ ($2<p$), $\nu^{\rm ssc}_{\rm c}\propto t^{-\frac{2-4\epsilon}{8-\epsilon}}$ and  $F^{\rm ssc}_{\rm max}\propto t^{\frac{2-4\epsilon}{8-\epsilon}}$, respectively.    During the post-jet-break decay phase, the bulk Lorentz factor becomes $\Gamma\propto t^{-\frac{3}{6-\epsilon}}$.  The minimum and the cooling electron Lorentz factor are given by $\gamma_m\propto t^{-\frac{3}{6-\epsilon}}$ and $\gamma_c\propto t^{\frac{3+\epsilon}{6-\epsilon}}$, respectively.  The synchrotron spectral breaks and maximum flux have the proportionalities  $\nu^{\rm sync}_{\rm m}\propto t^{-\frac{12}{6-\epsilon}}$, $\nu^{\rm sync}_{\rm c}\propto t^{\frac{2\epsilon}{6-\epsilon}}$ and $F^{\rm sync}_{\rm max}\propto t^{-\frac{3(2+\epsilon)}{6-\epsilon}}$, respectively. The SSC synchrotron spectral breaks and maximum flux for SSC emission evolve as $\nu^{\rm ssc}_{\rm m}\propto t^{-\frac{18}{6-\epsilon}}_{\rm }$, $\nu^{\rm ssc}_{\rm c} \propto t^{\frac{6+4\epsilon}{6-\epsilon}}_{\rm }$ and  $F^{\rm ssc}_{\rm max} \propto t^{-\frac{6+4\epsilon}{6-\epsilon}}_{\rm }$, respectively.

\subsubsection{Stellar-wind environment (${\rm k=2}$)}

Given the evolution of the bulk Lorentz factor before afterglow emission enters the observer's field of view $\Gamma\propto t^{-\frac{1}{4- \epsilon}}$ \citep{2000ApJ...532..281B, 2010MNRAS.403..926G}, the Lorentz factor of the electrons with the lowest energy and the Lorentz factor of the electrons with energy above which they cool effectively evolve as $\gamma_m\propto t^{-\frac{2(4-\epsilon)-p(3-\epsilon)}{2(p-1)(4-\epsilon)}}$ ($t^{-\frac{1}{4 - \epsilon}}$) for $1<p<2$ ($2<p$)  and $\gamma_c\propto t^{\frac{3-\epsilon}{4-\epsilon}}$, respectively. The synchrotron spectral breaks evolve as $\nu^{\rm sync}_{\rm m}\propto t^{-\frac{4+p-\epsilon}{(p-1)(4-\epsilon)}}$ ($t^{-\frac{6-\epsilon}{4-\epsilon}}$) for $1<p<2$ ($2<p$) and $\nu^{\rm sync}_{\rm c}\propto t^{\frac{2-\epsilon}{4-\epsilon}}$, and the maximum synchrotron flux as $F^{\rm sync}_{\rm max}\propto t^{-\frac{2}{4-\epsilon}}$. The SSC spectral breaks and maximum flux for SSC emission evolve as  $\nu^{\rm ssc}_{\rm m}\propto t^{-\frac{3(4-\epsilon)-p(2-\epsilon)}{(p-1)(4-\epsilon)}}$ ($t^{-\frac{8-\epsilon}{4-\epsilon}}$) for $1<p<2$ ($2<p$), $\nu^{\rm ssc}_{\rm c}\propto t^{\frac{8-3\epsilon}{4-\epsilon}}$ and  $F^{\rm ssc}_{\rm max}\propto t^{-1}$.  During the post-jet-break decay phase, the bulk Lorentz factor becomes $\Gamma\propto t^{-\frac{1}{2-\epsilon}}$.  The minimum and the cooling electron Lorentz factor are given by $\gamma_m\propto t^{-\frac{1}{2-\epsilon}}$ and $\gamma_c\propto t^{\frac{1-\epsilon}{2-\epsilon}}$, respectively.  The synchrotron spectral breaks and maximum flux are  $\nu^{\rm sync}_{\rm m}\propto t^{-\frac{4-\epsilon}{2-\epsilon}}$, $\nu^{\rm sync}_{\rm c}\propto t^{-\frac{\epsilon}{2-\epsilon}}$ and $F^{\rm sync}_{\rm max}\propto t^{-\frac{2}{2-\epsilon}}$, respectively. The SSC spectral breaks and maximum flux for SSC emission evolve as $\nu^{\rm ssc}_{\rm m}\propto t^{-\frac{6-\epsilon}{2-\epsilon}}_{\rm }$, $\nu^{\rm ssc}_{\rm c} \propto t^{\frac{2-3\epsilon}{2-\epsilon}}_{\rm }$ and  $F^{\rm ssc}_{\rm max} \propto t^{-1}_{\rm }$, respectively.

\subsubsection{Timescale Analysis}

The observed flux increases when the jet decelerates in the circumstellar medium until the bulk Lorentz factor becomes $\Gamma=\frac{b}{\Delta\theta}$; with $b$ a parameter close to unity ($b\lesssim 1$ or $b\gtrsim 1$), which depends on the model \citep[e.g., see][for a discussion]{2002ApJ...570L..61G, 2002ApJ...579..699N}. From the evolution of the Lorentz factor,  the maximum flux takes place at

{\small
\begin{eqnarray}
\label{t_pk_ep}
t_{\rm pk} = \cases{ 
3.7\times 10^3\,{\rm s}\, \left(\frac{1+z}{1.022}\right) n^{-\frac13} E_{\rm 0,51.7}^{\frac{1}{3}} (1+b^2)\, \Gamma_{0,2}^{-\frac{\epsilon}{3}} \left(\frac{\Delta\theta_{2}}{b}\right)^{\frac{8-\epsilon}{3}},\hspace{0.8cm} {\rm for\,\,\, k=0} \cr
9.6\times 10^2\,{\rm s}\, \left(\frac{1+z}{1.022}\right) A_{\rm w,-1}^{-1} E_{\rm 0,51.7} (1+b^2) \Gamma_{0,2}^{-\epsilon} \left(\frac{\Delta\theta_{2}}{b}\right)^{4-\epsilon},\,\,\,\,\, \hspace{0.5cm} {\rm for\,\,\, k=2}\,, \cr
}
\end{eqnarray}
}

For $\epsilon=0$ and $b=1$, the values reported in \cite{2002ApJ...579..699N} are recovered. The jet break transition  occurs when the bulk Lorentz factor becomes $\Gamma=1/\theta_j$  \citep{1999ApJ...525..737R, 1999ApJ...519L..17S}. In the most general case, it happens at

{\small
\begin{eqnarray}
\label{t_j_ep}
t_{\rm j} = \cases{ 
1.4\times 10^4\,{\rm s}\, \left(\frac{1+z}{1.022}\right) n^{-\frac13} E_{\rm 0,j,47.9}^{\frac{1}{3}} \left[1+\left(\frac{\Delta\theta_{2}}{\theta_{j,1}}\right)^2\right]   \, \Gamma_{0,2}^{-\frac{\epsilon}{3}} \theta_{j,1}^{\frac{6-\epsilon}{3}},\hspace{0.8cm} {\rm for\,\,\, k=0} \cr
1.8\times 10^5\,{\rm s}\, \left(\frac{1+z}{1.022}\right) A_{\rm w,-1}^{-1} E_{\rm 0,j,47.9} \left[1+\left(\frac{\Delta\theta_{2}}{\theta_{j,1}}\right)^2\right] \Gamma_{0,2}^{-\epsilon} \theta_{j,1}^{2-\epsilon},\,\,\,\,\, \hspace{0.5cm} {\rm for\,\,\, k=2}\,, \cr
}
\end{eqnarray}
}

where the term $1+\left(\frac{\Delta\theta}{\theta_j}\right)^2$ is obtained with the approximation $\delta_D\approx \frac{2\Gamma}{1+\left(\Gamma\Delta\theta\right)^2}$ and considering $\Gamma=1/\theta_j$. For the on-axis scenario ($\Delta\theta=0$), the Doppler factor $\delta_D\approx 2\Gamma$ is obtained, and for $\epsilon=0$, the values derived in \cite{1999ApJ...519L..17S} are recovered.   From Eqs. \ref{t_pk_ep} and \ref{t_j_ep}, the relationship between $t_{\rm pk}$ and $t_{\rm j}$ can be written as

{\small
\begin{eqnarray}
\label{t_peak_1}
t_{\rm pk} = \frac{1+b^2}{1 + \left( \frac{\Delta\theta}{\theta_{\rm j}}\right)^2}\times\cases{ 
\, \frac{1}{b^{\frac{8-\epsilon}{3}} } \left( \frac{\Delta\theta}{\theta_{\rm j}}\right)^{\frac{8-\epsilon}{3}}t_{\rm j},\hspace{0.8cm} {\rm for\,\,\, k=0} \cr
\, \frac{1}{b^{4-\epsilon} } \left( \frac{\Delta\theta}{\theta_{\rm j}}\right)^{4-\epsilon}t_{\rm j},\,\,\,\,\, \hspace{0.7cm} {\rm for\,\,\, k=2}\,. \cr
}
\end{eqnarray}
}
In Figure \ref{fig5}, we plot the time when the maximum flux is reached $t_{\rm pk}$ as a function of $\Delta\theta$ and the parameter $b$ according to Eq. \ref{t_pk_ep}. We explore two different values for the rest of the model parameters, namely in the two leftmost columns we choose $E_0=5\times10^{51}\ \mathrm{erg}$ and in the two rightmost columns we pick $E_0=5\times10^{53}\ \rm erg$, while the first row corresponds to $k=0$ (with ${\rm n=1\ \mathrm{cm}^{-3}}$) and the second to $k=2$ (with $A_{\rm w}=0.1$). Finally, the first and third columns take the value $\epsilon=0.8$, while columns two and four assume $\epsilon=0.2$.   In general, we notice that in all panels in this Figure an increase in the $b$-parameter leads to a decrease in $t_{\rm pk}$. We take note that small variations of $b$ lead to variations over orders of magnitude in $t_{\rm pk}$ and that this change is more apparent in the case where $k=2$. Another general feature found in all panels is that as $\Delta\theta$ grows, so does $t_{\rm pk}$. Upon comparison between columns one and two, and three and four, we note that changes in $\epsilon$ do not lead to substantial changes in $t_{\rm pk}$, other than an increase of $t_{\rm pk}$ for large $\Delta\theta$ and a decrease for small $\Delta\theta$. We remark that this behaviour is the same under change of density profile from $k=0$ (first row) to $k=2$ (second row). Finally, an increase in the kinetic energy $E_0$ results in an increase of $t_{\rm pk}$, in accordance with Eq. \ref{t_pk_ep}. This figure shows that depends on the parameter values, the maximum flux could occurs in timescale from seconds to years. It is worth noting that the jet opening angle $\theta_{\rm j}$ could be derived considering the kinetic energy $E_0$ and $\Delta\theta$, with the condition $\theta_{\rm obs}> 2 \theta_{\rm j}$. 

In Figure \ref{fig6}, we plot the ratio of the peak time and jet-break time $t_{\rm pk}/t_{\rm j}$ as a function of $\Delta\theta/\theta_{\rm j}$ and the parameter $b$ according to Eq. \ref{t_peak_1}. The first row corresponds to $k=0$, while the second to $k=2$. Columns one and two use the value $\epsilon=0.8$, while columns three and four take the value $\epsilon=0.2$. For each value of $\epsilon$ we consider the ranges of $0.3\leq b \leq 1$ (columns 1 and 3) and $1 < b \leq 3$ (columns 2 and 4).   The general behavior in all panels is that as the ratio of the angles $\Delta\theta/\theta_{\rm j}$ increases, the ratio of the timescales does as well. On the other hand, an enhancement of the $b$-parameter leads to a decrease in the ratio of the timescales. These variations are best observed in the panels of the second row ($k=2$) as in this case the power indexes of the relevant parameters are greater in magnitude (see Eq. \ref{t_peak_1}). It is important to note that, while columns one and three show that for small values of the $b$-parameter the peak time is larger than the jet break time, it is possible that the jet break happens after the time when the maximum flux is reached. This is exemplified by columns 2 and 4, where for the values $1 < b \leq 3$ we note several scenarios where this is the case, even for large values of $\Delta\theta/\theta_{\rm j}$. This figure shows that depending on the parameter values the peak of the afterglow light curve would occur before or after the time of the jet break.

\subsection{Injection of energy into the blastwave}

\subsubsection{Constant-density medium (${\rm k=0}$)}

Given the evolution of the bulk Lorentz factor before afterglow emission enters in the observer's field of view $\Gamma\propto t^{-\frac{2+q}{8}}$, the Lorentz factors of the electrons with the lowest energy and above which they cool effectively evolve as $\gamma_m\propto t^{-\frac{8+4q-p(2+q)}{16(p-1)}}$ ($t^{-\frac{2+q}{8}}$) for $1<p<2$ ($2<p$) and $\gamma_c\propto t^{-\frac{2-3q}{8}}$, respectively. In the synchrotron model, the characteristic and cooling spectral breaks evolve as $\nu^{\rm syn}_{\rm m}\propto t^{-\frac{4+2q+p(2+q)}{8(p-1)}}$ ($\nu^{\rm syn}_{\rm m}\propto t^{-\frac{2+q}{2}}$) for $1<p<2$ ($2<p$) and $\nu^{\rm syn}_{\rm c}\propto t^{\frac{q-2}{2}}$, respectively, and the maximum synchrotron flux as $F^{\rm syn}_{\rm max}\propto t^{1-q}$. In the SSC model, the characteristic and cooling spectral breaks evolve as $\nu^{\rm ssc}_{\rm m}\propto t^{-\frac{3(2+q)}{4(p-1)}}$ ($\nu^{\rm ssc}_{\rm m}\propto t^{-\frac{3(2+q)}{4}}$) for $1<p<2$ ($2<p$) and $\nu^{\rm ssc}_{\rm c}\propto t^{-\frac{6-5q}{4}}$, respectively, and the maximum synchrotron flux as $F^{\rm ssc}_{\rm max}\propto t^{\frac{2(6-5q)}{8}}$.    The bulk Lorentz factor $\Gamma\propto t^{-\frac{q+2}{6}}$.  The minimum and the cooling electron Lorentz factor are given by $\gamma_m\propto t^{-\frac{2+q}{6}}$ and $\gamma_c\propto t^{\frac{q}{2}}$.  The synchrotron spectral breaks and maximum flux are  $\nu_{\rm m}\propto t^{-\frac{2(2+q)}{3}}$, $\nu_{\rm c}\propto t^{-\frac{2(1-q)}{3}}$ and $F_{\rm max}\propto t^{\frac{(1-4q)}{3}}$.   The SSC synchrotron spectral breaks and maximum flux for SSC emission are given by $\nu^{\rm ssc}_{\rm m}\propto  t^{-(2+q)}_{\rm }$ $\nu^{\rm ssc}_{\rm c} \propto t^{-\frac{2-5q}{3}}_{\rm }$ and $F^{\rm ssc}_{\rm max} \propto t^{\frac{2-5q}{3}}_{\rm }$

\subsubsection{Stellar-wind environment (${\rm k=2}$)}

Given the evolution of the bulk Lorentz factor before afterglow emission enters in the observer's field of view $\Gamma\propto t^{-\frac{q}{4}}$, the Lorentz factors of the electrons with the lowest energy and above which they cool effectively evolve as $\gamma_m\propto t^{-\frac{8-p(4-q)}{8(p-1)}}$ ($t^{-\frac{q}{4}}$) for $1<p<2$ ($2<p$) and $\gamma_c\propto t^{\frac{4-q}{4}}$, respectively. In the synchrotron model, the characteristic and cooling spectral breaks evolve as $\nu^{\rm syn}_{\rm m}\propto t^{-\frac{4+pq}{4(p-1)}}$ ($\nu^{\rm syn}_{\rm m}\propto t^{-\frac{2+q}{2}}$) for $1<p<2$ ($2<p$) and $\nu^{\rm syn}_{\rm c}\propto t^{\frac{2-q}{2}}$, respectively, and the maximum synchrotron flux as $F^{\rm syn}_{\rm max}\propto t^{-\frac{q}{2}}$. In the SSC model, the characteristic and cooling spectral breaks evolve as $\nu^{\rm ssc}_{\rm m}\propto t_{2}^{-\frac{6-p(2-q)}{2(p-1)}}$ ($\nu^{\rm ssc}_{\rm m}\propto t^{-(1+q)}$) for $1<p<2$ ($2<p$) and $\nu^{\rm ssc}_{\rm c}\propto t^{3-q}$, respectively, and the maximum synchrotron flux as $F^{\rm ssc}_{\rm max}\propto t^{-1}$.   During the post-jet-break decay phase, the bulk Lorentz factor becomes $\Gamma\propto t^{-\frac{q}{2}}$.  The minimum and the cooling electron Lorentz factor are given by $\gamma_m\propto t^{-\frac{q}{2}}$ and $\gamma_c\propto t^{\frac{2-q}{2}}$.  The synchrotron spectral breaks and maximum flux are  $\nu_{\rm m}\propto t^{-(1+q)}$, $\nu_{\rm c}\propto t^{1-q}$ and $F_{\rm max}\propto t^{-q}$. The SSC synchrotron spectral breaks and maximum flux for SSC emission evolves as $\nu^{\rm ssc}_{\rm m}\propto t^{-(1+2q)}_{\rm }$, $\nu^{\rm ssc}_{\rm c} \propto t^{3-2q}_{\rm }$,  $F^{\rm ssc}_{\rm max} \propto t^{-1}_{\rm }$.

\subsubsection{Timescale Analysis}

Similarly to the radiative scenario, the observed flux increases when the jet decelerates in the circumstellar medium. The maximum flux takes place at

{\small
\begin{eqnarray}
\label{t_pk_q}
t_{\rm pk} = \cases{ 
1.6\times 10^3\,{\rm s}\, \left(\frac{1+z}{1.022}\right)^{\frac{3}{2+q}} n^{-\frac{1}{2+q}} E_{\rm 0,51.7}^{\frac{1}{2+q}} (1+b^2)^{\frac{3}{2+q}}\,  \left(\frac{\Delta\theta_{2}}{b}\right)^{\frac{8}{2+q}},\hspace{0.8cm} {\rm for\,\,\, k=0} \cr
2.1\times 10^2\,{\rm s}\, \left(\frac{1+z}{1.022}\right)^{\frac{1}{q}} A_{\rm w,-1}^{-\frac{1}{q}} E^{\frac{1}{q}}_{\rm 0,51.7} (1+b^2)^{\frac{1}{q}}  \left(\frac{\Delta\theta_{2}}{b}\right)^{\frac{4}{q}},\,\,\,\,\, \hspace{1.55cm} {\rm for\,\,\, k=2}\,. \cr
}
\end{eqnarray}
}

For $q=1$ and $b=1$, the values reported in \cite{2002ApJ...579..699N} are recovered. The jet break transition  occurs when the bulk Lorentz factor becomes $\Gamma=1/\theta_j$  \citep{1999ApJ...525..737R, 1999ApJ...519L..17S}. In general, it happens at

{\small
\begin{eqnarray}
\label{t_j_q}
t_{\rm j} = \cases{ 
4.5\times 10^3\,{\rm s}\, \left(\frac{1+z}{1.022}\right)^{\frac{3}{2+q}} n^{-\frac{1}{2+q}} E_{\rm 0,j,47.9}^{\frac{1}{2+q}}  \left[1 + \left( \frac{\Delta\theta_{2}}{\theta_{\rm j,1}}\right)^2 \right]^{\frac{3}{2+q}}\,  \theta_{j,1}^{\frac{6}{2+q}},\hspace{0.8cm} {\rm for\,\,\, k=0} \cr
4.4\times 10^3\,{\rm s}\, \left(\frac{1+z}{1.022}\right)^{\frac{1}{q}} A_{\rm w,-1}^{-\frac{1}{q}} E^{\frac{1}{q}}_{\rm 0,j,47.9} \left[1 + \left( \frac{\Delta\theta_{2}}{\theta_{\rm j,1}}\right)^2 \right]^{\frac{1}{q}}\theta_{j,1}^{\frac{2}{q}},\,\,\,\,\, \hspace{1.55cm} {\rm for\,\,\, k=2}\,, \cr
}
\end{eqnarray}
}
where the term $\Delta\theta/\theta_{\rm j}$ was discussed in the radiative scenario.  From Eqs. \ref{t_pk_q} and \ref{t_j_q}, the relationship between $t_{\rm pk}$ and $t_{\rm j}$ can be written as

\begin{figure*}
\centering{
\includegraphics[scale=0.65]{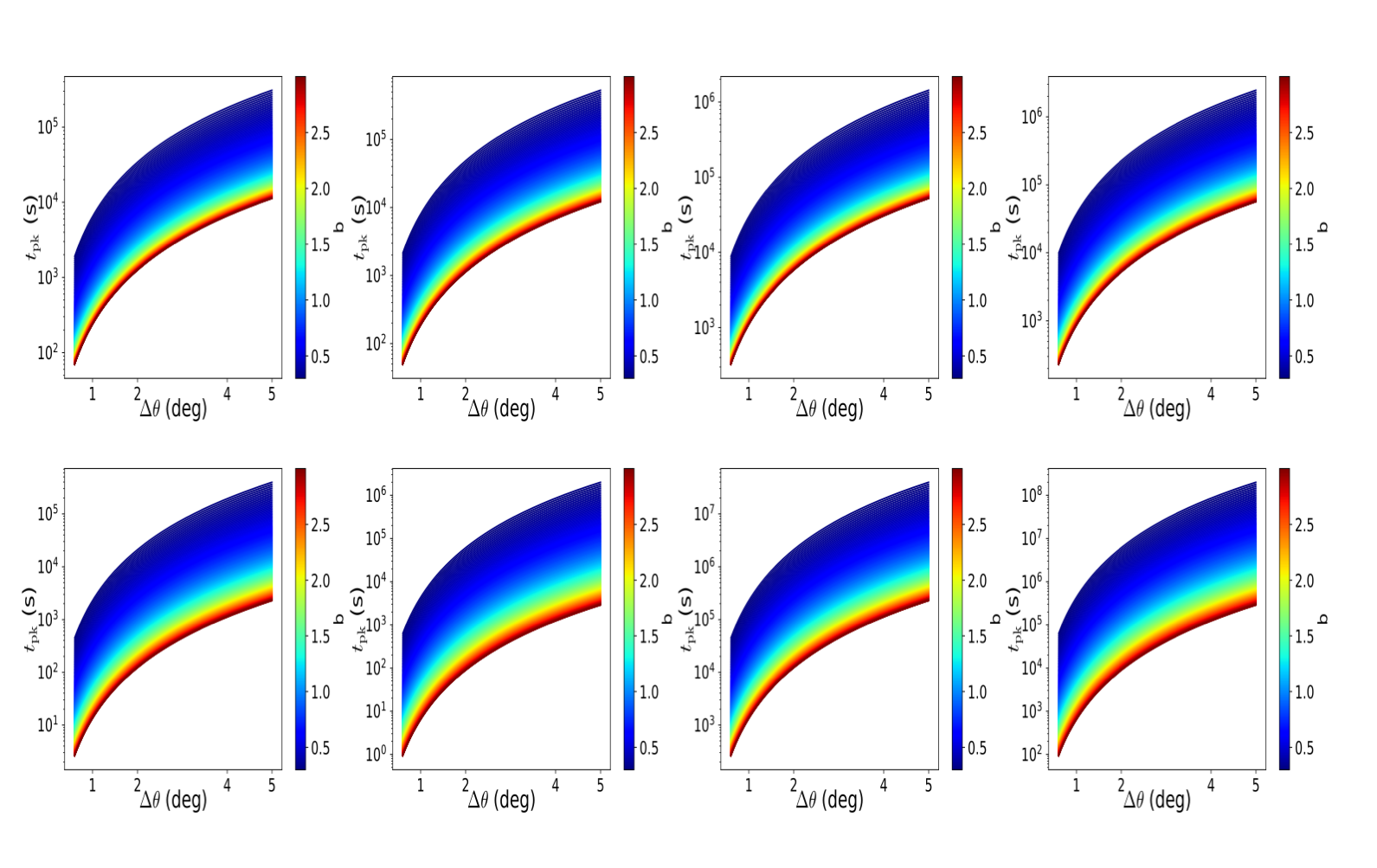}
}
\caption{The 3-D parameter space of the expected values of $t_{\rm pk}$ as a function of $\Delta\theta$ and the parameter ${\rm b}$ with two distinct values of $\epsilon$ and $E_{\rm 0}$ for $k=0$ (above) and $k=2$ (below).  Column one ($\epsilon=0.8$) and two ($\epsilon=0.2$) are for $E_{\rm 0}=5\times 10^{51}\, {\rm erg}$ and column three ($\epsilon=0.8$) and four ($\epsilon=0.2$) are for $E_{\rm 0}=5\times 10^{53}\, {\rm erg}$.  We consider the parameter $b$ in the range of $0.3\leq b \leq 3$ for ${\rm n=1\,{\rm cm^{-3}}}$ and $A_{\rm w}=0.1$.}\label{fig5}
\end{figure*}

{\small
\begin{eqnarray}
\label{t_peak_2}
t_{\rm pk} = \cases{ 
\, \frac{(1+b^2)^{\frac{3}{2+q}}}{b^{\frac{8}{2+q}} \left[1 + \left( \frac{\Delta\theta}{\theta_{\rm j}}\right)^2 \right]^{\frac{3}{2+q}}} \left( \frac{\Delta\theta}{\theta_{\rm j}}\right)^{\frac{8}{2+q}}  t_{\rm j},\hspace{0.8cm} {\rm for\,\,\, k=0} \cr
\, \frac{(1+b^2)^{\frac{1}{q}}}{b^{\frac{4}{q}} \left[1 + \left( \frac{\Delta\theta}{\theta_{\rm j}}\right)^2 \right]^{\frac{1}{q}}} \left( \frac{\Delta\theta}{\theta_{\rm j}}\right)^{\frac{4}{q}}  t_{\rm j},\,\,\,\,\, \hspace{1.4cm} {\rm for\,\,\, k=2}\,. \cr
}
\end{eqnarray}
}
In Figure \ref{fig7}, we plot the time when the maximum flux is reached $t_{\rm pk}$ as a function of $\Delta\theta$ and the parameter $b$ according to Eq. \ref{t_pk_q}. In the same way as in Figure \ref{fig5}, we consider two different values for the rest of the model parameters, with the only difference being that in the first and third columns we take the value $q=0.8$, while columns two and four use $q=0.2$.  In the same way as in Figure \ref{fig5}, we notice that in all panels an increase in the $b$-parameter leads to a decrease in $t_{\rm pk}$ and that as $\Delta\theta$ grows, so does $t_{\rm pk}$. Also, an increase in the kinetic energy $E_0$ results in an increase of $t_{\rm pk}$. Similarly, upon comparison between columns one and two, and three and four of the first row ($k=0$), we note that changes in $q$ do not lead to substantial changes in $t_{\rm pk}$, other than an overall increase of $t_{\rm pk}$. The case of the second row ($k=2$) is more interesting, however. Here, a change in $q$ greatly changes the gradient of $t_{\rm pk}$ as a function of both $\Delta\theta$ and $b$. This can be explained by examination of Eq. \ref{t_pk_q}, where we notice that for $k=0$ the power indexes of the relevant parameters have a factor of $2+q$ in the denominator, so as $q\to0$ the power indexes remain finite. For $k=2$, we have just the factor $q$ in the denominator, so as $q\to0$ the indexes grow to infinity, so even small changes in the relevant parameters will lead to huge differences in $t_{\rm pk}$. This figure shows that depends on the parameter values, the maximum flux could occurs in timescale from seconds (even before) to years. It is important to take into account that the jet opening angle $\theta_{\rm j}$ could be estimated considering the kinetic energy $E_0$ and $\Delta\theta$, with the approximation $\theta_{\rm obs}> 2 \theta_{\rm j}$.

\begin{figure*}
\centering{
\includegraphics[scale=0.65]{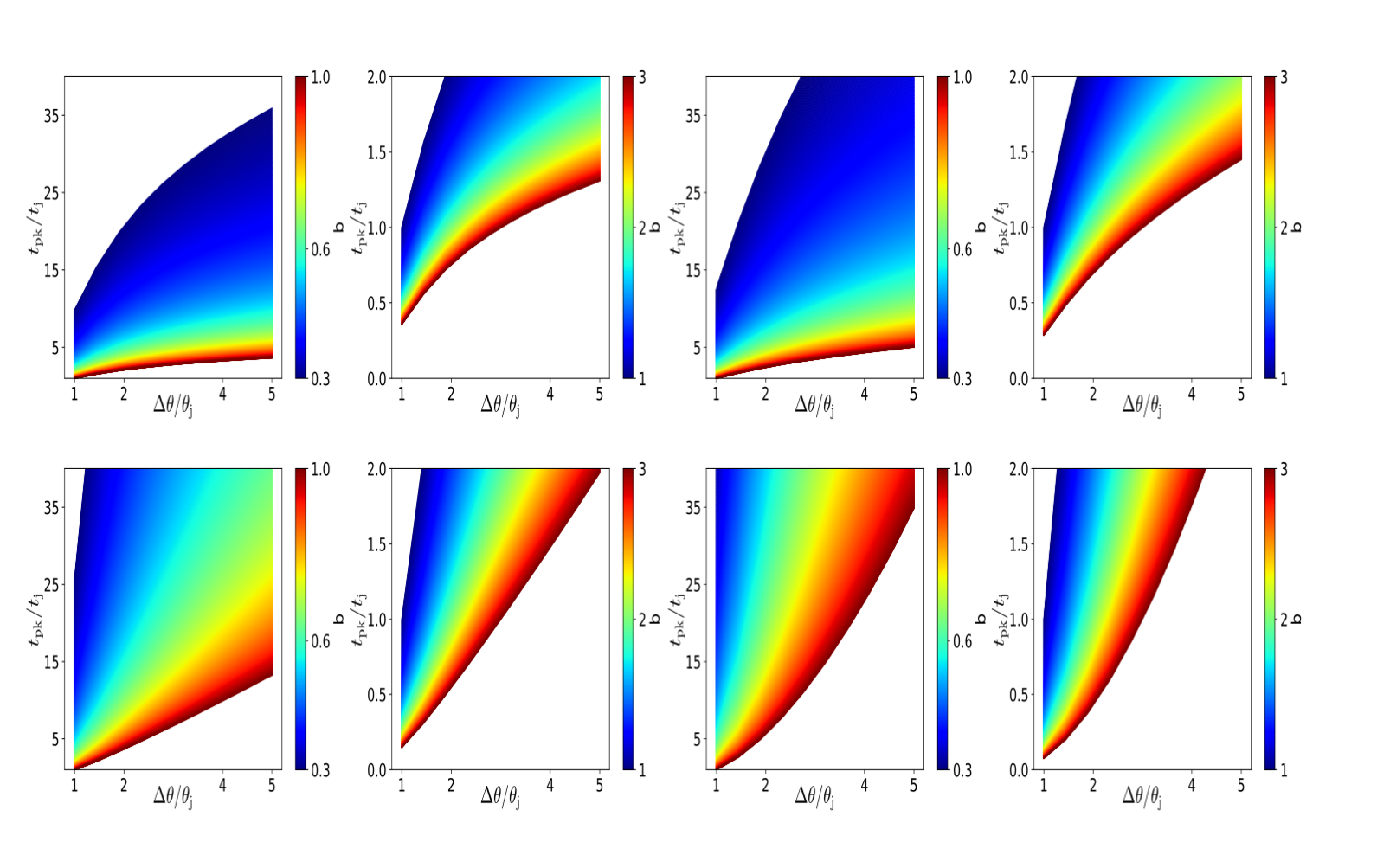}
}
\caption{The 3-D parameter space of the expected values of $t_{\rm pk}/t_{\rm j}$ as a function of $\Delta\theta/\theta_{\rm j}$ and the parameter ${\rm b}$ with two distinct values of $\epsilon$ for $k=0$ (above) and $k=2$ (below).  For each value of $\epsilon$ we consider the ranges of $0.3\leq b \leq 1$ and $1 < b \leq 3$. Columns one and two correspond to $\epsilon=0.8$ and columns three and four to $\epsilon=0.2$.}\label{fig6}
\end{figure*}

\begin{figure*}
\centering{
\includegraphics[scale=0.65]{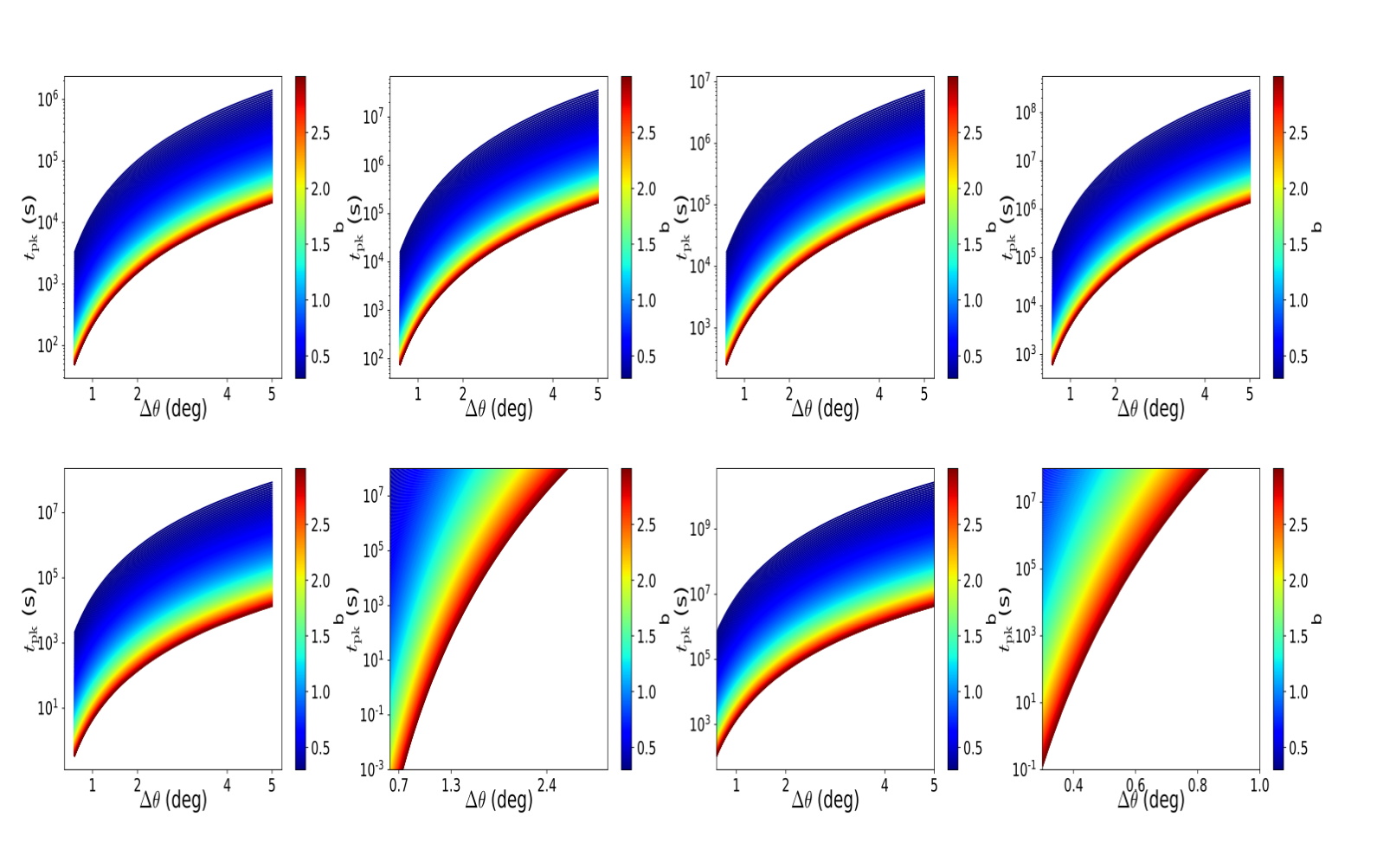}
}
\caption{The same as Fig. \ref{fig5}, but for two values of $q$.  Column one ($q=0.8$) and two ($q=0.2$) are for $E_{\rm 0}=5\times 10^{51}\, {\rm erg}$ and column three ($q=0.8$) and four ($q=0.2$) are for $E_{\rm 0}=5\times 10^{53}\, {\rm erg}$.}\label{fig7}
\end{figure*}

\begin{figure*}
\centering{
\includegraphics[scale=0.65]{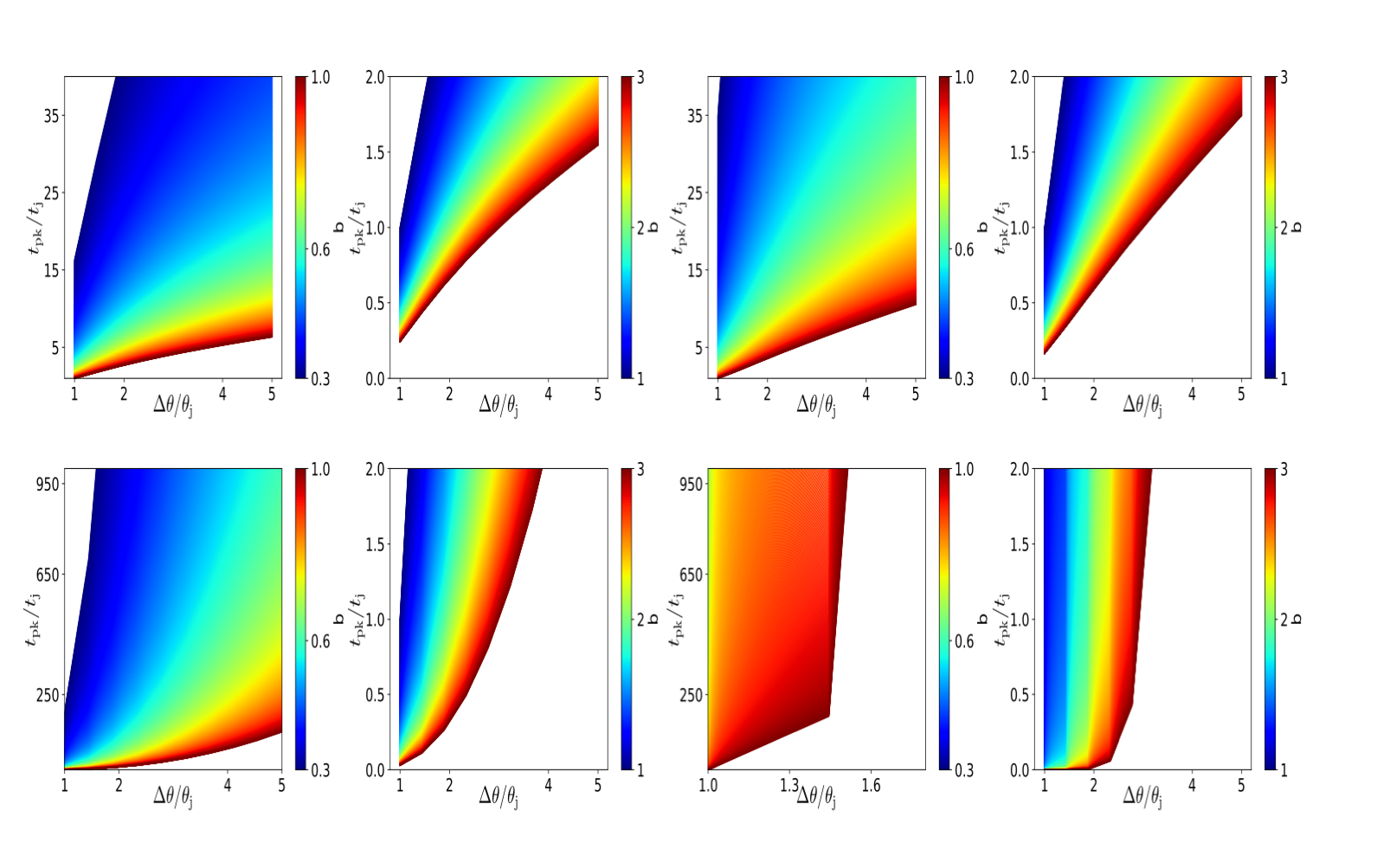}
}
\caption{The same as Fig. \ref{fig6}, but for two values of $q$. Columns one and two correspond to $q=0.8$ and columns three and four to $q=0.2$.}\label{fig8}
\end{figure*}

In Figure \ref{fig8}, we plot the ratio of the peak time and jet-break time $t_{\rm pk}/t_{\rm j}$ as a function of $\Delta\theta/\theta_{\rm j}$ and the parameter $b$ according to Eq. \ref{t_peak_2}. We make the same considerations as in Figure \ref{fig6}, with the only difference being that in the first and second columns we take the value $q=0.8$, while columns three and four use $q=0.2$.   In the same way as in Figure \ref{fig6}, the general behavior in all panels is that as the ratio of the angles $\Delta\theta/\theta_{\rm j}$ increases, the ratio of the timescales does as well; and as the $b$-parameter grows, the ratio of the timescales drops. These changes are best observed in the second row ($k=2$) as in this case the power indexes of the relevant parameters are greater in magnitude (see Eq. \ref{t_peak_2}). The case of energy injection is even better for the observation of this behavior because, as we explained previously, for $k=2$ the power index of the relevant parameters grows unimpeded as $q\to0$. Here we note that for energy injection it is also possible that the jet break happens after the time when the maximum flux is reached, which is exemplified by columns 2 and 4, where for the values $1 < b \leq 3$ we note several scenarios where this is the case, even for large values of $\Delta\theta/\theta_{\rm j}$.   This figure displays that depending on the parameter values the peak of the afterglow light curve would take place before or after the time of the jet break. 


\subsection{Application: GRB 160625B}

The Gamma-ray Burst Monitor (GBM) on board the Fermi satellite successfully detected and determined the location of GRB 160625B at 22:40:16.28 Universal Time (UT) on June 25, 2016 \citep{2016GCN..19581...1B}. The Fermi-LAT promptly followed up at 22:43:24.82 UT \citep{2016GCN..19586...1D}.  The temporal duration of the prompt emission, as determined by the Fermi GBM instrument, was evaluated to be $T_{90}=453.38\,{\rm s}$, which corresponds to a fluence and an total isotropic energy of $(2.5\pm 0.3)\times 10^{-5}\,{\rm erg\,cm^{-2}}$ and $E_{\rm \gamma, iso}=(1.5\pm 0.1)\times 10^{53}\, {\rm erg}$, respectively \citep{Ajello_2019}. The X-ray Telescope (XRT) instrument on board the Neil Gehrels Swift Observatory looked into this burst in the photon counting mode for almost $\sim 5\times 10^{6}\,{\rm s}$. The best-fit value of the intrinsic absorption column is $1.6\pm 0.6\times 10^{21}\,{\rm cm^{-2}}$, with a redshift $z=1.406$  \citep{2016GCN..19600...1X}.  A series of optical observations were conducted using different telescopes \citep{2016arXiv161203089Z, 2017Natur.547..425T}.\\

The data files used for the Fermi - LAT analysis were obtained from the online data website.\footnote{https://fermi.gsfc.nasa.gov/cgi-bin/ssc/LAT/LATDataQuery.cgi} These data were analyzed in the $0.1-100$~GeV  energy range with the Fermi Science tools.\footnote{https://fermi.gsfc.nasa.gov/ssc/data/analysis/software/}    In the current work, we use the answers documented for this burst as published by \cite{Ajello_2019}, using the unbinned likelihood analysis method reported by the Fermi-LAT team.\footnote{https://fermi.gsfc.nasa.gov/ssc/data/analysis/scitools/likelihood\_tutorial.html} In accordance with the likelihood methodology, we generate a lifetime cube using the software program \texttt{gtltcube}. The process involves employing a step size of $\delta \theta=0.025$, a bin size of 0.5, and a maximum zenith angle of 100$^{\circ}$. The exposure map was generated with the \texttt{gtexpmap} function, which took into account a 30-degree zone centered on the site of the gamma-ray burst (GRB). The map was constructed with 100 spatial bins in both longitude and latitude, as well as 50 energy bins.  The likelihood analysis is executed using the \texttt{pyLikelihood} tool, which may be found at the following link: https://fermi.gsfc.nasa.gov/ssc/data/analysis/scitools/python\_tutorial.html. Finally, we achieve a photon selection probability larger than 90\% by the use of the \texttt{gtsrcprob} tool.  The Swift-XRT observations are publicly available in the website science database. These data points were converted from 10 to 1 keV by applying the conversion factor given in  \cite{2010A&A...519A.102E}.  Optical data collected with Mini-Mega TORTOLA, TSHAO, AbAO, Mondy, CrAO, Maidanak, SAO RAS were taken from \cite{2017ApJ...848...15F}.

\begin{figure*}
\centering
\includegraphics[width=0.72\textwidth]{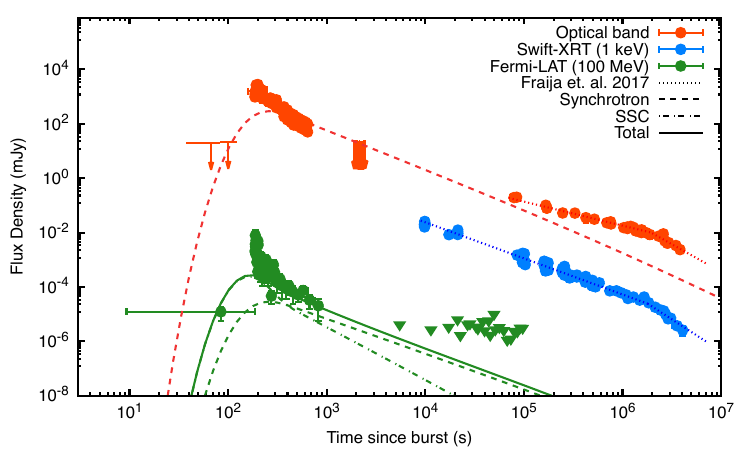}
\includegraphics[width=0.62\textwidth]{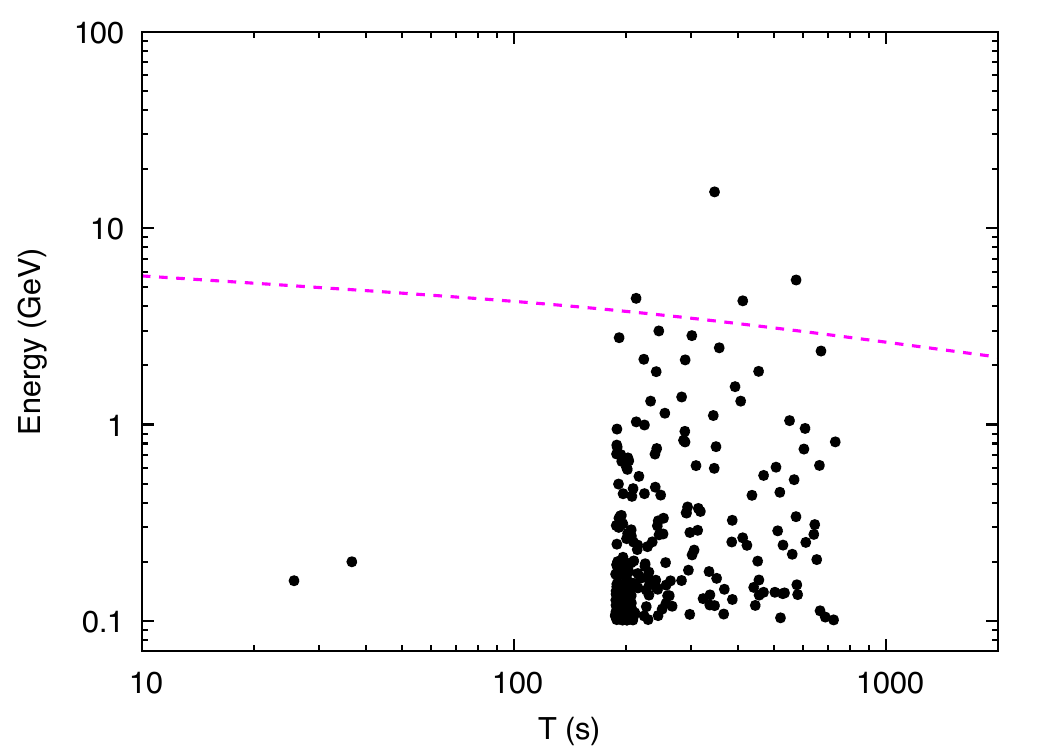}
\caption{Upper Panel: Multiwavelength observations of GRB 160625B with the best-fit light curves. The light curves that describe the early observations (LAT and optical for ${\rm  t<10^3\,s}$) are generated with the synchrotron and SSC off-axis afterglow model, and the light curves that explain the late observations (X-ray and optical for ${\rm t>10^3\, s}$) are taken from \cite{2017ApJ...848...15F}.  Lower Panel: All the photons with energies $> 100$~MeV and probabilities $>90$\% of being associated with GRB 160625B. The magneta line is the maximum photon energy radiated by the synchrotron afterglow model.}
\label{grb160625B}
\end{figure*}

\begin{figure*}
\centering
\includegraphics[width=0.82\textwidth]{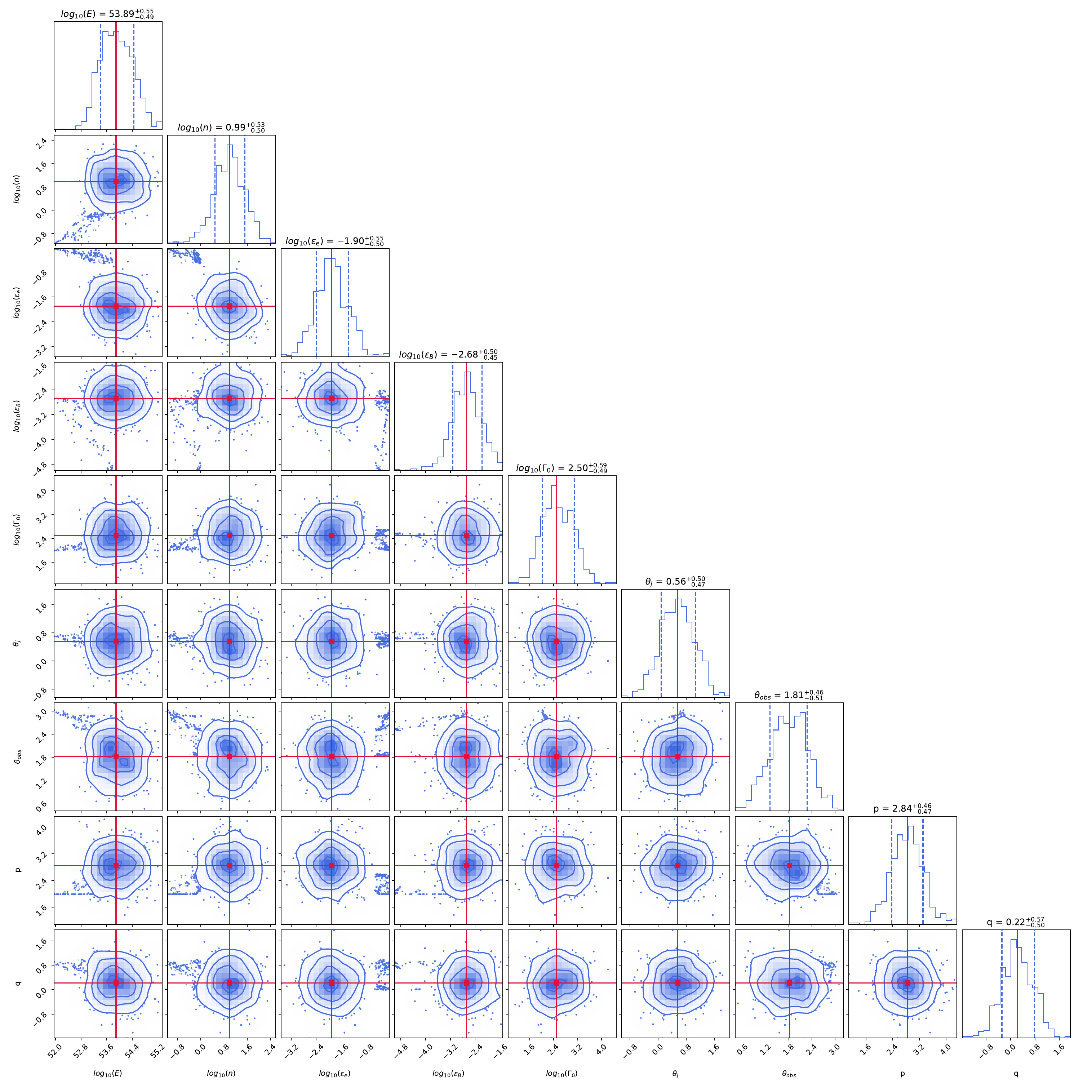}
\caption{The corner plot displays the outcomes of our MCMC parameter estimate for the synchrotron and SSC off-axis model applied to GRB 160625B. The histograms on the diagonal display the marginalized posterior densities for each parameter and the median values are shown by red lines.}
\label{mcmc}
\end{figure*}

Figure \ref{grb160625B} shows the multiwavelength afterglow observations of GRB 160625B with the best-fit curves (upper panel) and all photons with energies $>100$~MeV and probabilities $>90$\% of being associated with GRB 160625B (lower panel). The best-fit curves modelling the LAT and optical observations for ${\rm  t<10^3\,s}$ are generated with the synchrotron (dashed) and SSC (dot-dashed) off-axis model evolving in the constant-density environment. The curves that describe the X-ray and optical observations for ${\rm t>10^3\, s}$ with the standard synchrotron on-axis model were taken from \cite{2017ApJ...848...15F}.  The magenta line in the lower panel exhibits the maximum photon energies released by the synchrotron afterglow model.  Figure~\ref{mcmc} shows the corner plot of our Markov Chain Monte Carlo (MCMC) parameters estimated for the synchrotron and SSC off-axis model applied to GRB 160625B. The histograms on the diagonal display the marginalized posterior densities for each parameter and the median values are shown by red lines.  We use MCMC simulations with nine parameters used for GRB 160625B to find the fit values that explain the multiwavelength observations with the current model. To represent all the data, a total of 15900 samples and 4400 tuning steps are used.  Table \ref{Table:ISM_Fit} lists the best-fit values of each parameter.

The upper panel in Figure \ref{grb160625B} shows that a narrow off-axis jet, which has an initial bulk Lorentz factor and jet opening half-angle of $316$ and $0.56^\circ$, respectively, can explain the  early afterglow observations  (${\rm  t<10^3\,s}$) and a wide on-axis jet with  bulk Lorentz factor and jet opening half-angle of $25$ and $8.3^\circ$, respectively, can explain the late afterglow (${\rm t>10^3\, s}$) when they evolve in a homogeneous medium. Therefore, the entire afterglow observations are consistent with afterglow emission from a double jet that decelerates in the homogeneous medium. 

Analysis of the LAT-detected photons (the lower panel of Figure \ref{grb160625B}) with a probability greater than 90$\%$ to be associated to GRB 160625B shows that the energy range of the photons is large, with 255 photons over 100~MeV, 21 exceeding 1~GeV, and one over 10~GeV. It indicates that the first energetic photon was detected 25.6~s after the trigger time with an energy of 160.2~MeV, and the highest energy photon was observed 346.2~s after trigger and had an energy of 15.3~GeV.  The magenta line of the lower panel in Figure \ref{grb160625B} shows that the synchrotron off-axis model cannot explain the highest-energy photons displayed by GRB 160625B. Therefore, a different process such as the SSC scenario could successfully explain these energetic photons. For instance, the number of photons ($N_{\gamma}$) with energy $h\nu$ that reach the Fermi-LAT instrument during a specific time interval ($\Delta t$) can be estimated considering the effective area of LAT ($A$) and the observed flux at the determined time. In this case the number of photons is {\small $ N_{\gamma}\sim 1\,{\rm ph} \left(\frac{h\nu}{10\,{\rm~GeV}}\right) \left(\frac{10\,{\rm s}}{\Delta t} \right) \left(\frac{F^{\rm ssc}_{\nu,j}}{10^{-9}\,{\rm \frac{erg}{cm^{2}\,s}}}\right) \left(\frac{10^4\,{\rm cm^{2}}}{A}\right)$}. In this case, the most energetic photons could be explained by the SSC mechanism.

The best-fit value of circumstellar environment found together with the distance $z=1.406$ ratifies the hypothesis that the host galaxy is a dwarf-irregular galaxy with a conventional size of $L\simeq 0.1\,{\rm kpc}$ for an intrinsic column density  $N_{\rm H}\simeq 4.11\times 10^{21}\,{\rm cm^{-2}}$.     The optimal estimation of the microphysical parameter $\epsilon_{\rm B}=2.1\times10^{-3}$ is within the range of values used to characterize the multiwavelength afterglow observations in several GRBs; $10^{-5} \leq \epsilon_{B} \leq 10^{-1}$ \citep{1999ApJ...523..177W, 2002ApJ...571..779P, 2003ApJ...597..459Y, 2005MNRAS.362..921P, 2014ApJ...785...29S}.   The efficiency of the gamma-ray emission process is an essential factor that gives important information.   The best value of the equivalent kinetic energy $E=7.8 \times 10^{53}\,{\rm erg}$ and the isotropic energies in gamma-rays reported by the GBM instrument during the prompt episode in the range of $E_{\rm \gamma, iso}= (1.5\pm 0.1)\times 10^{53}\,{\rm erg}$ \citep{Ajello_2019} lead to a kinetic efficiency of $\eta = 0.16$, which is consistent with the values reported for the description of GRB afterglow observations \citep{2001ApJ...557..399G, 2007ApJ...655..989Z, 2015PhR...561....1K}.

\cite{2017ApJ...848...15F} described the non-thermal multiwavelength observations of GRB 160625B and proposed a transition phase from stellar wind to constant-density medium  between the early and the late afterglow  observations. The multiwavelength observations of the early afterglow were consistent with a stellar-wind environment, whereas the late observations were consistent with the afterglow evolution in a constant-density medium.   The transition phase was estimated to be at $\simeq 8\times 10^3\ {\rm s}$ when the relativistic jet was decelerating at $\sim 1\, {\rm pc}$ from the progenitor. In our description, we use a double jet that decelerates in the same circumstellar homogeneous medium with evoking a stratified environment.  

On the other hand,  GRB 190829A with $z = 0.0785 \pm 0.005$ \citep{2019GCN.25565....1V} was one of the closest  and least energetic bursts detected in TeV energies \citep[e.g., see][]{2021ApJ...918...12F, 2021arXiv210602510H}.  H.E.S.S. telescopes reported the detection of VHE gamma-rays, which coincided with the direction of GRB 190829A \citep{2021arXiv210602510H}.  \cite{2021MNRAS.504.5647S} proposed a narrow and a wide jet viewed off- and on-axis, respectively, to describe GRB 190829A. For the off-axis scenario, they used the approximation of $\theta_{\rm obs}\gtrsim 2\theta_j$ with the values of $\theta_j=0.86^\circ$, $\theta_{\rm obs}=1.78^\circ$, and with a $\Gamma=350$, the authors reported a maximum flux at $t_{\rm pk}\sim2\times10^3\,{\rm s}$. A comparable or equivalent description was performed for a successful description of GRB 160625B, suggesting that both bursts could share similar features.

\section{Summary}
\label{sec5}

For the purpose of studying the evolution of the spectral and temporal indexes of bursts reported in 2FLGC, which consists of a subset of 29 bursts with photon energies above a few GeV, we have derived the CRs of the SSC afterglow model in the adiabatic and radiative scenario, and when the central engine continuously injects energy into the blastwave.  We take into account the radiative parameter $\epsilon$, the energy injection index $q$, and the electron spectral index for $1<p<2$ and $ 2 < p$ in a model of the SSC afterglow developing in the ISM and the stellar-wind medium.
 
The analysis of the 2FLGC bursts has led us to the conclusion that in the case of no-energy injection for the synchrotron off-axis model, the most preferred scenario corresponds to stellar wind for slow cooling regime (${\rm \nu_m^{sync} < \nu_{\rm LAT} < \nu_c^{sync} }$) with an occurrence of 14 GRBs (16.09 \%),  followed by fast/slow cooling regime (${\rm max\{\nu_m^{sync},\nu_c^{sync} \} < \nu_{\rm LAT}}$) with an occurrence of 2 GRBs (2.29 \%).  For the SSC off-axis model, the most preferred scenario corresponds to ISM for slow cooling regime (${\rm \nu_m^{ssc} < \nu_{\rm LAT} < \nu_c^{ssc} }$) with an occurrence of 6 GRBs (6.89 \%) followed by stellar wind for fast/slow cooling regime (${\rm max\{\nu_m^{ssc},\nu_c^{ssc} \} < \nu_{\rm LAT}}$) with an occurrence of 2 GRBs (2.29 \%). In the case of energy injection,  for the synchrotron off-axis model the most preferred scenario corresponds to stellar wind for slow cooling regime (${\rm \nu_m^{sync} < \nu_{\rm LAT} < \nu_c^{sync} }$) with an occurrence of 19 bursts (21.83 \%) and for the SSC off-axis model the most preferred scenario corresponds to wind for fast/slow cooling regime (${\rm max\{\nu_m^{ssc},\nu_c^{ssc} \} < \nu_{\rm LAT}}$) with an occurrence of 3 bursts (3.34 \%). 

The afterglow of a GRB and the variation of its spectral and temporal indices are both well explained by the classic synchrotron forward-shock model, which is valid down to the synchrotron limit.  Photon energies larger than $10\, {\rm GeV}$ have been collected in 29 bursts for 2FLGC, which cannot be fully interpreted by the synchrotron scenario.  Since the synchrotron model predicts that the highest photon energy emitted during the deceleration phase would be about $\approx 1 \, {\rm GeV}$, we argue that the SSC afterglow model is more appropriate for interpreting this sample of bursts. Since neutrino non-coincidences with GRBs have been reported by the IceCube team, we rule out photo-hadronic interactions \citep{2012Natur.484..351A, 2016ApJ...824..115A, 2015ApJ...805L...5A}. Therefore, the SSC process may be the most appropriate, even if the CRs of the synchrotron standard model might fulfill this sample of 29 bursts.\\

In this study, we provide an analysis of the non-thermal multi-wavelength observations of GRB 160625B. The data used in our analysis for Fermi-LAT and  Swift-XRT were downloaded from their respective Science data center and for optical were collected from numerous  terrestrial telescopes. The multi-wavelength observations of the early afterglow (LAT and optical for ${\rm  t<10^3\,s}$) showed consistency with the off-axis afterglow model and the late afterglow (X-ray and optical for ${\rm t>10^3\, s}$) with the on-axis model. We have found that the entire afterglow observations are consistent with afterglow emission from a double jet that decelerates in the homogeneous medium; a narrow off-axis jet with a bulk Lorentz factor and jet opening half-angle of $316$ and $0.56^\circ$, respectively, can explain the  early afterglow observations and a wide on-axis jet with a bulk Lorentz factor and a jet opening half-angle of $25$ and $8.3^\circ$, respectively, can explain the late afterglow.

In the context of our model, we have constrained the value of $\varepsilon_{\rm B}$ that is required so that  observed LAT light curves can be described with an off-axis afterglow model. We have done this for both synchrotron and SSC in the adiabatic regime with energy injection and we have considered both an ISM and a wind-like medium. In general, we have found that the synchrotron and SSC afterglow models lead to values of $\varepsilon_{\rm B}$ in the range of given by \cite{2014ApJ...785...29S}. Based on the different parameter range,  we have also performed a full analysis of the timescales of the peak in the afterglow light curve and the jet break in the radiative/adiabatic and energy injection scenario. It is worth noting that for a large off-axis angle only nearby GRBs can be observed since for a top-hat jet the fluence declines very rapidly for off-axis jets. However, when an off-axis jet exhibits a very small opening angle with a small $\Delta \theta \gtrsim \theta_j$ a high redshift GRB can be observed. 

\cite{2019ApJ...883..134T} systematically analyzed the closure relations in a sample of 59 chosen LAT-detected bursts, taking into account temporal and spectral indexes. They discovered, for one thing, that while the conventional synchrotron emission describes the spectral and temporal ones in most situations, there is still a sizable proportion of bursts that can scarcely be represented with this model. However, they demonstrated that numerous GRBs fulfill the slow-cooling regime's CRs, but only when the magnetic microphysical parameter has an unusually tiny value of $\epsilon_B<10^{-7}$. We demonstrated here that the SSC afterglow model, with or without energy injection, yields a $\varepsilon_B$-value in the range of $3.5\times10^{-5}\leq\epsilon_B\leq0.33$ \citep[e.g., see][]{2014ApJ...785...29S}. Therefore, the CRs of SSC afterglow models are required to explain those bursts in the 2FLGC that cannot be explained in the synchrotron scenario, together with the study given by \cite{2019ApJ...883..134T} (e.g., those with photon energies above 10 GeV).

\section{acknowledgements}

We thank Peter Veres and Tanmoy Laskar for useful discussions.  NF acknowledges financial support  from UNAM-DGAPA-PAPIIT  through  grant IN106521.

\bibliography{main}

\begin{thebibliography}{}
\expandafter\ifx\csname natexlab\endcsname\relax\def\natexlab#1{#1}\fi
\providecommand{\url}[1]{\href{#1}{#1}}
\providecommand{\dodoi}[1]{doi:~\href{http://doi.org/#1}{\nolinkurl{#1}}}
\providecommand{\doeprint}[1]{\href{http://ascl.net/#1}{\nolinkurl{http://ascl.net/#1}}}
\providecommand{\doarXiv}[1]{\href{https://arxiv.org/abs/#1}{\nolinkurl{https://arxiv.org/abs/#1}}}

\bibitem[{{Aartsen} {et~al.}(2015){Aartsen}, {Ackermann}, {Adams}, {Aguilar},
  {Ahlers}, {Ahrens}, {Altmann}, {Anderson}, {Arguelles}, {Arlen}, \&
  et~al.}]{2015ApJ...805L...5A}
{Aartsen}, M.~G., {Ackermann}, M., {Adams}, J., {et~al.} 2015, \apjl, 805, L5,
  \dodoi{10.1088/2041-8205/805/1/L5}

\bibitem[{{Aartsen} {et~al.}(2016){Aartsen}, {Abraham}, {Ackermann}, {Adams},
  {Aguilar}, {Ahlers}, {Ahrens}, {Altmann}, {Anderson}, {Ansseau}, \&
  et~al.}]{2016ApJ...824..115A}
{Aartsen}, M.~G., {Abraham}, K., {Ackermann}, M., {et~al.} 2016, \apj, 824,
  115, \dodoi{10.3847/0004-637X/824/2/115}

\bibitem[{{Abbasi} {et~al.}(2012){Abbasi}, {Abdou}, {Abu-Zayyad}, {Ackermann},
  {Adams}, {Aguilar}, {Ahlers}, {Altmann}, {Andeen}, {Auffenberg}, \&
  et~al.}]{2012Natur.484..351A}
{Abbasi}, R., {Abdou}, Y., {Abu-Zayyad}, T., {et~al.} 2012, \nat, 484, 351,
  \dodoi{10.1038/nature11068}

\bibitem[{Abbott {et~al.}(2017)Abbott, Abbott, Abbott, \&
  et~al.}]{2041-8205-848-2-L12}
Abbott, B.~P., Abbott, R., Abbott, T.~D., \& et~al. 2017, The Astrophysical
  Journal Letters, 848, L12.
\newblock \url{http://stacks.iop.org/2041-8205/848/i=2/a=L12}

\bibitem[{{Ackermann} \& et~al.(2013)}]{2013ApJ...763...71A}
{Ackermann}, M., \& et~al. 2013, \apj, 763, 71,
  \dodoi{10.1088/0004-637X/763/2/71}

\bibitem[{{Ajello} {et~al.}(2019){Ajello}, {Arimoto}, {Axelsson}, {Baldini},
  {Barbiellini}, \& {et al.}}]{Ajello_2019}
{Ajello}, M., {Arimoto}, M., {Axelsson}, M., {et~al.} 2019, \apj, 878, 52,
  \dodoi{10.3847/1538-4357/ab1d4e}

\bibitem[{{Band} {et~al.}(1993){Band}, {Matteson}, {Ford}, {Schaefer},
  {Palmer}, {Teegarden}, {Cline}, {Briggs}, {Paciesas}, {Pendleton}, {Fishman},
  {Kouveliotou}, {Meegan}, {Wilson}, \& {Lestrade}}]{1993ApJ...413..281B}
{Band}, D., {Matteson}, J., {Ford}, L., {et~al.} 1993, \apj, 413, 281,
  \dodoi{10.1086/172995}

\bibitem[{{Blandford} \& {McKee}(1976)}]{1976PhFl...19.1130B}
{Blandford}, R.~D., \& {McKee}, C.~F. 1976, Physics of Fluids, 19, 1130,
  \dodoi{10.1063/1.861619}

\bibitem[{{B{\"o}ttcher} \& {Dermer}(2000)}]{2000ApJ...532..281B}
{B{\"o}ttcher}, M., \& {Dermer}, C.~D. 2000, \apj, 532, 281,
  \dodoi{10.1086/308580}

\bibitem[{{Burns}(2016)}]{2016GCN..19581...1B}
{Burns}, E. 2016, GRB Coordinates Network, 19581

\bibitem[{{Burns} {et~al.}(2018){Burns}, {Veres}, {Connaughton}, {Racusin},
  {Briggs}, {Christensen}, {Goldstein}, {Hamburg}, {Kocevski}, {McEnery},
  {Bissaldi}, {Dal Canton}, {Cleveland}, {Gibby}, {Hui}, {von Kienlin},
  {Mailyan}, {Paciesas}, {Roberts}, {Siellez}, {Stanbro}, \&
  {Wilson-Hodge}}]{2018ApJ...863L..34B}
{Burns}, E., {Veres}, P., {Connaughton}, V., {et~al.} 2018, \apjl, 863, L34,
  \dodoi{10.3847/2041-8213/aad813}

\bibitem[{{Burrows} {et~al.}(2005){Burrows}, {Romano}, {Falcone}, {Kobayashi},
  {Zhang}, \& {et al.}}]{2005Sci...309.1833B}
{Burrows}, D.~N., {Romano}, P., {Falcone}, A., {et~al.} 2005, Science, 309,
  1833, \dodoi{10.1126/science.1116168}

\bibitem[{{Chevalier} \& {Li}(2000)}]{2000ApJ...536..195C}
{Chevalier}, R.~A., \& {Li}, Z.-Y. 2000, \apj, 536, 195, \dodoi{10.1086/308914}

\bibitem[{{Dai} \& {Lu}(1998)}]{1998MNRAS.298...87D}
{Dai}, Z.~G., \& {Lu}, T. 1998, \mnras, 298, 87,
  \dodoi{10.1046/j.1365-8711.1998.01681.x}

\bibitem[{{Dai} {et~al.}(2006){Dai}, {Wang}, {Wu}, \&
  {Zhang}}]{2006Sci...311.1127D}
{Dai}, Z.~G., {Wang}, X.~Y., {Wu}, X.~F., \& {Zhang}, B. 2006, Science, 311,
  1127, \dodoi{10.1126/science.1123606}

\bibitem[{{Dainotti} {et~al.}(2021{\natexlab{a}}){Dainotti}, {Lenart},
  {Fraija}, {Nagataki}, {Warren}, {De Simone}, {Srinivasaragavan}, \&
  {Mata}}]{2021PASJ...73..970D}
{Dainotti}, M.~G., {Lenart}, A.~{\L}., {Fraija}, N., {et~al.}
  2021{\natexlab{a}}, \pasj, 73, 970, \dodoi{10.1093/pasj/psab057}

\bibitem[{{Dainotti} {et~al.}(2021{\natexlab{b}}){Dainotti}, {Omodei},
  {Srinivasaragavan}, {Vianello}, {Willingale}, {O'Brien}, {Nagataki},
  {Petrosian}, {Nuygen}, {Hernandez}, {Axelsson}, {Bissaldi}, \&
  {Longo}}]{2021ApJS..255...13D}
{Dainotti}, M.~G., {Omodei}, N., {Srinivasaragavan}, G.~P., {et~al.}
  2021{\natexlab{b}}, \apjs, 255, 13, \dodoi{10.3847/1538-4365/abfe17}

\bibitem[{{Dichiara} {et~al.}(2020){Dichiara}, {Troja}, {O'Connor}, {Marshall},
  {Beniamini}, {Cannizzo}, {Lien}, \& {Sakamoto}}]{2020MNRAS.492.5011D}
{Dichiara}, S., {Troja}, E., {O'Connor}, B., {et~al.} 2020, \mnras, 492, 5011,
  \dodoi{10.1093/mnras/staa124}

\bibitem[{{Dirirsa} {et~al.}(2016){Dirirsa}, {Vianello}, {Racusin}, \&
  {Axelsson}}]{2016GCN..19586...1D}
{Dirirsa}, F., {Vianello}, G., {Racusin}, J., \& {Axelsson}, M. 2016, GRB
  Coordinates Network, 19586

\bibitem[{{Duncan} \& {Thompson}(1992)}]{1992ApJ...392L...9D}
{Duncan}, R.~C., \& {Thompson}, C. 1992, \apjl, 392, L9, \dodoi{10.1086/186413}

\bibitem[{{Eichler} \& {Levinson}(2004)}]{2004ApJ...614L..13E}
{Eichler}, D., \& {Levinson}, A. 2004, \apjl, 614, L13, \dodoi{10.1086/425310}

\bibitem[{{Evans} {et~al.}(2010){Evans}, {Willingale}, {Osborne}, {O'Brien},
  {Page}, \& {et al.}}]{2010A&A...519A.102E}
{Evans}, P.~A., {Willingale}, R., {Osborne}, J.~P., {et~al.} 2010, \aap, 519,
  A102, \dodoi{10.1051/0004-6361/201014819}

\bibitem[{{Fraija}(2015)}]{2015ApJ...804..105F}
{Fraija}, N. 2015, \apj, 804, 105, \dodoi{10.1088/0004-637X/804/2/105}

\bibitem[{{Fraija} {et~al.}(2019{\natexlab{a}}){Fraija}, {Barniol Duran},
  {Dichiara}, \& {Beniamini}}]{2019ApJ...883..162F}
{Fraija}, N., {Barniol Duran}, R., {Dichiara}, S., \& {Beniamini}, P.
  2019{\natexlab{a}}, \apj, 883, 162, \dodoi{10.3847/1538-4357/ab3ec4}

\bibitem[{{Fraija} {et~al.}(2022{\natexlab{a}}){Fraija}, {Dainotti}, {Ugale},
  {Jyoti}, \& {Warren}}]{2022ApJ...934..188F}
{Fraija}, N., {Dainotti}, M.~G., {Ugale}, S., {Jyoti}, D., \& {Warren}, D.~C.
  2022{\natexlab{a}}, \apj, 934, 188, \dodoi{10.3847/1538-4357/ac7a9c}

\bibitem[{{Fraija} {et~al.}(2019{\natexlab{b}}){Fraija}, {Dichiara},
  {Pedreira}, {Galvan-Gamez}, {Becerra}, {Barniol Duran}, \&
  {Zhang}}]{2019ApJ...879L..26F}
{Fraija}, N., {Dichiara}, S., {Pedreira}, A.~C. C. d. E.~S., {et~al.}
  2019{\natexlab{b}}, \apjl, 879, L26, \dodoi{10.3847/2041-8213/ab2ae4}

\bibitem[{{Fraija} {et~al.}(2022{\natexlab{b}}){Fraija}, {Galvan-Gamez},
  {Betancourt Kamenetskaia}, {Dainotti}, {Dichiara}, {Veres}, {Becerra}, \& {do
  E.~S. Pedreira}}]{2022ApJ...940..189F}
{Fraija}, N., {Galvan-Gamez}, A., {Betancourt Kamenetskaia}, B., {et~al.}
  2022{\natexlab{b}}, \apj, 940, 189, \dodoi{10.3847/1538-4357/ac68e1}

\bibitem[{{Fraija} {et~al.}(2020){Fraija}, {Laskar}, {Dichiara}, {Beniamini},
  {Duran}, {Dainotti}, \& {Becerra}}]{2020ApJ...905..112F}
{Fraija}, N., {Laskar}, T., {Dichiara}, S., {et~al.} 2020, \apj, 905, 112,
  \dodoi{10.3847/1538-4357/abc41a}

\bibitem[{{Fraija} {et~al.}(2016){Fraija}, {Lee}, \&
  {Veres}}]{2016ApJ...818..190F}
{Fraija}, N., {Lee}, W., \& {Veres}, P. 2016, \apj, 818, 190,
  \dodoi{10.3847/0004-637X/818/2/190}

\bibitem[{{Fraija} {et~al.}(2017{\natexlab{a}}){Fraija}, {Lee}, {Araya},
  {Veres}, {Barniol Duran}, \& {Guiriec}}]{2017ApJ...848...94F}
{Fraija}, N., {Lee}, W.~H., {Araya}, M., {et~al.} 2017{\natexlab{a}}, \apj,
  848, 94, \dodoi{10.3847/1538-4357/aa8d65}

\bibitem[{{Fraija} {et~al.}(2019{\natexlab{c}}){Fraija}, {Lopez-Camara},
  {Pedreira}, {Betancourt Kamenetskaia}, {Veres}, \&
  {Dichiara}}]{2019ApJ...884...71F}
{Fraija}, N., {Lopez-Camara}, D., {Pedreira}, A.~C. C. d. E.~S., {et~al.}
  2019{\natexlab{c}}, \apj, 884, 71, \dodoi{10.3847/1538-4357/ab40a9}

\bibitem[{{Fraija} {et~al.}(2019{\natexlab{d}}){Fraija}, {Pedreira}, \&
  {Veres}}]{2019ApJ...871..200F}
{Fraija}, N., {Pedreira}, A.~C.~C.~d.~E.~S., \& {Veres}, P. 2019{\natexlab{d}},
  \apj, 871, 200, \dodoi{10.3847/1538-4357/aaf80e}

\bibitem[{{Fraija} {et~al.}(2021){Fraija}, {Veres}, {Beniamini},
  {Galvan-Gamez}, {Metzger}, {Barniol Duran}, \&
  {Becerra}}]{2021ApJ...918...12F}
{Fraija}, N., {Veres}, P., {Beniamini}, P., {et~al.} 2021, \apj, 918, 12,
  \dodoi{10.3847/1538-4357/ac0aed}

\bibitem[{{Fraija} {et~al.}(2017{\natexlab{b}}){Fraija}, {Veres}, {Zhang},
  {Barniol Duran}, {Becerra}, {Zhang}, {Lee}, {Watson}, {Ordaz-Salazar}, \&
  {Galvan-Gamez}}]{2017ApJ...848...15F}
{Fraija}, N., {Veres}, P., {Zhang}, B.~B., {et~al.} 2017{\natexlab{b}}, \apj,
  848, 15, \dodoi{10.3847/1538-4357/aa8a72}

\bibitem[{{Fraija} {et~al.}(2019{\natexlab{e}}){Fraija}, {Dichiara},
  {Pedreira}, {Galvan-Gamez}, {Becerra}, {Montalvo}, {Montero}, {Betancourt
  Kamenetskaia}, \& {Zhang}}]{2019ApJ...885...29F}
{Fraija}, N., {Dichiara}, S., {Pedreira}, A.~C. C. d. E.~S., {et~al.}
  2019{\natexlab{e}}, \apj, 885, 29, \dodoi{10.3847/1538-4357/ab3e4b}

\bibitem[{{Gao} {et~al.}(2015){Gao}, {Ding}, {Wu}, {Dai}, \&
  {Zhang}}]{2015ApJ...807..163G}
{Gao}, H., {Ding}, X., {Wu}, X.-F., {Dai}, Z.-G., \& {Zhang}, B. 2015, \apj,
  807, 163, \dodoi{10.1088/0004-637X/807/2/163}

\bibitem[{{Ghisellini} {et~al.}(2010){Ghisellini}, {Ghirlanda}, {Nava}, \&
  {Celotti}}]{2010MNRAS.403..926G}
{Ghisellini}, G., {Ghirlanda}, G., {Nava}, L., \& {Celotti}, A. 2010, \mnras,
  403, 926, \dodoi{10.1111/j.1365-2966.2009.16171.x}

\bibitem[{{Granot} {et~al.}(2018){Granot}, {Gill}, {Guetta}, \& {De
  Colle}}]{2018MNRAS.481.1597G}
{Granot}, J., {Gill}, R., {Guetta}, D., \& {De Colle}, F. 2018, \mnras, 481,
  1597, \dodoi{10.1093/mnras/sty2308}

\bibitem[{{Granot} {et~al.}(2017){Granot}, {Guetta}, \&
  {Gill}}]{2017ApJ...850L..24G}
{Granot}, J., {Guetta}, D., \& {Gill}, R. 2017, \apjl, 850, L24,
  \dodoi{10.3847/2041-8213/aa991d}

\bibitem[{{Granot} {et~al.}(2002){Granot}, {Panaitescu}, {Kumar}, \&
  {Woosley}}]{2002ApJ...570L..61G}
{Granot}, J., {Panaitescu}, A., {Kumar}, P., \& {Woosley}, S.~E. 2002, \apjl,
  570, L61, \dodoi{10.1086/340991}

\bibitem[{{Guetta} {et~al.}(2001){Guetta}, {Spada}, \&
  {Waxman}}]{2001ApJ...557..399G}
{Guetta}, D., {Spada}, M., \& {Waxman}, E. 2001, \apj, 557, 399,
  \dodoi{10.1086/321543}

\bibitem[{{H.~E.~S.~S. Collaboration}(2021)}]{2021arXiv210602510H}
{H.~E.~S.~S. Collaboration}. 2021, arXiv e-prints, arXiv:2106.02510.
\newblock \doarXiv{2106.02510}

\bibitem[{{Hajela} {et~al.}(2021){Hajela}, {Margutti}, {Bright}, {Alexander},
  {Metzger}, {Nedora}, {Kathirgamaraju}, {Margalit}, {Radice}, {Berger},
  {MacFadyen}, {Giannios}, {Chornock}, {Heywood}, {Sironi}, {Gottlieb},
  {Coppejans}, {Laskar}, {Cendes}, {Barniol Duran}, {Eftekhari}, {Fong},
  {McDowell}, {Nicholl}, {Xie}, {Zrake}, {Bernuzzi}, {Broekgaarden},
  {Kilpatrick}, {Terreran}, {Villar}, {Blanchard}, {Gomez}, {Hosseinzadeh},
  {Matthews}, \& {Rastinejad}}]{2021arXiv210402070H}
{Hajela}, A., {Margutti}, R., {Bright}, J.~S., {et~al.} 2021, arXiv e-prints,
  arXiv:2104.02070.
\newblock \doarXiv{2104.02070}

\bibitem[{{Hotokezaka} {et~al.}(2018){Hotokezaka}, {Kiuchi}, {Shibata},
  {Nakar}, \& {Piran}}]{2018ApJ...867...95H}
{Hotokezaka}, K., {Kiuchi}, K., {Shibata}, M., {Nakar}, E., \& {Piran}, T.
  2018, \apj, 867, 95, \dodoi{10.3847/1538-4357/aadf92}

\bibitem[{{Huang} {et~al.}(1999){Huang}, {Dai}, \& {Lu}}]{1999MNRAS.309..513H}
{Huang}, Y.~F., {Dai}, Z.~G., \& {Lu}, T. 1999, \mnras, 309, 513,
  \dodoi{10.1046/j.1365-8711.1999.02887.x}

\bibitem[{{Huang} {et~al.}(2000){Huang}, {Gou}, {Dai}, \&
  {Lu}}]{2000ApJ...543...90H}
{Huang}, Y.~F., {Gou}, L.~J., {Dai}, Z.~G., \& {Lu}, T. 2000, \apj, 543, 90,
  \dodoi{10.1086/317076}

\bibitem[{{Ioka} \& {Nakamura}(2018)}]{2018PTEP.2018d3E02I}
{Ioka}, K., \& {Nakamura}, T. 2018, Progress of Theoretical and Experimental
  Physics, 2018, 043E02, \dodoi{10.1093/ptep/pty036}

\bibitem[{{Izzo} {et~al.}(2020){Izzo}, {Auchettl}, {Hjorth}, {De Colle},
  {Gall}, {Angus}, {Raimundo}, \& {Ramirez-Ruiz}}]{2020A&A...639L..11I}
{Izzo}, L., {Auchettl}, K., {Hjorth}, J., {et~al.} 2020, \aap, 639, L11,
  \dodoi{10.1051/0004-6361/202038152}

\bibitem[{{Kasliwal} {et~al.}(2017){Kasliwal}, {Nakar}, {Singer}, {Kaplan},
  {Cook}, {Van Sistine}, {Lau}, {Fremling}, {Gottlieb}, {Jencson}, {Adams},
  {Feindt}, {Hotokezaka}, {Ghosh}, {Perley}, {Yu}, {Piran}, {Allison},
  {Anupama}, {Balasubramanian}, {Bannister}, {Bally}, {Barnes}, {Barway},
  {Bellm}, {Bhalerao}, {Bhattacharya}, {Blagorodnova}, {Bloom}, {Brady},
  {Cannella}, {Chatterjee}, {Cenko}, {Cobb}, {Copperwheat}, {Corsi}, {De},
  {Dobie}, {Emery}, {Evans}, {Fox}, {Frail}, {Frohmaier}, {Goobar}, {Hallinan},
  {Harrison}, {Helou}, {Hinderer}, {Ho}, {Horesh}, {Ip}, {Itoh}, {Kasen},
  {Kim}, {Kuin}, {Kupfer}, {Lynch}, {Madsen}, {Mazzali}, {Miller}, {Mooley},
  {Murphy}, {Ngeow}, {Nichols}, {Nissanke}, {Nugent}, {Ofek}, {Qi}, {Quimby},
  {Rosswog}, {Rusu}, {Sadler}, {Schmidt}, {Sollerman}, {Steele}, {Williamson},
  {Xu}, {Yan}, {Yatsu}, {Zhang}, \& {Zhao}}]{2017Sci...358.1559K}
{Kasliwal}, M.~M., {Nakar}, E., {Singer}, L.~P., {et~al.} 2017, Science, 358,
  1559, \dodoi{10.1126/science.aap9455}

\bibitem[{{Kouveliotou} {et~al.}(1993){Kouveliotou}, {Meegan}, {Fishman},
  {Bhat}, {Briggs}, {Koshut}, {Paciesas}, \& {Pendleton}}]{1993ApJ...413L.101K}
{Kouveliotou}, C., {Meegan}, C.~A., {Fishman}, G.~J., {et~al.} 1993, \apjl,
  413, L101, \dodoi{10.1086/186969}

\bibitem[{{Kumar} \& {Barniol Duran}(2009)}]{2009MNRAS.400L..75K}
{Kumar}, P., \& {Barniol Duran}, R. 2009, \mnras, 400, L75,
  \dodoi{10.1111/j.1745-3933.2009.00766.x}

\bibitem[{{Kumar} \& {Barniol Duran}(2010)}]{2010MNRAS.409..226K}
---. 2010, \mnras, 409, 226, \dodoi{10.1111/j.1365-2966.2010.17274.x}

\bibitem[{{Kumar} \& {Zhang}(2015)}]{2015PhR...561....1K}
{Kumar}, P., \& {Zhang}, B. 2015, \physrep, 561, 1,
  \dodoi{10.1016/j.physrep.2014.09.008}

\bibitem[{{Lamb} \& {Kobayashi}(2017)}]{2017MNRAS.472.4953L}
{Lamb}, G.~P., \& {Kobayashi}, S. 2017, \mnras, 472, 4953,
  \dodoi{10.1093/mnras/stx2345}

\bibitem[{{Lamb} \& {Kobayashi}(2018)}]{2018MNRAS.478..733L}
---. 2018, \mnras, 478, 733, \dodoi{10.1093/mnras/sty1108}

\bibitem[{{Lazzati} {et~al.}(2017){Lazzati}, {L{\'o}pez-C{\'a}mara},
  {Cantiello}, {Morsony}, {Perna}, \& {Workman}}]{2017ApJ...848L...6L}
{Lazzati}, D., {L{\'o}pez-C{\'a}mara}, D., {Cantiello}, M., {et~al.} 2017,
  \apjl, 848, L6, \dodoi{10.3847/2041-8213/aa8f3d}

\bibitem[{{L{\"u}} {et~al.}(2017){L{\"u}}, {Zhang}, {Zhong}, {Hou}, {Sun},
  {Rice}, \& {Liang}}]{2017ApJ...835..181L}
{L{\"u}}, H.-J., {Zhang}, H.-M., {Zhong}, S.-Q., {et~al.} 2017, \apj, 835, 181,
  \dodoi{10.3847/1538-4357/835/2/181}

\bibitem[{{Margutti} {et~al.}(2017){Margutti}, {Berger}, {Fong}, {Guidorzi},
  {Alexander}, {Metzger}, {Blanchard}, {Cowperthwaite}, {Chornock},
  {Eftekhari}, {Nicholl}, {Villar}, {Williams}, {Annis}, {Brown}, {Chen},
  {Doctor}, {Frieman}, {Holz}, {Sako}, \&
  {Soares-Santos}}]{2017ApJ...848L..20M}
{Margutti}, R., {Berger}, E., {Fong}, W., {et~al.} 2017, \apjl, 848, L20,
  \dodoi{10.3847/2041-8213/aa9057}

\bibitem[{{M{\'e}sz{\'a}ros} \& {Rees}(1997)}]{1997ApJ...476..232M}
{M{\'e}sz{\'a}ros}, P., \& {Rees}, M.~J. 1997, \apj, 476, 232

\bibitem[{{Metzger} {et~al.}(2011){Metzger}, {Giannios}, {Thompson},
  {Bucciantini}, \& {Quataert}}]{2011MNRAS.413.2031M}
{Metzger}, B.~D., {Giannios}, D., {Thompson}, T.~A., {Bucciantini}, N., \&
  {Quataert}, E. 2011, \mnras, 413, 2031,
  \dodoi{10.1111/j.1365-2966.2011.18280.x}

\bibitem[{{Moderski} {et~al.}(2000){Moderski}, {Sikora}, \&
  {Bulik}}]{2000ApJ...529..151M}
{Moderski}, R., {Sikora}, M., \& {Bulik}, T. 2000, \apj, 529, 151,
  \dodoi{10.1086/308257}

\bibitem[{{Mooley} {et~al.}(2018){Mooley}, {Deller}, {Gottlieb}, {Nakar},
  {Hallinan}, {Bourke}, {Frail}, {Horesh}, {Corsi}, \& {Hotokezaka}}]{mooley}
{Mooley}, K.~P., {Deller}, A.~T., {Gottlieb}, O., {et~al.} 2018, \nat, 561,
  355, \dodoi{10.1038/s41586-018-0486-3}

\bibitem[{{Nakar} {et~al.}(2009){Nakar}, {Ando}, \&
  {Sari}}]{2009ApJ...703..675N}
{Nakar}, E., {Ando}, S., \& {Sari}, R. 2009, \apj, 703, 675,
  \dodoi{10.1088/0004-637X/703/1/675}

\bibitem[{{Nakar} \& {Piran}(2018)}]{2018MNRAS.478..407N}
{Nakar}, E., \& {Piran}, T. 2018, \mnras, 478, 407,
  \dodoi{10.1093/mnras/sty952}

\bibitem[{{Nakar} {et~al.}(2002){Nakar}, {Piran}, \&
  {Granot}}]{2002ApJ...579..699N}
{Nakar}, E., {Piran}, T., \& {Granot}, J. 2002, \apj, 579, 699,
  \dodoi{10.1086/342791}

\bibitem[{{Paczy{\'n}ski}(1998)}]{1998ApJ...494L..45P}
{Paczy{\'n}ski}, B. 1998, \apjl, 494, L45, \dodoi{10.1086/311148}

\bibitem[{{Palatiello} {et~al.}(2017){Palatiello}, {Noda}, {Inoue}, {Colin},
  {Moretti}, {Longo}, {MAGIC Collaboration}, \& {Fermi
  Collaboration}}]{2017ifs..confE..84P}
{Palatiello}, M., {Noda}, K., {Inoue}, S., {et~al.} 2017, in Proceedings of the
  7th International Fermi Symposium, 84, \dodoi{10.22323/1.312.0084}

\bibitem[{{Panaitescu}(2005)}]{2005MNRAS.362..921P}
{Panaitescu}, A. 2005, \mnras, 362, 921,
  \dodoi{10.1111/j.1365-2966.2005.09352.x}

\bibitem[{{Panaitescu} \& {Kumar}(2002)}]{2002ApJ...571..779P}
{Panaitescu}, A., \& {Kumar}, P. 2002, \apj, 571, 779, \dodoi{10.1086/340094}

\bibitem[{{Perley} {et~al.}(2009){Perley}, {Metzger}, {Granot}, {Butler},
  {Sakamoto}, {Ramirez-Ruiz}, {Levan}, {Bloom}, {Miller}, {Bunker}, {Chen},
  {Filippenko}, {Gehrels}, {Glazebrook}, {Hall}, {Hurley}, {Kocevski}, {Li},
  {Lopez}, {Norris}, {Piro}, {Poznanski}, {Prochaska}, {Quataert}, \&
  {Tanvir}}]{2009ApJ...696.1871P}
{Perley}, D.~A., {Metzger}, B.~D., {Granot}, J., {et~al.} 2009, \apj, 696,
  1871, \dodoi{10.1088/0004-637X/696/2/1871}

\bibitem[{{Perna} {et~al.}(2006){Perna}, {Armitage}, \&
  {Zhang}}]{2006ApJ...636L..29P}
{Perna}, R., {Armitage}, P.~J., \& {Zhang}, B. 2006, \apjl, 636, L29,
  \dodoi{10.1086/499775}

\bibitem[{{Piran}(1999)}]{1999PhR...314..575P}
{Piran}, T. 1999, \physrep, 314, 575, \dodoi{10.1016/S0370-1573(98)00127-6}

\bibitem[{{Piran} \& {Nakar}(2010)}]{2010ApJ...718L..63P}
{Piran}, T., \& {Nakar}, E. 2010, \apjl, 718, L63,
  \dodoi{10.1088/2041-8205/718/2/L63}

\bibitem[{{Planck Collaboration} {et~al.}(2016){Planck Collaboration}, {Ade},
  {Aghanim}, {Arnaud}, {Ashdown}, {Aumont}, \& {et al.}}]{2016A&A...594A..13P}
{Planck Collaboration}, {Ade}, P.~A.~R., {Aghanim}, N., {et~al.} 2016, \aap,
  594, A13, \dodoi{10.1051/0004-6361/201525830}

\bibitem[{{Proga} \& {Zhang}(2006)}]{2006MNRAS.370L..61P}
{Proga}, D., \& {Zhang}, B. 2006, \mnras, 370, L61,
  \dodoi{10.1111/j.1745-3933.2006.00189.x}

\bibitem[{{Ramirez-Ruiz} {et~al.}(2005){Ramirez-Ruiz}, {Garcia-Segura},
  {Salmonson}, \& {Perez-Rendon}}]{2005ApJ...631..435R}
{Ramirez-Ruiz}, E., {Garcia-Segura}, G., {Salmonson}, J.~D., \& {Perez-Rendon},
  B. 2005, \apjl, 631, 435, \dodoi{10.1086/432433}

\bibitem[{{Resmi} {et~al.}(2018){Resmi}, {Schulze}, {Ishwara-Chandra}, {Misra},
  {Buchner}, {De Pasquale}, {S{\'a}nchez-Ram{\'\i}rez}, {Klose}, {Kim},
  {Tanvir}, \& {O'Brien}}]{2018ApJ...867...57R}
{Resmi}, L., {Schulze}, S., {Ishwara-Chandra}, C.~H., {et~al.} 2018, \apj, 867,
  57, \dodoi{10.3847/1538-4357/aae1a6}

\bibitem[{{Rhoads}(1999)}]{1999ApJ...525..737R}
{Rhoads}, J.~E. 1999, \apj, 525, 737, \dodoi{10.1086/307907}

\bibitem[{{Santana} {et~al.}(2014){Santana}, {Barniol Duran}, \&
  {Kumar}}]{2014ApJ...785...29S}
{Santana}, R., {Barniol Duran}, R., \& {Kumar}, P. 2014, \apj, 785, 29,
  \dodoi{10.1088/0004-637X/785/1/29}

\bibitem[{{Sari} \& {Esin}(2001)}]{2001ApJ...548..787S}
{Sari}, R., \& {Esin}, A.~A. 2001, \apj, 548, 787, \dodoi{10.1086/319003}

\bibitem[{{Sari} {et~al.}(1999){Sari}, {Piran}, \&
  {Halpern}}]{1999ApJ...519L..17S}
{Sari}, R., {Piran}, T., \& {Halpern}, J.~P. 1999, \apjl, 519, L17,
  \dodoi{10.1086/312109}

\bibitem[{{Sari} {et~al.}(1998){Sari}, {Piran}, \&
  {Narayan}}]{1998ApJ...497L..17S}
{Sari}, R., {Piran}, T., \& {Narayan}, R. 1998, \apjl, 497, L17,
  \dodoi{10.1086/311269}

\bibitem[{{Sato} {et~al.}(2021){Sato}, {Obayashi}, {Yamazaki}, {Murase}, \&
  {Ohira}}]{2021MNRAS.504.5647S}
{Sato}, Y., {Obayashi}, K., {Yamazaki}, R., {Murase}, K., \& {Ohira}, Y. 2021,
  \mnras, 504, 5647, \dodoi{10.1093/mnras/stab1273}

\bibitem[{{Srinivasaragavan} {et~al.}(2020){Srinivasaragavan}, {Dainotti},
  {Fraija}, {Hernandez}, {Nagataki}, {Lenart}, {Bowden}, \&
  {Wagner}}]{2020ApJ...903...18S}
{Srinivasaragavan}, G.~P., {Dainotti}, M.~G., {Fraija}, N., {et~al.} 2020,
  \apj, 903, 18, \dodoi{10.3847/1538-4357/abb702}

\bibitem[{{Tak} {et~al.}(2019){Tak}, {Omodei}, {Uhm}, {Racusin}, {Asano}, \&
  {McEnery}}]{2019ApJ...883..134T}
{Tak}, D., {Omodei}, N., {Uhm}, Z.~L., {et~al.} 2019, \apj, 883, 134,
  \dodoi{10.3847/1538-4357/ab3982}

\bibitem[{{Thompson}(1994)}]{1994MNRAS.270..480T}
{Thompson}, C. 1994, \mnras, 270, 480, \dodoi{10.1093/mnras/270.3.480}

\bibitem[{{Troja} {et~al.}(2017){Troja}, {Lipunov}, {Mundell}, \&
  et~al.}]{2017Natur.547..425T}
{Troja}, E., {Lipunov}, V.~M., {Mundell}, C.~G., \& et~al. 2017, \nat, 547, 425

\bibitem[{Troja {et~al.}(2017)Troja, Piro, van Eerten, \& et~al.}]{troja2017a}
Troja, E., Piro, L., van Eerten, H., \& et~al. 2017, Nature, 000, 1,
  \dodoi{10.1038/nature24290}

\bibitem[{{Troja} {et~al.}(2016){Troja}, {Sakamoto}, {Cenko}, {Lien},
  {Gehrels}, {Castro-Tirado}, {Ricci}, {Capone}, {Toy}, {Kutyrev}, {Kawai},
  {Cucchiara}, {Fruchter}, {Gorosabel}, {Jeong}, {Levan}, {Perley},
  {Sanchez-Ramirez}, {Tanvir}, \& {Veilleux}}]{2016ApJ...827..102T}
{Troja}, E., {Sakamoto}, T., {Cenko}, S.~B., {et~al.} 2016, \apj, 827, 102,
  \dodoi{10.3847/0004-637X/827/2/102}

\bibitem[{{Troja} {et~al.}(2018){Troja}, {Ryan}, {Piro}, {van Eerten}, {Cenko},
  {Yoon}, {Lee}, {Im}, {Sakamoto}, {Gatkine}, {Kutyrev}, \&
  {Veilleux}}]{2018NatCo...9.4089T}
{Troja}, E., {Ryan}, G., {Piro}, L., {et~al.} 2018, Nature Communications, 9,
  4089, \dodoi{10.1038/s41467-018-06558-7}

\bibitem[{{Troja} {et~al.}(2019){Troja}, {Castro-Tirado}, {Becerra
  Gonz{\'a}lez}, {Hu}, {Ryan}, {Cenko}, {Ricci}, {Novara},
  {S{\'a}nchez-R{\'a}mirez}, {Acosta-Pulido}, {Ackley}, {Caballero
  Garc{\'\i}a}, {Eikenberry}, {Guziy}, {Jeong}, {Lien}, {M{\'a}rquez},
  {Pandey}, {Park}, {Sakamoto}, {Tello}, {Sokolov}, {Sokolov}, {Tiengo},
  {Valeev}, {Zhang}, \& {Veilleux}}]{2019MNRAS.489.2104T}
{Troja}, E., {Castro-Tirado}, A.~J., {Becerra Gonz{\'a}lez}, J., {et~al.} 2019,
  \mnras, 489, 2104, \dodoi{10.1093/mnras/stz2255}

\bibitem[{{Usov}(1992)}]{1992Natur.357..472U}
{Usov}, V.~V. 1992, \nat, 357, 472, \dodoi{10.1038/357472a0}

\bibitem[{{Valeev} \& {et al.}(2019)}]{2019GCN.25565....1V}
{Valeev}, A., \& {et al.} 2019, GRB Coordinates Network, Circular Service,
  No.~25565, \#1 (2019), 25565

\bibitem[{{van Marle} {et~al.}(2006){van Marle}, {Langer}, {Achterberg}, \&
  {Garcia-Segura}}]{2006A&A...460..105V}
{van Marle}, A.~J., {Langer}, N., {Achterberg}, A., \& {Garcia-Segura}, G.
  2006, \aaps, 460, 105, \dodoi{10.1051/0004-6361:20065709}

\bibitem[{{von Kienlin} {et~al.}(2019){von Kienlin}, {Veres}, {Roberts},
  {Hamburg}, {Bissaldi}, {Briggs}, {Burns}, {Goldstein}, {Kocevski}, {Preece},
  {Wilson-Hodge}, {Hui}, {Mailyan}, \& {Malacaria}}]{2019ApJ...876...89V}
{von Kienlin}, A., {Veres}, P., {Roberts}, O.~J., {et~al.} 2019, \apj, 876, 89,
  \dodoi{10.3847/1538-4357/ab10d8}

\bibitem[{{Wang} {et~al.}(2010){Wang}, {He}, {Li}, {Wu}, \&
  {Dai}}]{2010ApJ...712.1232W}
{Wang}, X.-Y., {He}, H.-N., {Li}, Z., {Wu}, X.-F., \& {Dai}, Z.-G. 2010, \apj,
  712, 1232, \dodoi{10.1088/0004-637X/712/2/1232}

\bibitem[{{Waxman} \& {Bahcall}(1997)}]{1997PhRvL..78.2292W}
{Waxman}, E., \& {Bahcall}, J. 1997, \prl, 78, 2292,
  \dodoi{10.1103/PhysRevLett.78.2292}

\bibitem[{{Weinberg}(1972)}]{1972gcpa.book.....W}
{Weinberg}, S. 1972, {Gravitation and Cosmology}

\bibitem[{{Wijers} \& {Galama}(1999)}]{1999ApJ...523..177W}
{Wijers}, R.~A.~M.~J., \& {Galama}, T.~J. 1999, \apj, 523, 177,
  \dodoi{10.1086/307705}

\bibitem[{{Woosley}(1993)}]{1993ApJ...405..273W}
{Woosley}, S.~E. 1993, \apj, 405, 273, \dodoi{10.1086/172359}

\bibitem[{{Wu} {et~al.}(2005){Wu}, {Dai}, {Huang}, \&
  {Lu}}]{2005ApJ...619..968W}
{Wu}, X.~F., {Dai}, Z.~G., {Huang}, Y.~F., \& {Lu}, T. 2005, \apj, 619, 968,
  \dodoi{10.1086/426666}

\bibitem[{{Xu} {et~al.}(2016){Xu}, {Malesani}, {Fynbo}, {Tanvir}, {Levan}, \&
  {Perley}}]{2016GCN..19600...1X}
{Xu}, D., {Malesani}, D., {Fynbo}, J.~P.~U., {et~al.} 2016, GRB Coordinates
  Network, 19600

\bibitem[{{Yost} {et~al.}(2003){Yost}, {Harrison}, {Sari}, \&
  {Frail}}]{2003ApJ...597..459Y}
{Yost}, S.~A., {Harrison}, F.~A., {Sari}, R., \& {Frail}, D.~A. 2003, \apj,
  597, 459, \dodoi{10.1086/378288}

\bibitem[{{Zhang}(2019)}]{2019arXiv191109862Z}
{Zhang}, B. 2019, arXiv e-prints, arXiv:1911.09862.
\newblock \doarXiv{1911.09862}

\bibitem[{{Zhang} {et~al.}(2006){Zhang}, {Fan}, {Dyks}, {Kobayashi},
  {M{\'e}sz{\'a}ros}, {Burrows}, {Nousek}, \& {Gehrels}}]{2006ApJ...642..354Z}
{Zhang}, B., {Fan}, Y.~Z., {Dyks}, J., {et~al.} 2006, \apj, 642, 354,
  \dodoi{10.1086/500723}

\bibitem[{{Zhang} {et~al.}(2007){Zhang}, {Liang}, {Page}, {Grupe}, {Zhang},
  {Barthelmy}, {Burrows}, {Campana}, {Chincarini}, {Gehrels}, {Kobayashi},
  {M{\'e}sz{\'a}ros}, {Moretti}, {Nousek}, {O'Brien}, {Osborne}, {Roming},
  {Sakamoto}, {Schady}, \& {Willingale}}]{2007ApJ...655..989Z}
{Zhang}, B., {Liang}, E., {Page}, K.~L., {et~al.} 2007, \apj, 655, 989,
  \dodoi{10.1086/510110}

\bibitem[{{Zhang} {et~al.}(2016){Zhang}, {Zhang}, {Castro-Tirado}, {Dai},
  {Tam}, {Wang}, {Hu}, {Karpov}, {Pozanenko}, {Zhang}, {Mazaeva}, {Minaev},
  {Volnova}, {Oates}, {Gao}, {Wu}, {Shao}, {Tang}, {Beskin}, {Biryukov},
  {Bondar}, {Ivanov}, {Katkova}, {Orekhova}, {Perkov}, {Sasyuk}, {Mankiewicz},
  {{\.Z}arnecki}, {Cwiek}, {Opiela}, {Zadro{\.z}ny}, {Aptekar}, {Frederiks},
  {Svinkin}, {Kusakin}, {Inasaridze}, {Burhonov}, {Rumyantsev}, {Klunko},
  {Moskvitin}, {Fatkhullin}, {Sokolov}, {Valeev}, {Jeong}, {Park},
  {Caballero-Garc{\'{\i}}a}, {Cunniffe}, {Tello}, {Ferrero}, {Pandey},
  {Jel{\'{\i}}nek}, {S{\'a}nchez-Ram{\'{\i}}rez}, \&
  {Castell{\'o}n}}]{2016arXiv161203089Z}
{Zhang}, B.-B., {Zhang}, B., {Castro-Tirado}, A.~J., {et~al.} 2016, ArXiv
  e-prints.
\newblock \doarXiv{1612.03089}

\bibitem[{{Zhang} {et~al.}(2017){Zhang}, {Jin}, {Wang}, \&
  {Wei}}]{2017ApJ...835...73Z}
{Zhang}, S., {Jin}, Z.-P., {Wang}, Y.-Z., \& {Wei}, D.-M. 2017, \apj, 835, 73,
  \dodoi{10.3847/1538-4357/835/1/73}

\end{thebibliography}
\bibliographystyle{aasjournal}

\addcontentsline{toc}{chapter}{Bibliography}

\begin{table}[h!]
	\centering \renewcommand{\arraystretch}{1.55}\addtolength{\tabcolsep}{1pt}
	\caption{CRs of the synchrotron off-axis scenario in stellar-wind and constant-density medium for an adiabatic and radiative regime without energy injection.}
	\label{table1}
	\begin{tabular}{ccccccc}
		\hline
		&  & $\beta$ & $\alpha(p)$ & $\alpha(p)$ & $\alpha(\beta)$ & $\alpha(\beta)$ \\
		&  & & $(1<p<2)$ & $(2 < p )$ & $(1<p<2)$ & $(2 < p)$ \\
		\hline
		&  &  & ISM, Fast cooling &  &  &  \\
		\hline
		1 & $\nu^{\rm sync}_c<\nu<\nu^{\rm sync}_m$ & $\frac{1}{2}$ & $\frac{2\epsilon-13}{2-\epsilon}$ & $\frac{2\epsilon-13}{2-\epsilon}$ & $\frac{2(2\epsilon-13)\beta}{2-\epsilon}$ & $\frac{2(2\epsilon - 13)\beta}{2-\epsilon}$ \\
		2 & $\nu^{\rm sync}_m<\nu$ & $\frac{p}{2}$ & $-\frac{14+3p-4\epsilon}{2(2-\epsilon)}$ & $\frac{2\epsilon+3p-16}{2-\epsilon}$ & $-\frac{7+3\beta-2\epsilon}{2-\epsilon}$ & $\frac{2(\epsilon+3\beta-8)}{2-\epsilon}$ \\
		\hline
		&  &  & ISM, Slow cooling &  &  &  \\
		\hline
		3 & $\nu^{\rm sync}_{\rm m}<\nu<\nu^{\rm sync}_{\rm c}$ & $\frac{p - 1}{2}$ & $-\frac{3(4+p-2\epsilon)}{2(2-\epsilon)}$ & $\frac{3(p+\epsilon-5)}{2-\epsilon}$ & $-\frac{3[5+2(\beta-\epsilon)]}{2(2-\epsilon)}$ & $\frac{3[2(\beta-2)+\epsilon]}{2-\epsilon}$ \\
		4 & $\nu^{\rm sync}_{\rm c}<\nu$ & $\frac{p}{2}$ & $-\frac{14+3p-4\epsilon}{2(2-\epsilon)}$ & $\frac{2\epsilon+3p-16}{2-\epsilon}$ & $-\frac{7+3\beta-2\epsilon}{2-\epsilon}$ & $\frac{2(\epsilon+3\beta-8)}{2-\epsilon}$ \\
		\hline
		&  &  & Wind, Fast cooling &  &  &  \\
		\hline
		5 & $\nu^{\rm sync}_{\rm c}<\nu<\nu^{\rm sync}_{\rm m}$ & $\frac{1}{2}$ &$\frac{\epsilon - 8}{2(2-\epsilon)}$ & $\frac{\epsilon - 8}{2(2-\epsilon)}$ & $\frac{(\epsilon-8)\beta}{2-\epsilon}$ & $\frac{(\epsilon-8)\beta}{2-\epsilon}$ \\
		6 & $\nu^{\rm sync}_{\rm m}<\nu$ & $\frac{p}{2}$ & $-\frac{2+p}{2(2-\epsilon)}$ & $\frac{4(p-3)+ \epsilon(2-p)}{2(2-\epsilon)}$ & $-\frac{1+\beta}{2-\epsilon}$ & $\frac{2(2\beta-3)+\epsilon(1-\beta)}{2-\epsilon}$ \\
		\hline
		&  &  & Wind, Slow cooling &  &  &  \\
		\hline
		7 & $\nu^{\rm sync}_{\rm m}<\nu<\nu^{\rm sync}_{\rm c}$ & $\frac{p-1}{2}$ & $\frac{2-p-\epsilon}{2(2-\epsilon)}$ & $\frac{4(p-2) + \epsilon(1-p)}{2(2-\epsilon)}$ & $\frac{1-2\beta-\epsilon}{2(2-\epsilon)}$ & $\frac{2(2\beta-1)-\beta\epsilon}{2-\epsilon}$ \\
		8 & $\nu^{\rm sync}_{\rm c}<\nu$ & $\frac{p}{2}$ & $-\frac{2+p}{2(2-\epsilon)}$ & $\frac{4(p-3) + \epsilon(2-p)}{2(2-\epsilon)}$ & $-\frac{1+\beta}{2-\epsilon}$ & $\frac{2(2\beta-3)+\epsilon(1-\beta)}{2-\epsilon}$ \\
		\hline
	\end{tabular}
\end{table}

\begin{table}[h!]
\centering \renewcommand{\arraystretch}{1.55}\addtolength{\tabcolsep}{1pt}
\caption{The same as Table \ref{table1}, but for SSC off-axis scenario.}
\label{table2}
\begin{tabular}{ccccccc}
\hline
 &  & $\beta$ & $\alpha(p)$ & $\alpha(p)$ & $\alpha(\beta)$ & $\alpha(\beta)$ \\
 &  & & $(1<p<2)$ & $(2 < p) $ & $(1<p<2)$ & $(2 < p)$ \\
\hline
 &  &  & Wind, Slow cooling &  &  &  \\
\hline
1 & $\nu^{\rm ssc}_m<\nu<\nu^{\rm ssc}_c$ & $\frac{p-1}{2}$ & $\frac{14-4p-5\epsilon+p\epsilon}{2(2-\epsilon)}$ & $\frac{6(p-1)-\epsilon(p+1)}{2(2-\epsilon)}$ & $\frac{5-2\epsilon-\beta(4-\epsilon)}{2-\epsilon}$ & $\frac{6\beta-\epsilon(\beta+1)}{2-\epsilon}$ \\
2 & $\nu^{\rm ssc}_c<\nu$ & $\frac{p}{2}$ & $\frac{4-4p-2\epsilon+p\epsilon}{2(2-\epsilon)}$ & $\frac{6p-16+2\epsilon-p\epsilon}{2(2-\epsilon)}$ & $\frac{2-\epsilon-\beta(4-\epsilon)}{2-\epsilon}$ & $\frac{\beta(6-\epsilon)-8+\epsilon}{2-\epsilon}$ \\
\hline
 &  &  & Wind, Fast cooling &  &  &  \\
\hline
3 & $\nu^{\rm ssc}_c<\nu<\nu^{\rm ssc}_m$ & $\frac{1}{2}$ & $\frac{\epsilon-10}{2(2-\epsilon)}$ & $\frac{\epsilon-10}{2(2-\epsilon)}$ & $\frac{(\epsilon-10)\beta}{2-\epsilon}$ & $\frac{(\epsilon-10)\beta}{2-\epsilon}$ \\
4 & $\nu^{\rm ssc}_m<\nu$ & $\frac{p}{2}$ & $\frac{4-4p-2\epsilon+p\epsilon}{2(2-\epsilon)}$ & $\frac{6p-16+2\epsilon-p\epsilon}{2(2-\epsilon)}$ & $\frac{2-\epsilon-\beta(4-\epsilon)}{2-\epsilon}$ & $\frac{\beta(6-\epsilon)-8+\epsilon}{2-\epsilon}$ \\
\hline
&  &  & ISM, Slow cooling &  &  &  \\
\hline
5 & $\nu^{\rm ssc}_m<\nu<\nu^{\rm ssc}_c$ & $\frac{p-1}{2}$ &$\frac{4\epsilon-2-3p}{2-\epsilon}$ & $\frac{6p-4(5-\epsilon)}{2-\epsilon}$ & $\frac{4\epsilon-5-6\beta}{2-\epsilon}$ & $\frac{2(6\beta-7+2\epsilon)}{2-\epsilon}$ \\
6 & $\nu^{\rm ssc}_c<\nu$ & $\frac{p}{2}$ & $\frac{2\epsilon-3p-4}{2-\epsilon}$ & $\frac{2(3p+\epsilon-11)}{2-\epsilon}$ & $\frac{2(\epsilon-3\beta-2)}{2-\epsilon}$ & $\frac{2(6\beta+\epsilon-11)}{2-\epsilon}$ \\
\hline
 &  &  & ISM, Fast cooling &  &  &  \\
\hline
7 & $\nu^{\rm ssc}_c<\nu<\nu^{\rm ssc}_m$ & $\frac{1}{2}$ & $\frac{2(\epsilon-8)}{2-\epsilon}$ & $\frac{2(\epsilon-8)}{2-\epsilon}$ & $\frac{\epsilon-8}{(2-\epsilon)\beta}$ & $\frac{\epsilon-8}{(2-\epsilon)\beta}$ \\
8 & $\nu^{\rm ssc}_m<\nu$ & $\frac{p}{2}$ & $\frac{2\epsilon-3p-4}{2-\epsilon}$ & $\frac{2(3p+\epsilon-11)}{2-\epsilon}$ & $\frac{2(\epsilon-3\beta-2)}{2-\epsilon}$ & $\frac{2(6\beta+\epsilon-11)}{2-\epsilon}$ \\
\hline
\end{tabular}
\end{table}


\begin{table}[h!]
\centering \renewcommand{\arraystretch}{1.55}\addtolength{\tabcolsep}{1pt}
\caption{CRs of the synchrotron off-axis scenario in stellar-wind and constant-density medium for an adiabatic regime with energy injection.}
\label{table3}
\begin{tabular}{ccccccc}
\hline
 &  & $\beta$ & $\alpha(p)$ & $\alpha(p)$ & $\alpha(\beta)$ & $\alpha(\beta)$ \\
 &  & & $(1<p<2)$ & $(2 < p )$ & $(1<p<2)$ & $(2 < p )$ \\
\hline
 &  &  & ISM, Fast cooling &  &  &  \\
\hline
1 & $\nu^{\rm sync}_c<\nu<\nu^{\rm sync}_m$ & $\frac{1}{2}$ & $-\frac{10 + 3q}{2}$ & $-\frac{10 + 3q}{2}$ & $-(10+3q)\beta$ & $-(10+3q)\beta$ \\
2 & $\nu^{\rm sync}_m<\nu$ & $\frac{p}{2}$ & $-\frac{2(6+q) + p(2+q) }{4}$ & $\frac{p(2+q)-4(3+q)}{2}$ & $-\frac{6+q+\beta(2+q)}{2}$ & $\beta(2+q)-2(3+q)$ \\
\hline
 &  &  & ISM, Slow cooling &  &  &  \\
\hline
3 & $\nu^{\rm sync}_{\rm m}<\nu<\nu^{\rm sync}_{\rm c}$ & $\frac{p - 1}{2}$ & $-\frac{12+p(2+q)}{4}$  & $\frac{p(2+q)- 3(q+4)}{2}$  & $-\frac{14+q+2\beta(2+q)}{4}$ & $\beta(2+q)-5-q$ \\
4 & $\nu^{\rm sync}_{\rm c}<\nu$ & $\frac{p}{2}$ & $-\frac{2(6+q) + p(2+q) }{4}$ & $\frac{p(2+q)-4(3+q)}{2}$ & $-\frac{6+q+\beta(2+q)}{2}$ & $\beta(2+q)-2(3+q)$ \\
\hline
&  &  & Wind, Fast cooling &  &  &  \\
\hline
5 & $\nu^{\rm sync}_{\rm c}<\nu<\nu^{\rm sync}_{\rm m}$ & $\frac{1}{2}$ &$-\frac{1+3q}{2}$ & $-\frac{1+3q}{2}$ & $-(1+3q)\beta$ & $-(1+3q)\beta$ \\
6 & $\nu^{\rm sync}_{\rm m}<\nu$ & $\frac{p}{2}$ & $-\frac{q(2+p)}{4}$ & $\frac{(p-2)+ q(p-4)}{2}$ & $-\frac{q(1+\beta)}{2}$ & $(\beta-1)+q(\beta-2)$ \\
\hline
 &  &  & Wind, Slow cooling &  &  &  \\
\hline
7 & $\nu^{\rm sync}_{\rm m}<\nu<\nu^{\rm sync}_{\rm c}$ & $\frac{p-1}{2}$ & $\frac{2-pq}{4}$ & $\frac{(p-1) + q(p-3)}{2}$ & $\frac{2-(2\beta+1)q}{4}$ & $\beta(1+q)-q$ \\
8 & $\nu^{\rm sync}_{\rm c}<\nu$ & $\frac{p}{2}$ & $-\frac{q(2+p)}{4}$ & $\frac{(p-2)+ q(p-4)}{2}$ & $-\frac{q(1+\beta)}{2}$ & $(\beta-1)+q(\beta-2)$ \\
\hline
\end{tabular}
\end{table}

\begin{table}[h!]
\centering \renewcommand{\arraystretch}{1.55}\addtolength{\tabcolsep}{1pt}
\caption{The same as Table \ref{table3}, but for SSC off-axis scenario.}
\label{table4}
\begin{tabular}{ccccccc}
\hline
 &  & $\beta$ & $\alpha(p)$ & $\alpha(p)$ & $\alpha(\beta)$ & $\alpha(\beta)$ \\
 &  & & $(1<p<2)$ & $(2<p)$ & $(1<p<2)$ & $(2 < p)$ \\
\hline
 &  &  & Wind, Slow cooling &  &  &  \\
\hline
1 & $\nu^{\rm ssc}_m<\nu<\nu^{\rm ssc}_c$ & $\frac{p-1}{2}$ & $\frac{5-p+2q-pq}{2}$ & $\frac{1+p-4q+2pq}{2}$ & $\frac{4+q-2\beta(1+q)}{2}$ & $1-q+\beta(1+2q)$ \\
2 & $\nu^{\rm ssc}_c<\nu$ & $\frac{p}{2}$ & $\frac{2-p-pq}{2}$ & $\frac{p-2-6q+2pq}{2}$ & $1-\beta(1+q)$ & $\beta(1+2q)-(1+3q)$ \\
\hline
 &  &  & Wind, Fast cooling &  &  &  \\
\hline
3 & $\nu^{\rm ssc}_c<\nu<\nu^{\rm ssc}_m$ & $\frac{1}{2}$ & $-\frac{1+4q}{2}$ & $-\frac{1+4q}{2}$ & $-(1+4q)\beta$ & $-(1+4q)\beta$ \\
4 & $\nu^{\rm ssc}_m<\nu$ & $\frac{p}{2}$ & $\frac{2-p-pq}{2}$ & $\frac{p-2-6q+2pq}{2}$ & $1-\beta(1+q)$ & $\beta(1+2q)-(1+3q)$  \\
\hline
&  &  & ISM, Slow cooling &  &  &  \\
\hline
5 & $\nu^{\rm ssc}_m<\nu<\nu^{\rm ssc}_c$ & $\frac{p-1}{2}$ &$\frac{2q-pq-2p-4}{2}$ & $p(2+q)-2(4+q)$ & $\frac{q-6-2\beta(2+q)}{2}$ & $2\beta(2+q)-(6+q)$ \\
6 & $\nu^{\rm ssc}_c<\nu$ & $\frac{p}{2}$ & $-\frac{4+2p+pq}{2}$ & $p(2+q)-(8+3q)$ & $-[2+\beta(2+q)]$ & $2\beta(2+q)-(8+3q)$ \\
\hline
 &  &  & ISM, Fast cooling &  &  &  \\
\hline
7 & $\nu^{\rm ssc}_c<\nu<\nu^{\rm ssc}_m$ & $\frac{1}{2}$ & $-2(q+3)$ & $-2(q+3)$ & $-\frac{q+3}{\beta}$ & $-\frac{q+3}{\beta}$ \\
8 & $\nu^{\rm ssc}_m<\nu$ & $\frac{p}{2}$ & $-\frac{4+2p+pq}{2}$ & $p(2+q)-(8+3q)$ & $-[2+\beta(2+q)]$ & $2\beta(2+q)-(8+3q)$ \\
\hline
\end{tabular}
\end{table}


\begin{table}[h]
    \centering\renewcommand{\arraystretch}{1.4}
    \caption{The number of GRBs that fulfill each of the CRs of synchrotron off-axis scenario evolving in a stellar-wind and constant-density medium, together with their proportion.   The partially radiative scenario ($\epsilon=0.5$) with no energy injection was required.}
    \label{table5}
    \begin{tabular}{c c c c c c c}
    \toprule[1.2pt]
    \toprule[1.2pt]
     Type & Cooling & $\nu$ Range &  CR: 1 $<$ p $<$ 2  & CR:   $2 < p $   & GRBs Satisfying Relation & Proportion Satisfying Relation \\
     \midrule
    Wind & slow & ${\rm \nu_m^{\rm sync} < \nu < \nu_{c}^{\rm sync}}$ & $\frac{4-\epsilon+\beta(8-\epsilon)}{4-\epsilon}$ & $\frac{9-2\epsilon+\beta(\epsilon-2)}{4-\epsilon}$ & 14 & 16.09\% \\\cline{3-7}
     &  & ${\rm \nu_c^{\rm sync} < \nu}$ & $\frac{\epsilon-4+\beta(8-\epsilon)}{4-\epsilon}$ & $\frac{6-\epsilon+\beta(\epsilon-2)}{4-\epsilon}$ & 2 & 2.29\% \\
    \hline
     & fast & ${\rm \nu_c^{\rm sync} < \nu < \nu_{m}^{\rm sync}}$ & $\frac{\epsilon\beta}{4-\epsilon}$  & $\frac{\epsilon\beta}{4-\epsilon}$  & 0 & 0\% \\\cline{3-7}
    &  & ${\rm \nu_m^{\rm sync} < \nu }$ & $\frac{\epsilon-4+\beta(8-\epsilon)}{4-\epsilon}$  & $\frac{6-\epsilon+\beta(\epsilon-2)}{4-\epsilon}$  & 2 & 2.29\% \\
    \hline
    \hline
    ISM & slow & ${\rm \nu_m^{\rm sync} < \nu < \nu_{c}^{\rm sync}}$ & $\frac{2(2\epsilon+9\beta-1)}{8-\epsilon}$ & $-$ & 0 & 0.00\% \\\cline{3-7}
     &  & ${\rm \nu_c^{\rm sync} < \nu}$ & $\frac{2(\epsilon+9\beta-5)}{8-\epsilon}$ & $-$ & 0 & 0.00\% \\
    \hline
      & fast & ${\rm \nu_c^{\rm sync} < \nu < \nu_{m}^{sync}}$ & $\frac{2(2\epsilon-1)\beta}{8-\epsilon}$ & $\frac{2(2\epsilon-1)\beta}{8-\epsilon}$ & 0 & 0.00\% \\\cline{3-7}
    &  & ${\rm \nu_m^{\rm sync} < \nu}$ & $\frac{2(\epsilon+9\beta-5)}{8-\epsilon}$ & $-$ & 0 & 0.00\% \\
    \bottomrule
    \bottomrule
     \end{tabular}

\end{table}

\begin{table}[h]
    \centering\renewcommand{\arraystretch}{1.4}
    \caption{The same as Table \ref{table5}, but for SSC off-axis scenario.}
    \label{table6}
    \begin{tabular}{c c c c c c c}
    \toprule[1.2pt]
    \toprule[1.2pt]
     Type & Cooling & $\nu$ Range &  CR: 1 $<$ p $<$ 2  & CR:   $2 < p $   & GRBs Satisfying Relation & Proportion Satisfying Relation \\
     \midrule
 Wind & slow & ${\rm \nu_m^{\rm ssc} < \nu < \nu_{c}^{\rm ssc}}$ & $\frac{4-\epsilon+\beta(8-\epsilon)}{4-\epsilon}$ & $\frac{9-2\epsilon+\beta(\epsilon-2)}{4-\epsilon}$ & 1 & 1.14\% \\\cline{3-7}
     &  & ${\rm \nu_c^{\rm ssc} < \nu}$ & $\frac{\epsilon-4+\beta(8-\epsilon)}{4-\epsilon}$ & $\frac{6-\epsilon+\beta(\epsilon-2)}{4-\epsilon}$ & 2 & 2.29\% \\
    \hline
     & fast & ${\rm \nu_c^{\rm ssc} < \nu < \nu_{m}^{\rm ssc}}$ & $\frac{\epsilon\beta}{4-\epsilon}$  & $\frac{\epsilon\beta}{4-\epsilon}$  & 0 & 0.00\% \\\cline{3-7}
    &  & ${\rm \nu_m^{\rm ssc} < \nu }$ & $\frac{\epsilon-4+\beta(8-\epsilon)}{4-\epsilon}$  & $\frac{6-\epsilon+\beta(\epsilon-2)}{4-\epsilon}$  & 2 & 2.29\% \\
    \hline
    \hline
    ISM & slow & ${\rm \nu_m^{\rm ssc} < \nu < \nu_{c}^{\rm ssc}}$ & $\frac{2(2\epsilon+9\beta-1)}{8-\epsilon}$ & $-$ & 6 & 6.89\% \\\cline{3-7}
     &  & ${\rm \nu_c^{\rm ssc} < \nu}$ & $\frac{2(\epsilon+9\beta-5)}{8-\epsilon}$ & $-$ & 1 & 1.14\% \\
    \hline
      & fast & ${\rm \nu_c^{\rm ssc} < \nu < \nu_{m}^{ssc}}$ & $\frac{2(2\epsilon-1)\beta}{8-\epsilon}$ & $\frac{2(2\epsilon-1)\beta}{8-\epsilon}$ & 0 & 0.00\% \\\cline{3-7}
    &  & ${\rm \nu_m^{\rm ssc} < \nu}$ & $\frac{2(\epsilon+9\beta-5)}{8-\epsilon}$ & $-$ & 1 & 1.14
    \% \\
    \bottomrule
    \bottomrule
     \end{tabular}
\end{table}

\begin{table}[h]
    \centering\renewcommand{\arraystretch}{1.4}
    \caption{The number of GRBs that fulfill each of the CRs of synchrotron off-axis scenario evolving in a stellar-wind and constant-density medium, together with their proportion.   The adiabatic scenario with energy injection ($q=0.5$) was required.}
    \label{table7}
    \begin{tabular}{c c c c c c c}
    \toprule[1.2pt]
    \toprule[1.2pt]
     Type & Cooling & $\nu$ Range & CR: 1 $<$ p $<$ 2  & CR:  $2 \leq p $  & GRBs Satisfying Relation & Proportion Satisfying Relation \\
     \midrule
    Wind & slow & ${\rm \nu_m^{\rm sync} < \nu < \nu_{c}^{\rm sync}}$ & $\beta(q+1)+1$ & $\frac{2\beta(q-2)+q+8}{4}$ & 19 & 21.83\% \\\cline{3-7}
     & & ${\rm \nu_c^{\rm sync} < \nu}$ & $\beta(q+1)-1$ & $\frac{\beta(q-2)+q+2}{2}$  & 0 & 0.00\% \\
    \hline
     & fast & ${\rm \nu_c^{\rm sync} < \nu < \nu_{m}^{ssc}}$ & $\beta(q-1)$  & $\beta(q-1)$  & 0 & 0.00\% \\\cline{3-7}
     & & ${\rm \nu_m^{\rm sync} < \nu }$ & $\beta(q+1)-1$ & $\frac{\beta(q-2)+q+2}{2}$  & 0 & 0.00\% \\
    \hline
    \hline
     ISM & slow & ${\rm \nu_m^{\rm sync} < \nu < \nu_{c}^{\rm syn}}$ & $\frac{3\beta(q+2)+5q-6}{4}$ & $-$ & 0 & 0.00\% \\\cline{3-7}
     & & ${\rm \nu_c^{\rm sync} < \nu}$ & $\frac{3\beta(q+2)+q-6}{4}$  & $-$ & 0 & 0.00\% \\
    \hline
     & fast & ${\rm \nu_c^{\rm sync} < \nu < \nu_{m}^{ssc}}$ & $\frac{\beta(5q-6)}{4}$  & $\frac{\beta(5q-6)}{4}$ & 0 & 0.00\% \\\cline{3-7}
     &  & ${\rm \nu_m^{\rm sync} < \nu}$ & $\frac{3\beta(q+2)+q-6}{4}$ & $-$ & 0 & 0.00\% \\
    \bottomrule
    \bottomrule
    \end{tabular}
\end{table}

\begin{table}[h]
    \centering\renewcommand{\arraystretch}{1.4}
    \caption{The same as Table \ref{table7}, but for SSC off-axis scenario.}
    \label{table8}
    \begin{tabular}{c c c c c c c}
    \toprule[1.2pt]
    \toprule[1.2pt]
     Type & Cooling & $\nu$ Range &  CR: 1 $<$ p $<$ 2  & CR:  $2 < p $   & GRBs Satisfying Relation & Proportion Satisfying Relation \\
     \midrule
    Wind & slow & ${\rm \nu_m^{\rm ssc} < \nu < \nu_{c}^{\rm ssc}}$ & $\beta(q+1)+1$ & $\frac{2\beta(q-2)+q+8}{4}$ & 0 & 0\% \\\cline{3-7}
     & & ${\rm \nu_c^{\rm ssc} < \nu}$ & $\beta(q+1)-1$ & $\frac{\beta(q-2)+q+2}{2}$  & 3 & 3.44\% \\
    \hline
     & fast & ${\rm \nu_c^{\rm ssc} < \nu < \nu_{m}^{\rm ssc}}$ & $\beta(q-1)$  & $\beta(q-1)$  & 0 & 0.00\% \\\cline{3-7}
     & & ${\rm \nu_m^{\rm ssc} < \nu }$ & $\beta(q+1)-1$ & $\frac{\beta(q-2)+q+2}{2}$  & 3 & 3.44\% \\
    \hline
    \hline
     ISM & slow & ${\rm \nu_m^{\rm ssc} < \nu < \nu_{c}^{ssc}}$ & $\frac{3\beta(q+2)+5q-6}{4}$ & $-$ & 0 & 0.00\% \\\cline{3-7}
     & & ${\rm \nu_c^{\rm ssc} < \nu}$ & $\frac{3\beta(q+2)+q-6}{4}$  & $-$ & 0 & 0.00\% \\
    \hline
     & fast & ${\rm \nu_c^{\rm ssc} < \nu < \nu_{m}^{\rm ssc}}$ & $\frac{\beta(5q-6)}{4}$  & $\frac{\beta(5q-6)}{4}$ & 0 & 0.00\% \\\cline{3-7}
     &  & ${\rm \nu_m^{\rm ssc} < \nu}$ & $\frac{3\beta(q+2)+q-6}{4}$ & $-$ & 0 & 0.00\% \\
    \bottomrule
    \bottomrule
    \end{tabular}
\end{table}

\begin{table}[]
    \centering \renewcommand{\arraystretch}{1.5}\addtolength{\tabcolsep}{2pt}
    \caption{Results from MCMC simulations of an off-axis afterglow model evolving in a constant circumstellar medium.}
    \label{Table:ISM_Fit}
    \begin{tabular}{l c  }
    \hline
    \hline
      & GRB160625B \\ \hline \hline
     $\mathrm{log_{10}(E/erg)}$                & $53.89_{-0.49}^{+0.55}$  \\
     $\mathrm{log_{10}(n/cm^{-3})}$            & $0.99_{-0.50}^{+0.53}$   \\     
     $\mathrm{log_{10}(\varepsilon_{e})}$      & $-1.90_{-0.50}^{+0.55}$  \\     
     $\mathrm{log_{10}(\varepsilon_{B})}$      & $-2.68_{-0.45}^{+0.50}$  \\         
     $\mathrm{log_{10}(\Gamma_{0})}$           & $2.50_{-0.49}^{+0.59}$   \\         
     $\mathrm{\theta_{j}}$                     & $0.56_{-0.47}^{+0.50}$   \\
     $\mathrm{\theta_{obs}}$                   & $1.81_{-0.51}^{+0.46}$   \\          
     $\mathrm{p}$                              & $2.84_{-0.47}^{+0.46}$   \\
     $\mathrm{q}$                              & $0.22_{-0.50}^{+0.57}$   \\  \hline\hline
    \end{tabular}
\end{table}


\clearpage
\newpage

\newpage

\newpage

%

%

%

%

\end{document}